\pdfoutput=1
\newcommand*{\ATLASLATEXPATH}{}
\documentclass[atlasdraft=false, texlive=2016, UKenglish,LANGEDIT=false,cernpreprint]{\ATLASLATEXPATH atlasdoc}
 
\usepackage{\ATLASLATEXPATH atlaspackage}
\usepackage{\ATLASLATEXPATH atlasbiblatex}
\usepackage{\ATLASLATEXPATH atlasphysics}
 
\graphicspath{{logos/}{figures/}}
 
\def\approxprop{
\def\p{
\setbox0=\vbox{\hbox{$\propto$}}
\ht0=0.6ex \box0 }
\def\s{
\vbox{\hbox{$\sim$}}
}
\mathrel{\raisebox{0.7ex}{
\mbox{$\underset{\s}{\p}$}
}}
}
 
\usepackage{atlas-paper-defs}
 
\newcommand*{\pTrel}{\ensuremath{\pT^\text{rel}}\xspace}
\newcommand*{\pTjet}{\ensuremath{\pT^\text{jet}}\xspace}
 
\hypersetup{pdftitle={ATLAS document},pdfauthor={The ATLAS Collaboration}}
 
\PreprintIdNumber{CERN-EP-2019-090}
\AtlasJournalRef{Phys. Rev. D 100 (2019) 052011}
\AtlasDOI{10.1103/PhysRevD.100.052011}
\AtlasTitle{Properties of jet fragmentation using charged particles measured with the ATLAS detector in $pp$ collisions at $\sqrt{s}=13$~\TeV}
\AtlasAbstract{
This paper presents a measurement of quantities related to the formation of jets from high-energy quarks and gluons (fragmentation). Jets with transverse momentum $100$~\GeV~$<\pT<2.5$~\TeV~and pseudorapidity $|\eta| < 2.1$~from an integrated luminosity of 33 fb$^{-1}$ of $\sqrt{s}=13$~\TeV~proton--proton collisions are reconstructed with the ATLAS detector at the Large Hadron Collider.  Charged-particle tracks with $\pT > 500$~\MeV~and $|\eta| < 2.5$~are used to probe the detailed structure of the jet.  The fragmentation properties of the more forward and the more central of the two leading jets from each event are studied.  The data are unfolded to correct for detector resolution and acceptance effects.  Comparisons with parton shower Monte Carlo generators indicate that existing models provide a reasonable description of the data across a wide range of phase space, but there are also significant differences.  Furthermore, the data are interpreted in the context of quark- and gluon-initiated jets by exploiting the rapidity dependence of the jet flavor fraction.    A first measurement of the charged-particle multiplicity using model-independent jet labels (topic modeling) provides a promising alternative to traditional quark and gluon extractions using input from simulation.  The simulations provide a reasonable description of the quark-like data across the jet $\pT$ range presented in this measurement, but the gluon-like data have systematically fewer charged particles than the simulation}
 
\begin{document}
 
\maketitle
 
\section{Introduction}
\label{sec:intro}
 
Jets are collimated sprays of particles resulting from high-energy quark and gluon production.  The details of the process that underlies the \textit{fragmentation} of quarks and gluons with net quantum chromodynamic (QCD) charge into net neutral hadrons is not fully understood.  Jet formation is a complex multi-scale problem, including important contributions from QCD effects that cannot be described by perturbation theory.   Measuring basic quantities related to fragmentation is therefore essential to furthering our understanding of the emergent properties of QCD.
 
Perturbative and non-perturbative physically inspired models have free parameters that are tuned to data in order to best describe the radiation pattern inside jets~\cite{Buckley:2011ms}.  This is in turn an important input to all analyses at the Large Hadron Collider (LHC) due to the ubiquity of jets.  Measurements of jet substructure in proton--proton ($pp$) collisions at a center-of-mass energy of $\sqrt{s}=7$~\TeV~\cite{Aad:2011gn,Aad:2011kq,Aad:2013fba,ATLAS:2012am} have already been used by the ATLAS Collaboration for parameter optimizations (tunes) of the \PYTHIA~8 Monte Carlo (MC) generator~\cite{ATL-PHYS-PUB-2014-021}.  A measurement of the average number of charged particles inside jets at $\sqrt{s}=8$~\TeV~\cite{Aad:2016oit} was also used as input to recent developments in the \HERWIG~7 MC program~\cite{Reichelt:2017hts}.  Further measurements of jet constituent multiplicity and energy sharing will provide powerful constraints for future generator optimizations.
 
Quark- and gluon-initiated jets (henceforth quark and gluon jets) have different radiation patterns (see e.g.\ Ref.~\cite{Gallicchio:2012ez}).  As many analyses at the LHC target either quark-enriched or gluon-enriched processes, these radiation-pattern differences can be useful for jet tagging~\cite{Asquith:2018igt,Larkoski:2017jix}.  Measurements of jet structure can be used to calibrate quark-versus-gluon jet taggers.  By exploiting the rapidity dependence of the relative quark and gluon jet rates, ATLAS~\cite{Aad:2016oit} extracted the average charged-particle multiplicity for quark and gluon jets separately.  This was then combined with detector-level systematic uncertainties to provide quark/gluon tagger uncertainties at $\sqrt{s}=13$~\TeV~\cite{ATL-PHYS-PUB-2017-009}.  A more complex tagger based on several jet shapes could be calibrated in a similar manner using extended results.  The benefit of a particle-level measurement is that a portion of the calibration can be independent of ATLAS and LHC operating conditions.  Uncertainties in detector effects can be updated with the changing detector environment.  Adding more observables and measuring their differential distributions will improve this calibration.
 
Although the full radiation pattern inside jets is not calculable from first principles, the energy dependence of many observables can be calculated in perturbation theory. There have been significant theoretical advances in soft-collinear effective theory (SCET)~\cite{Bauer:2000ew,Bauer:2000yr,Bauer:2001ct,Bauer:2001yt} to derive factorization theorems that describe the evolution of universal non-perturbative functions~\cite{Chang:2013rca,Waalewijn2012sv,Krohn:2012fg,Elder:2017bkd}.  This was applied to the measurement of jet charge at $\sqrt{s}=8$~\TeV~\cite{Aad:2015cua}.  There have also been predictions and comparisons with the jet transverse momentum ($\pT$) dependence of the average number of charged particles inside jets (see Ref.~\cite{Aad:2016oit} and references therein).  This quantity does not have a perturbative expansion in the usual sense (as a series in $\alphas$); instead there is a series expansion in $\sqrt{\alphas}$~\cite{Capella:1999ms,Dremin:1999ji}.  This behavior is predicted for a wide class of \textit{Sudakov safe} observables~\cite{Larkoski:2015lea}.  At least for the case of charged-particle multiplicity, this non-standard expansion seems to be an excellent model of the data~\cite{Aad:2016oit}.
 
The goal of this paper is to measure properties of jet fragmentation using charged-particle tracks.  Such properties have been measured at many colliders at various center-of-mass energies, including the SPS~\cite{Arnison:167874,BAGNAIA1984291,ua22},  PETRA~\cite{Bartel:1982ub,zphys}, PEP~\cite{Derrick:163229,PhysRevLett.55.1954,PhysRevLett.72.3145,Abe:1996zi}, TRISTAN~\cite{PhysRevLett.63.1772}, CESR~\cite{PhysRevD.56.17}, LEP~\cite{Alexander:1991ce,Acton:1993jm,xyz,Buskulic1995389,Alexander1996659,Buskulic1996353,delphintrack,refId0,Abreu1999383,refId1,Akrawy1990617,opallep2}, HERA~\cite{Aid:1995up,Derrick:1995ca}, and the Tevatron~\cite{Affolder:2001jx,PhysRevD.68.012003,Aaltonen:2008ae,PhysRevLett.102.232002}.  Previous measurements by the ATLAS and CMS Collaborations were performed at $\sqrt{s}=2.76$~\TeV~\cite{Aaboud:2017bzv,Chatrchyan:2012gw}, $\sqrt{s}=5.02$~\TeV~\cite{Aaboud:2017tke,Aaboud:2018hpb,Aaboud:2019oac}, $\sqrt{s}=7$~\TeV~\cite{Aad:2011sc,Aad:2011gn,Chatrchyan:2012mec} and $\sqrt{s}=8$~\TeV~\cite{Aad:2016oit,Aad:2015cua,Sirunyan:2017tyr} in $pp$ collisions and are also compared with jet fragmentation measured in Pb+Pb collisions~\cite{Aaboud:2017bzv,Chatrchyan:2014ava,Aaboud:2017bzv,Aaboud:2017tke,Aaboud:2018hpb,Aaboud:2019oac,Aad:2014wha} and p+Pb collisions~\cite{Aaboud:2017tke}.  The measurement presented here represents a significant extension of previous work.  In particular, the accessible jet energy range is increased due to the larger $\sqrt{s}=13$~\TeV.  There are enough events in the 2016 dataset to probe the substructure of jets with $\pT$ up to 2.5~\TeV.
Next, the precision of the measurement has improved due to advances in track reconstruction inside jets during the long shutdown between LHC Runs 1 and 2, including the additional insertable B-layer (IBL) detector~\cite{ATLAS-TDR-2010-19,PIX-2018-001} and new algorithms for tracking inside dense environments~\cite{Aad:2014yva,Aaboud:2017all,myATL-PHYS-PUB-2017-016}.  Furthermore, detailed experimental studies to derive uncertainties in all aspects of tracking inside jets extend the capabilities of previous measurements to a wider region of phase space and also allow differential analyses~\cite{Aaboud:2017all,ATL-PHYS-PUB-2016-007}.  These new data therefore probe broader and deeper aspects of the radiation pattern inside jets across an extended phase space.
 
The paper is organized as follows.  Section~\ref{sec:obs} introduces the observables to be measured.  Then, following a brief description of the ATLAS detector in Section~\ref{sec:detector}, the data and simulation samples are documented in Section~\ref{sec:analysis}.  Charged-particle track, jet, and event reconstruction are detailed in Section~\ref{sec:obj}.  Corrections for detector effects (unfolding) are documented in Section~\ref{sec:unfold}.  A description of the corresponding systematic uncertainties can be found in Section~\ref{sec:systs} and the results are presented in Section~\ref{sec:result}.  Section~\ref{sec:conclusion} provides conclusions and future outlook.

\section{Observables}
\label{sec:obs}
 
This analysis builds upon the previous ATLAS jet structure measurements presented in Refs.~\cite{Aad:2011sc,Aad:2016oit,Aad:2015cua}.  The fundamental quantity is the fragmentation function $D_p^h(z,E)$, which describes the probability of finding a hadron $h$ with energy fraction $z$ of the parton $p$ that has energy $E$.  At a hadron collider, the jet  transverse momentum, $\pT$,\footnote{ATLAS uses a right-handed coordinate system with its origin at the nominal interaction point (IP)
in the center of the detector and the $z$-axis along the beam pipe.
The $x$-axis points from the IP to the center of the LHC ring,
and the $y$-axis points upwards.
Cylindrical coordinates $(r,\phi)$ are used in the transverse plane,
$\phi$ being the azimuthal angle around the $z$-axis.
The pseudorapidity is defined in terms of the polar angle $\theta$ as $\eta = -\ln \tan(\theta/2)$.
Angular distance is measured in units of $\Delta R \equiv \sqrt{(\Delta\eta)^{2} + (\Delta\phi)^{2}}$.} is a better proxy for the starting scale ($\mu$) of jet evolution.  To avoid confusion with previous measurements of similar observables, the transverse momentum fraction is denoted in this paper by the symbol $\zeta=\pT^\text{particle}/\pT^\text{jet}$.  The fragmentation function itself, like parton distribution functions (PDFs), cannot be calculated from first principles in perturbation theory.  However, it has a DGLAP~\cite{Altarelli:1977zs,Dokshitzer:1977sg,Gribov:1972ri} evolution and so the $\pT$ dependence of many observables can be calculated. In particular:
 
\begin{equation}
\label{eq:dglap}
\mu\frac{\partial }{\partial \mu}D_p^h(\zeta,\mu)=\sum_{p'}\int_\zeta^1\frac{d\zeta'}{\zeta'}\frac{\alphas(\mu) P_{p'\leftarrow p}(\zeta',\mu)}{\pi}D_{p'}^h\left(\frac{\zeta}{\zeta'},\mu\right),
\end{equation}
 
\noindent where $P_{p'\leftarrow p}(\zeta,\mu)$ are the Altarelli--Parisi splitting functions~\cite{Altarelli:1977zs} and depend on the scale $\mu$ through $\alphas$.  Charged particles are studied because they provide a way to measure single hadrons inside the jet (as opposed to calorimeter energy deposits, which can result from multiple particles) which gives access to $\sum_h D_p^h$.  A basic quantity related to the fragmentation function is the charged-particle multiplicity.  The average charged-particle multiplicity is an integral over $\zeta$ and a sum over $h$ and $p$ of $D_p^h(\zeta)$.  An extension of the multiplicity is the set of $\zeta$ moments of $D$.  The zeroth moment is the average multiplicity.   The full distribution of multiplicity depends on (multi-hadron) fragmentation functions in a complicated way; a more direct probe of $D$ is to measure hadron production as a function of $\zeta$, which is a sum of $D$ over $p$ and $h$ (but no integral over $\zeta$).  Additional observables are also studied in order to probe the angular spread of jet fragmentation beyond the collinear limit.  All of the observables are described below.
 
\begin{description}
\item \textbf{Charged-particle multiplicity} ($n_\text{ch}$): The number of charged particles inside a jet with $\pT$ above some threshold.  In terms of the fragmentation function:
 
\begin{equation*}
\langle n_\text{ch}\rangle(\pT^\text{jet})=\sum_pf_p(\pT^\text{jet})\sum_\text{$h$ charged}\int_{\text{threshold}/\pT^\text{jet}}^1d\zeta D_p^h(\zeta,\pT^\text{jet}),
\end{equation*}
 
\noindent where $f_p$ is the fraction of parton type $p$ at a given jet $\pT$.  The multiplicity is not calculable in perturbation theory, but to lowest order in $\sqrt{\alphas}$, the ratio of the multiplicity for quark-initiated jets to that for gluon-initiated jets is the ratio of color factors $C_A/C_F=9/4$.  The fraction of quark jets increases with $\pT$, which decreases the inclusive multiplicity.  However, this is compensated by an inherent increase in the multiplicity with $\pT$ for both quark and gluon jets~\cite{pinkbook}.   In addition to the mean, the full $(1/N_\text{jet})dN_\text{jet}/dn_\text{ch}$ distribution is measured.
 
\item \textbf{Summed fragmentation function}:  The distribution of the momentum fraction $\zeta$ is studied inside jets summed over charged-hadron types.  The quantity that is measured is $F(\zeta,\pT^\text{jet})=(1/N_\text{jet})dn_\text{ch}/d\zeta$.  In terms of the fragmentation function:
 
\begin{equation*}
F(\zeta,\pT^\text{jet})=\sum_pf_p(\pT^\text{jet})\sum_\text{$h$ charged} D_p^h(\zeta,\pT^\text{jet}).
\end{equation*}
 
\noindent By definition, $\int d\zeta F(\zeta)=\langle n_\text{ch}\rangle$.    In addition to measuring the distribution $F(\zeta)$ in bins of jet $\pT$, summary statistics of the $F(\zeta)$ distribution are extracted to show how the distribution evolves with jet $\pT$.  The following properties of the $\zeta$ distribution are extracted:
 
\begin{itemize}
\item Partial fractions of $F(\zeta)$: $\int_0^X F(\zeta)d\zeta/\int F(\zeta)d\zeta = n_\text{ch}(\zeta < X)/n_\text{ch}$ to show how much of the jet energy is carried by particles of a given $\pT$ fraction.  For illustration, the values considered are $X\in\{0.1,0.01,0.001\}$.  As $X\rightarrow 1$, these partial fractions become a constant value of $1.0$, independent of the jet $\pT$.
\item Moments of $F(\zeta)$: $\langle \zeta^\kappa\rangle = \int \zeta^\kappa F(\zeta)d\zeta/\int F(\zeta)d\zeta$.  The distribution of $F(\zeta)$ is nearly normally distributed in $\log(\zeta)$, which means that it is defined by its first two moments~\cite{pinkbook}.  For this reason, $\kappa=2$ is measured as a function of the jet $\pT$.  For illustration, the case $\kappa=1/2$ is also considered.
\item Weighted sums over the jet: $\langle \sum_{i\in\text{jet}} \zeta_i^\kappa\rangle = \int \zeta^\kappa F(\zeta)d\zeta$.  The values considered are $\kappa\in\{1/2,2\}$.  The observable $\sum_{i\in\text{jet}} \zeta_i^2$ is often called $p_\text{T}^\text{D}$ and can be used for quark/gluon jet tagging~\cite{Chatrchyan:2012sn}.  For a given jet type, these observables increase monotonically with increasing jet $p_\text{T}$ for $\kappa \lesssim 1$ and decrease  monotonically for $\kappa \gtrsim 1$ (see Section~\ref{sec:results:qg}); the $\kappa$ values chosen are representative of these trends.
\end{itemize}
 
Each of these derived quantities is extracted from the measured $F(\zeta)$ distribution.  More details about the procedure for unfolding these derived quantities are presented in Section~\ref{sec:unfold}.
 
\item \textbf{Transverse momentum}: $\pTrel\equiv \pT^\text{charged particle}\sin\Delta\phi$, where $\Delta\phi$ is the angle between the momentum of the constituent charged particle and the jet axis in the transverse plane.  The quantity that is measured is $f(\pTrel,\pTjet)=(1/N_\text{jet})dn_\text{ch}/d\pTrel$.  The average value is defined by $\langle p_\text{T}^\text{rel}\rangle=\int p_\text{T}^\text{rel} f(\pTrel)/\int f(\pTrel)$.
 
\item \textbf{Radial profile}: The number of charged particles in various annuli around the jet axis.  The quantity that is measured is $\rho_\text{ch}(r,\pT^\text{jet})=(1/N_\text{jet})dn_\text{ch}/2\pi rdr$, where $r=\Delta R(\text{charged particle},\text{jet})$.  The average value is defined by $\langle r\rangle=\int r\rho_\text{ch}(r)/\int\rho_\text{ch}(r)$.
\end{description}
 
The last two quantities are not simple derivatives of the fragmentation function as they additionally depend on finite opening angles encoded in the $d\theta/\theta$ emission phase space.  Since quantities are measured as a function of jet $\pt$ which is defined using charged and neutral particles, the observables are sensitive to the charged-to-neutral fraction inside jets. However, this fraction is robust to mis-modelling as isospin is an approximate symmetry of the strong force.
 
\section{ATLAS detector}
\label{sec:detector}
 
The ATLAS detector~\cite{PERF-2007-01} at the LHC covers nearly the entire solid angle around the collision point.  It consists of an inner tracking detector surrounded by a thin superconducting solenoid, electromagnetic and hadronic calorimeters,
and a muon spectrometer incorporating three large superconducting toroidal magnets.
The inner-detector system (ID) is immersed in a \SI{2}{\tesla} axial magnetic field
and provides charged-particle tracking in the range \(|\eta| < 2.5\).
 
The high-granularity silicon pixel detector covers the vertex region and typically provides four measurements per track,
the first hit being normally in the IBL.
It is followed by the silicon microstrip tracker (SCT) which usually provides eight measurements per track.
These silicon detectors are complemented by the transition radiation tracker (TRT),
which enables radially extended track reconstruction up to \(|\eta| = 2.0\).
 
The calorimeter system covers the pseudorapidity range \(|\eta| < 4.9\).
Within the region \(|\eta|< 3.2\), electromagnetic calorimetry is provided by barrel and
endcap high-granularity lead/liquid-argon (LAr) calorimeters,
with an additional thin LAr presampler covering \(|\eta| < 1.8\),
to correct for energy loss in material upstream of the calorimeters.
Hadronic calorimetry is provided by the steel/scintillating-tile calorimeter,
segmented into three barrel structures within \(|\eta| < 1.7\), and two copper/LAr hadronic endcap calorimeters.
The solid angle coverage is completed with forward copper/LAr and tungsten/LAr calorimeter modules
optimised for electromagnetic and hadronic measurements respectively.
 
Interesting events are selected to be recorded by the first-level trigger system implemented in custom hardware,
followed by selections made by algorithms implemented in software in the high-level trigger~\cite{TRIG-2016-01}.
The first-level trigger reduces the \SI{40}{\MHz} bunch crossing rate to below \SI{100}{\kHz},
which the high-level trigger further reduces in order to record events to disk at about \SI{1}{\kHz}.
 
\section{Datasets and simulated samples}
\label{sec:analysis}
 
These measurements use the dataset of $pp$ collisions recorded by the ATLAS detector in 2016, corresponding to an integrated luminosity of 33 \ifb~at a center-of-mass-energy of $\sqrt{s}=13$~\TeV.   Events are only considered if they are collected during stable beam conditions and satisfy all data quality requirements. Due to the high instantaneous luminosity and the large total inelastic proton--proton cross section, on average there are about 25 simultaneous (pileup) collisions in each bunch crossing.
 
The measurements presented in this paper use a variety of MC samples for estimating correction factors as well as for comparison with the corrected data. Dijet events were generated at leading order (LO) with \PYTHIA 8.186~\cite{Sjostrand:2007gs}, with the $2 \rightarrow 2$ matrix element convolved with the NNPDF2.3LO PDF set~\cite{Ball:2012cx} and using the A14 tune of multiple-parton-interaction and shower parameters~\cite{ATL-PHYS-PUB-2014-021}.  \PYTHIA uses a $\pT$-ordered parton shower model.  Additional dijet events were simulated using different generators, in order to study the impact of modeling uncertainties. \SHERPA 2.1~\cite{Gleisberg:2008ta} events were generated using multi-leg $2\rightarrow 2$ and $2\rightarrow 3$ matrix elements, which were matched to parton showers following the CKKW prescription~\cite{Catani:2001cc}. These \SHERPA events were simulated using the CT10 PDF set~\cite{Lai:2010vv} and the default \SHERPA event tune. \HERWIGpp 2.7~\cite{Bahr:2008pv,Corcella:2000bw} was used to provide a sample of events with an angle-ordered parton shower model. These events were generated with the $2\rightarrow 2$ matrix element, convolved with the CTEQ6L1 PDF set~\cite{Pumplin:2002vw} and configured with the UE-EE-5 tune~\cite{Seymour:2013qka}.
 
All simulated events were passed through a full simulation of the ATLAS detector~\cite{SOFT-2010-01} implemented in \GEANT 4~\cite{Agostinelli2003250}, which describes the interactions of particles with the detector and the subsequent digitization of analog signals. The effects of multiple simultaneous $pp$ collisions were simulated with inelastic $pp$ collisions using the \PYTHIA 8.186 generator with the A2~\cite{ATL-PHYS-PUB-2012-003} set of tuned parameters and the MSTW2008LO~\cite{Martin:2009iq} PDF set; these events were overlaid on the nominal dijet events.

\section{Object and event selection}
\label{sec:obj}
 
Since the data are unfolded to particle level, it is necessary to define both the particle-level and detector-level objects used in the measurement.   The former are chosen to be as close as possible to the latter in order to minimize the model dependence caused by an extrapolation from the measured phase space at detector level to the phase space at particle level.  Section~\ref{sec:objectdef} describes the definition of charged-particle tracks and jets.  Following the discussion of objects, Section~\ref{sec:events} describes the particle-level and detector-level event selection criteria.
 
\subsection{Object reconstruction}
\label{sec:objectdef}
 
While it is not possible to separate the underlying event from the hard scatter at particle level, it is possible to remove the contribution from pileup.  Therefore, the unfolding target is particle-level distributions produced in single proton--proton interactions.  However, at detector level, there is ambiguity about which $pp$ collision vertex corresponds to the hard-scatter event.  Collision vertices are reconstructed from tracks in the inner detector. Each vertex is required to be associated with at least two tracks with $\pT>0.4$~\GeV. The primary hard-scattering vertex of the event is chosen to be the one with the highest $\sum \pT^{2}$ calculated using all tracks associated with the vertex.
 
Particle-level jets are built from MC-simulated stable particles ($c\tau > 10$ mm) excluding muons and neutrinos.  By definition, particles from pileup and from interactions with the detectors are not included.  These jets are clustered using the anti-$k_t$~\cite{Cacciari:2008gp} algorithm with radius parameter $R=0.4$ as implemented in FastJet~\cite{Cacciari:2011ma}.  Detector-level jets are built from topological calorimeter-cell energy clusters~\cite{PERF-2014-07} using the same algorithm as is used at particle level.  A series of simulation- and data-based correction and calibration factors are applied to ensure that the resulting jet $\pT$ is the same as the particle-level value on average~\cite{Aaboud:2017jcu}.  Jets are required to have $\pT > 60$~\GeV\ so that the rate of jets originating from pileup is negligible.  The detector-level phase space includes one bin at low jet $\pT$ (60--100~\GeV) which is not in the fiducial phase space of the measurement due to the large impact of migrations into and out of the acceptance.
 
Charged particles are used to compute the particle-level definitions of all observables if they are clustered within a particle-level jet and have $\pT>500$~\MeV\ and $|\eta| < 2.5$.  The detector-level analog to charged particles is tracks.  Tracks are reconstructed from hits in the inner detector (see e.g. Ref.~\cite{Aaboud:2017all}) and a series of quality criteria are applied to the selected tracks to reject those originating from hits due to multiple charged particles (fake tracks) and from pileup.  The transverse momentum resolution is approximately $\sigma(\pT) /\pT \approx 0.05\% \times \pT/ \GeV \oplus 1\%$, with a significant degradation in the core of high-$\pT$ jets due to challenges associated with pattern recognition.\footnote{For example, the $\pT$ resolution is approximately 30\% at 100~\GeV~when five or more particles are within $\Delta R<0.015$.}  Tracks are required to pass the tight primary selection as well as the loose track-to-vertex association~\cite{ATL-PHYS-PUB-2015-051}.  In particular, they must have $\pT > 500$~\MeV\ and $|\eta| < 2.5$, and the number of pixel and strip clusters associated with the track is required to be at least 9 (11) for $|\eta| < 1.65$ ($\geq 1.65$).  In addition, the transverse impact parameter $d_0$ relative to the beamline must be less than 2~mm and the longitudinal impact parameter, $z_0$, is required to satisfy $|z_0\sin\theta|<3$~mm.  Tracks are matched to jets via ghost association ~\cite{Cacciari:2008gn}.  This matching procedure creates ghost versions of the tracks with the same direction but infinitesimal $\pT$.  Jet clustering is repeated and tracks are assigned to the jet that contains their ghosted version.  For the isolated, high $\pT$ jets used in this measurement, ghost association is nearly identical to a geometric matching based on $\Delta R < 0.4$.
 
\subsection{Event selection}
\label{sec:events}
 
Particle-level events are required to have at least two jets with $|\eta| < 2.1$ (within the tracking detector acceptance) and the leading two such jets must satisfy $\pT^\text{lead}/\pT^\text{sublead} < 1.5$.  This jet-$\pT$ balance requirement simplifies the interpretation of the final state in terms of a $2\rightarrow 2$ scattering process.
 
Detector-level events are selected using single-jet triggers. Due to the large cross section for jet production, most of the jet triggers are \textit{prescaled}: events that pass the trigger are randomly discarded with a fixed probability.  The trigger used for a particular jet $\pT$ is chosen to ensure that the trigger is 100\% efficient (for the measurement phase space and prior to prescaling) and has the lowest prescale factor.   Events in data are weighted by the prescale.  The lowest-threshold unprescaled jet trigger is used for jets with $\pT>600$~\GeV.  Detector-level events are required to pass the same selection requirements as particle-level events: there must be at least two offline calibrated jets with $|\eta| < 2.1$ and the leading two of these jets must satisfy $\pT^\text{lead}/\pT^\text{sublead} < 1.5$.  Figure~\ref{fig:pteta} shows the basic kinematic properties of the two leading jets passing this event selection compared with various MC predictions at detector level.
 
\begin{figure}[htbp]
\centering
\includegraphics[width=0.49\textwidth]{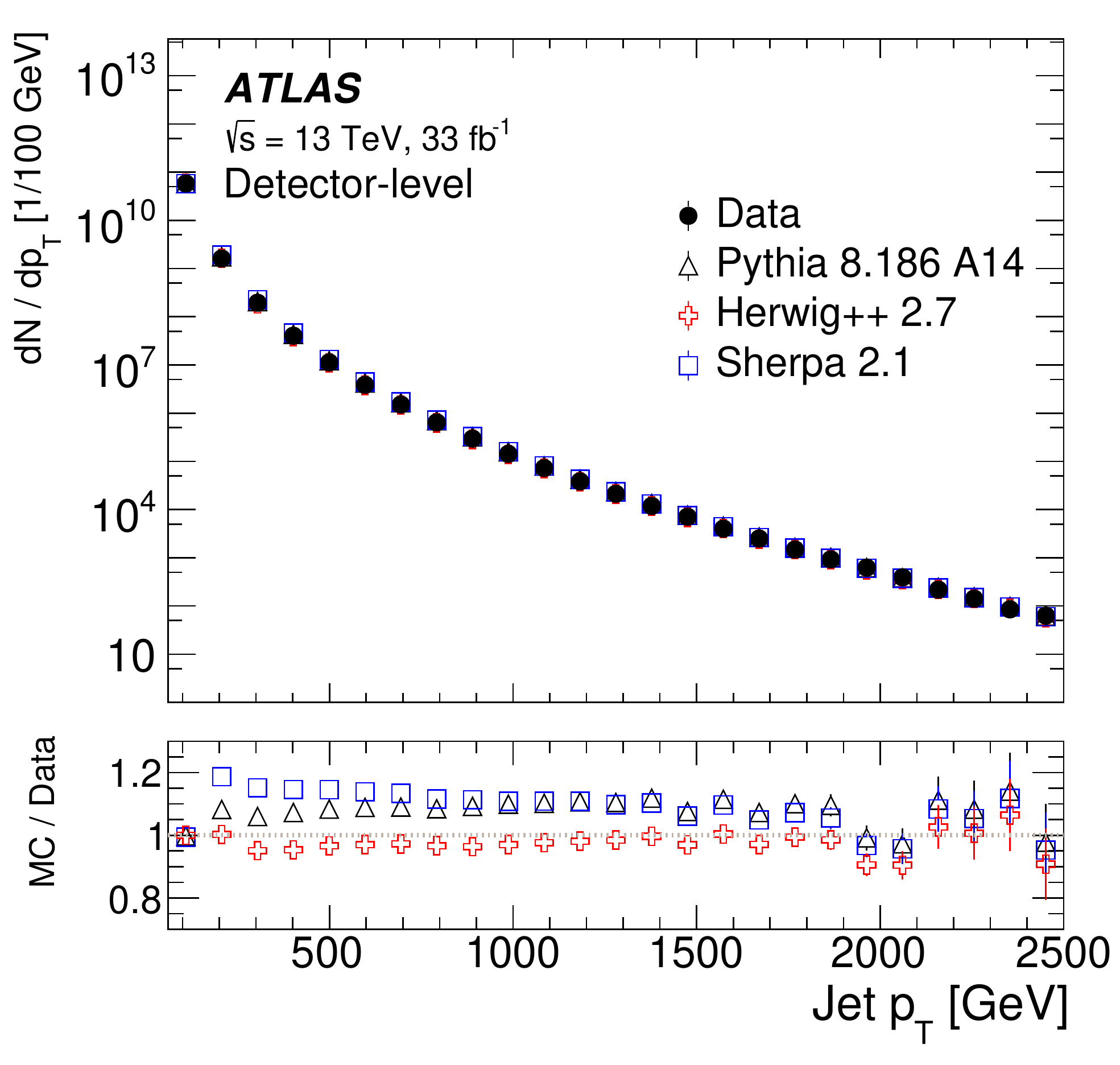}  \includegraphics[width=0.49\textwidth]{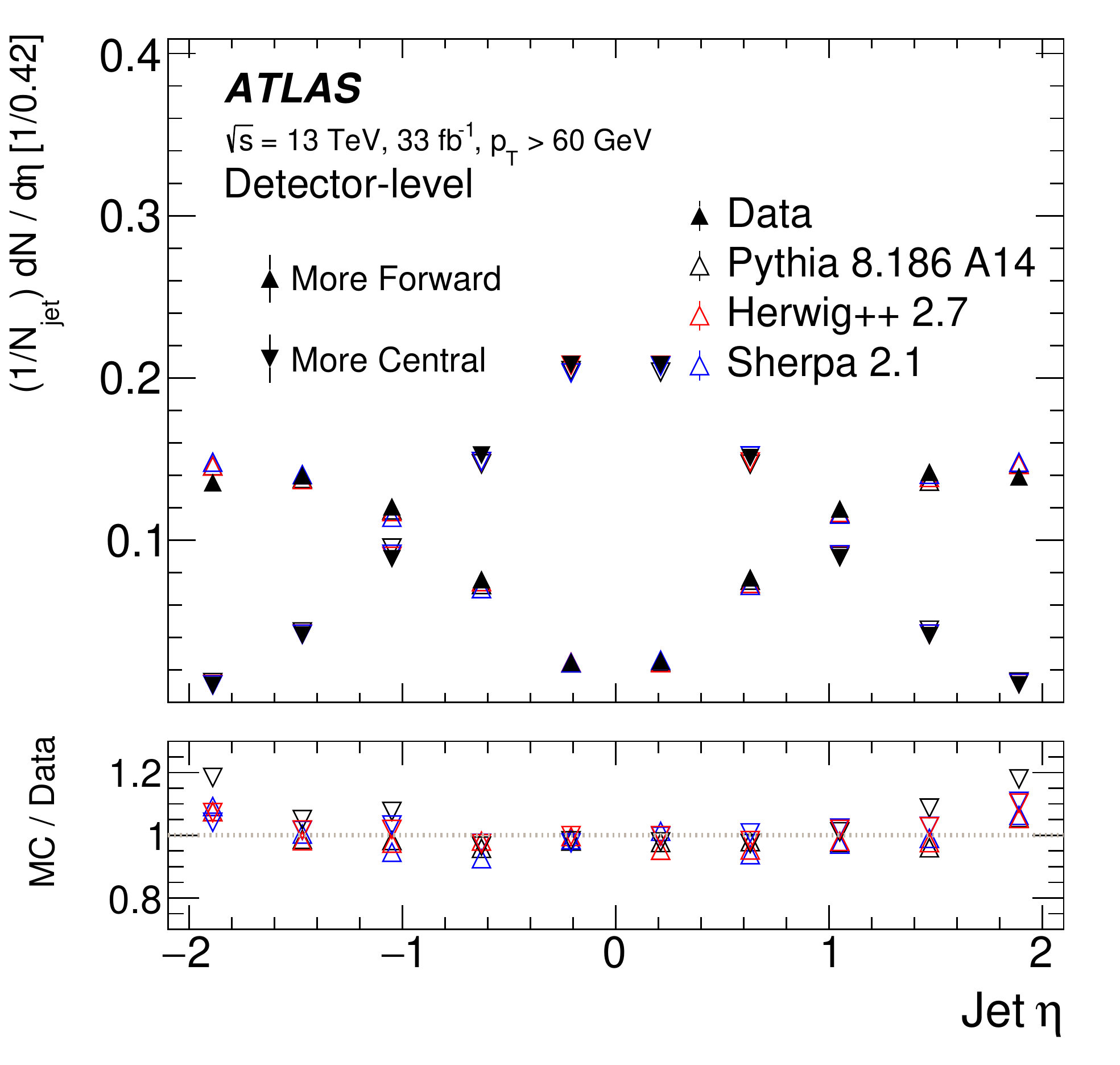}
\caption{Left: The transverse momentum, $\pT$, spectrum for the selected jets; the simulation is normalized to the data.  The normalization is dominated by the first bin, which accounts for the overall offset in the other bins for some of the predictions.  Right: The pseudorapidity, $\eta$, distribution for the selected jets, split into the more forward and the more central of the two jets.  Error bars only include the statistical uncertainty.}
\label{fig:pteta}
\end{figure}
 
The substructure of the two leading jets is used in the analysis.  Figure~\ref{fig:subs} shows detector-level distributions for a selection of the observables that were introduced in Section~\ref{sec:obs}.  For the jets with $\pT\sim 1$~\TeV\ shown in Figure~\ref{fig:subs}, the most probable number of tracks is about 15 and the most probable momentum fraction is about 1\%.  The radiation pattern is peaked at the center of the jet, so both the $\pTrel$ and $r$ distributions are peaked at zero.  The \PYTHIA, \HERWIGpp, and \SHERPA distributions generally bracket the data and are accurate to within about 20\%.
 
\begin{figure}[htbp]
\centering
\includegraphics[width=0.45\textwidth]{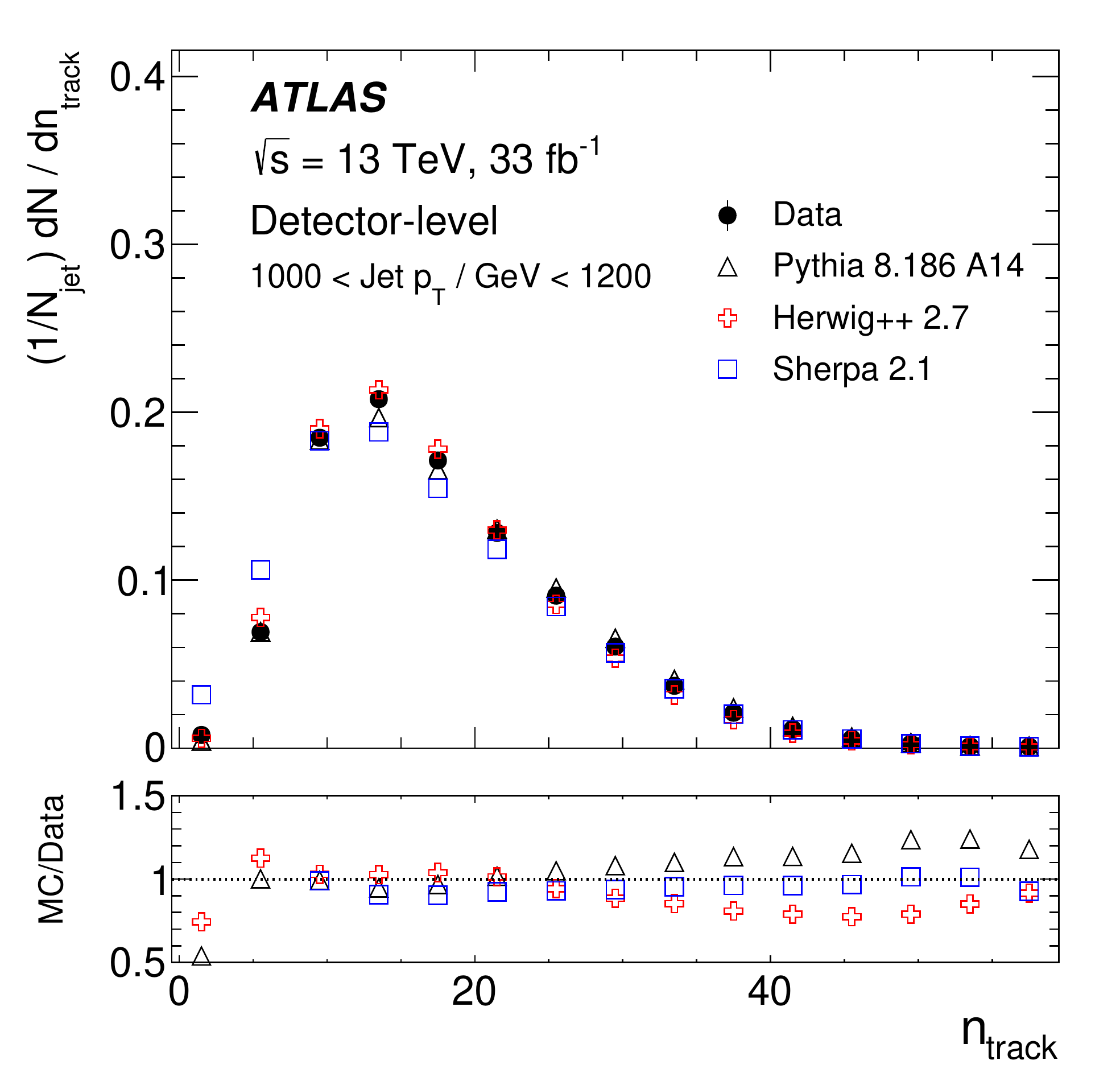}\includegraphics[width=0.45\textwidth]{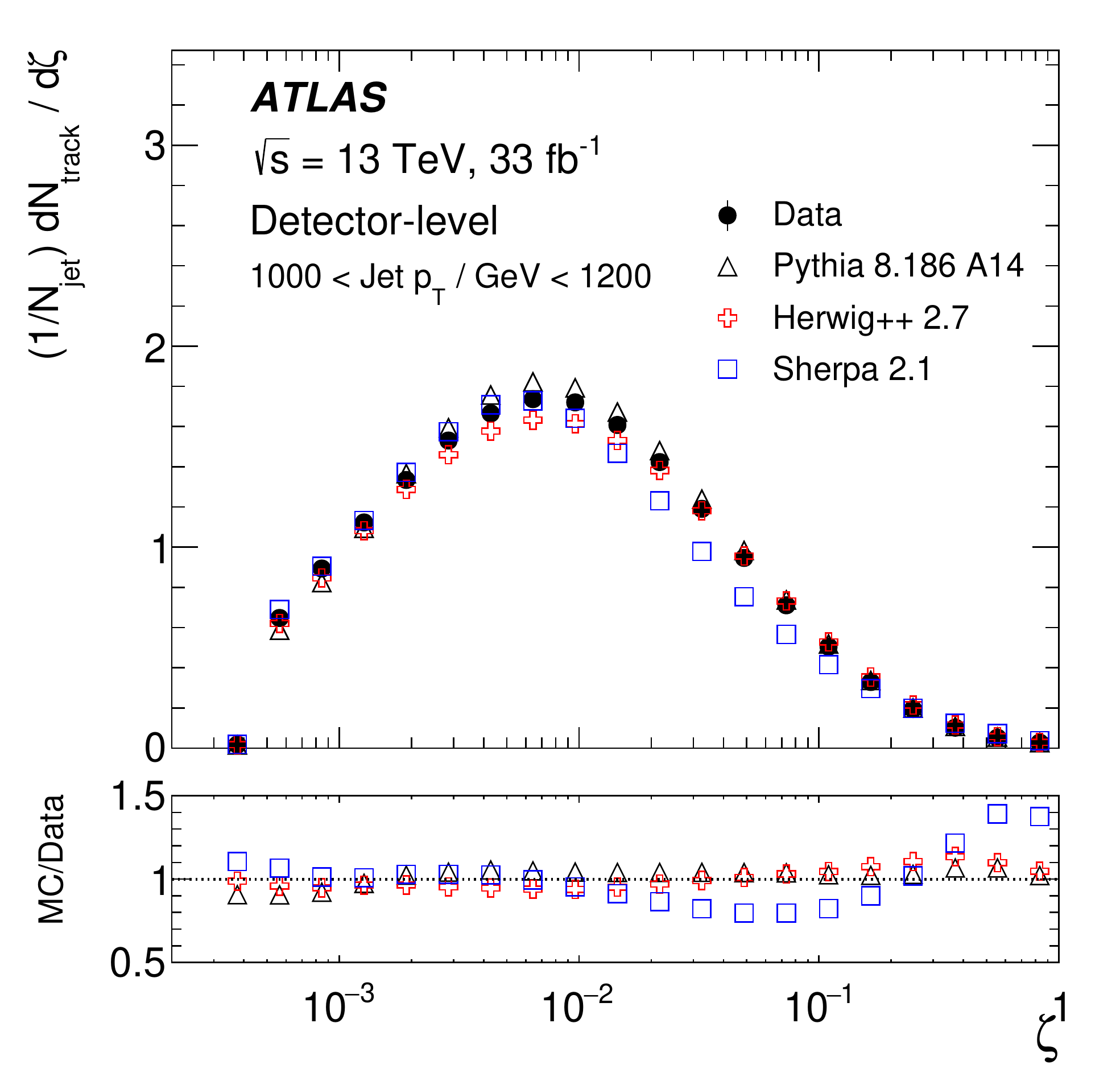}\\
\includegraphics[width=0.45\textwidth]{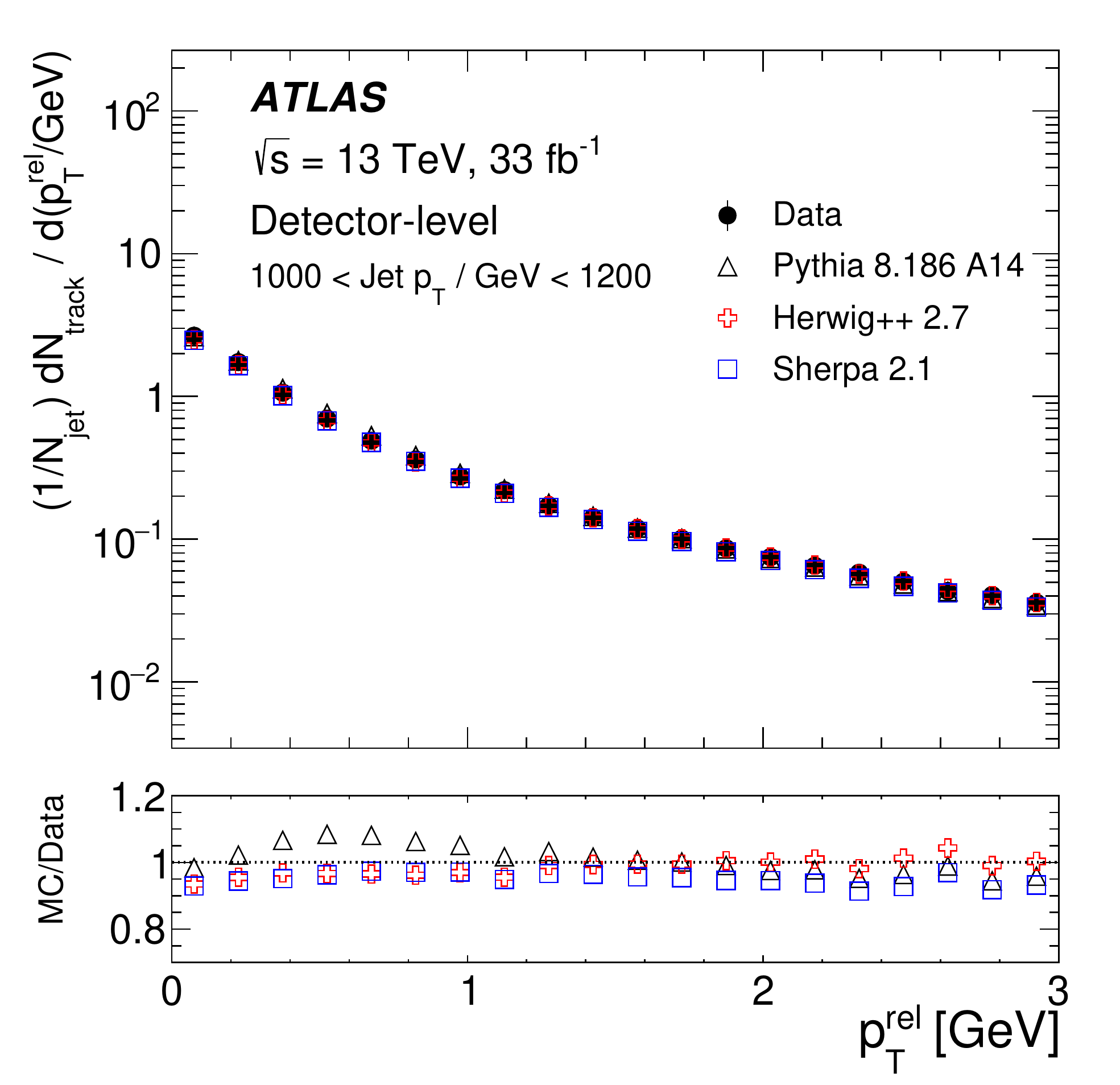}\includegraphics[width=0.45\textwidth]{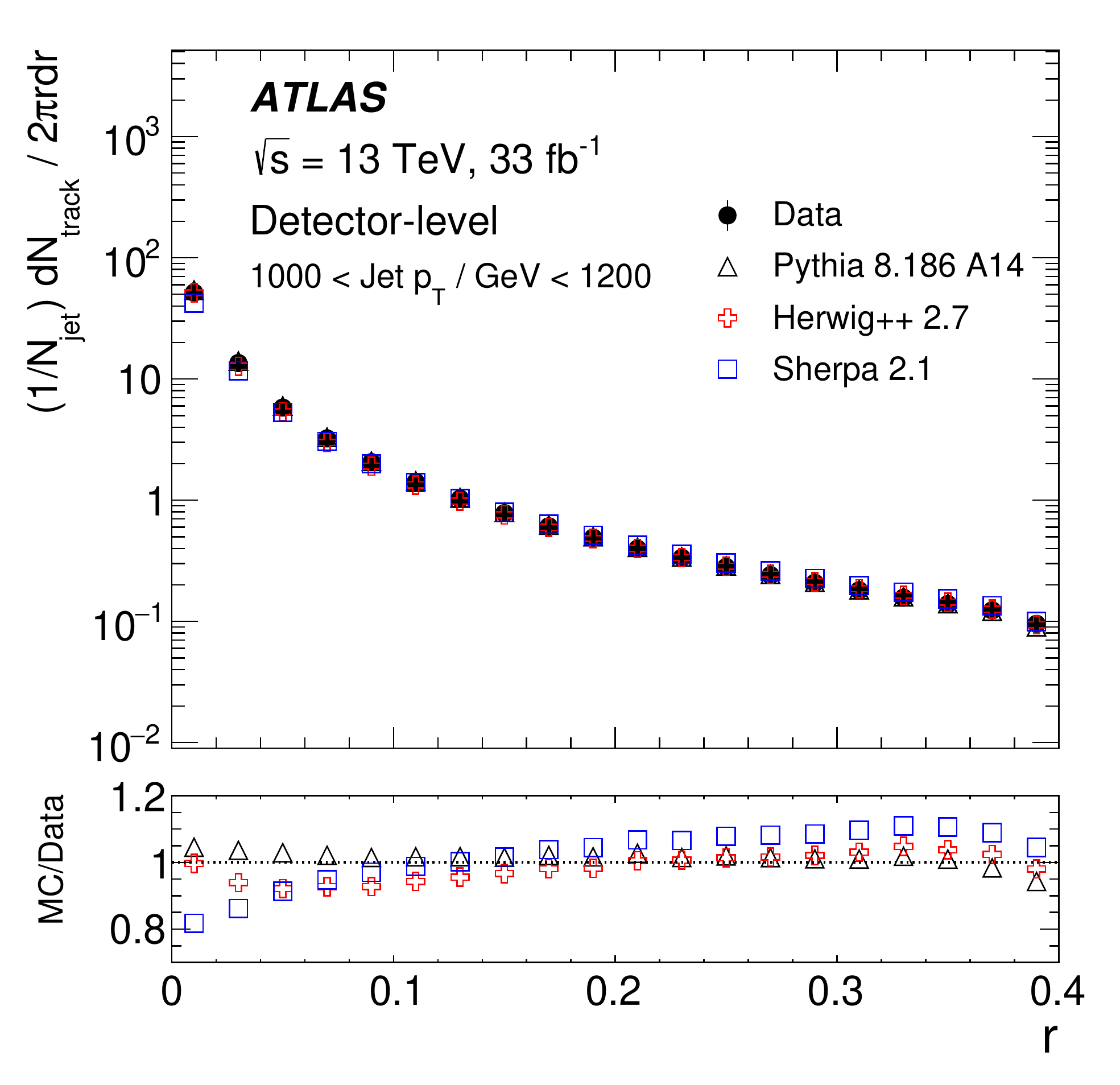}
\caption{The detector-level distributions of (top left) the number of charged-particle tracks $n_\text{track}$, (top right) the transverse momentum fraction $\zeta$, (bottom left) the transverse momentum $\pTrel$, and (bottom right) the radial profile in bins of the distance $r$ from the jet axis for jets with transverse momentum 1~\TeV~$< \pTjet < 1.2$~\TeV. Error bars include the statistical uncertainty only.}
\label{fig:subs}
\end{figure}
 
In order to expose differences between quark and gluon jets, the more forward and more central of the two jets are distinguished and measured separately.  Figure~\ref{fig:qg} shows the gluon-jet fraction as a function of jet $\pT$ and jet $\eta$ (more details about quark/gluon definitions are given in Section~\ref{sec:results:qg}). For a fixed jet $\pT$, higher-$|\eta|$ jets are more often quark-initiated due to valence quarks scattering off gluons.  For a fixed $\eta$, the quark fraction increases with jet $\pT$ due to the relative increase in valence-quark scattering off a quark or gluon compared with gluon--gluon scattering.
 
\begin{figure}[htbp]
\centering
\includegraphics[width=0.45\textwidth]{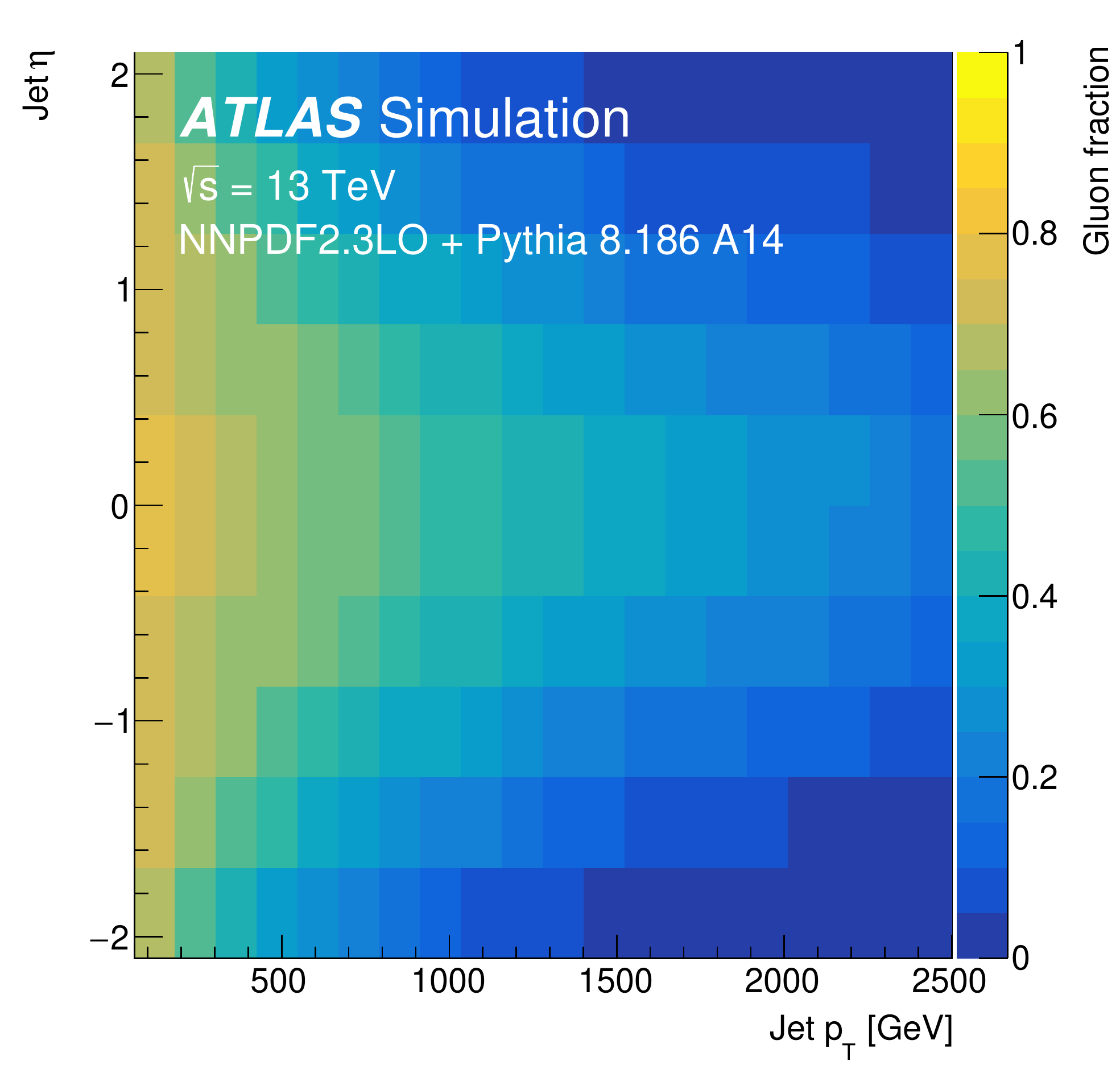}  \includegraphics[width=0.45\textwidth]{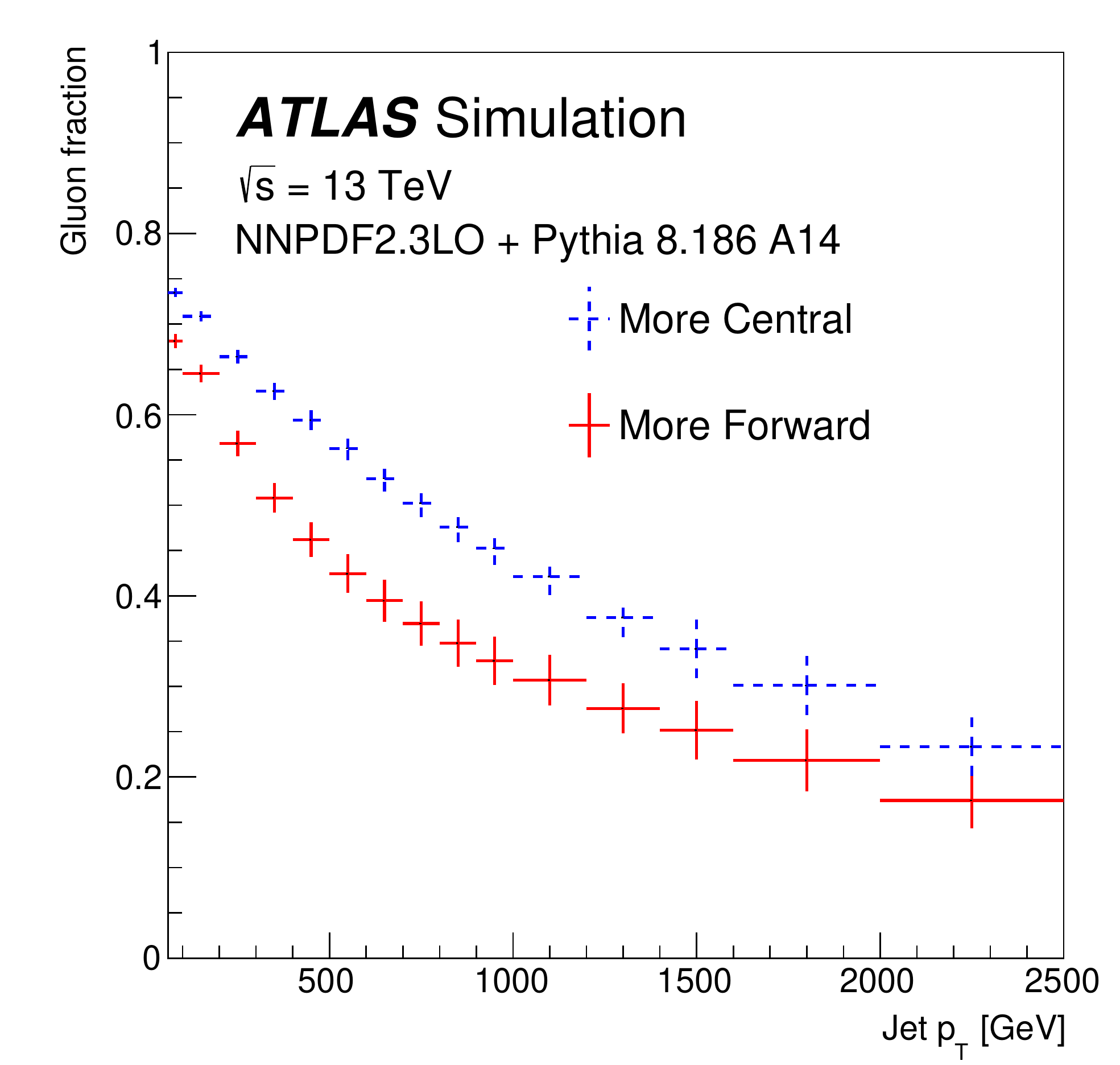}
\caption{Left: The gluon-jet fraction as a function of jet transverse momentum $\pT$ and jet pseudorapidity $\eta$.  Right: the fraction of the more forward and the more central jets that are gluon-initiated.  The error bars in the right plot represent the uncertainty computed from the 100 NNPDF2.3LO replicas~\cite{Ball:2012cx}.  See Section~\ref{sec:results:qg} for more details about quark/gluon definitions and uncertainties.}
\label{fig:qg}
\end{figure}
 
Table~\ref{tab:fiducial} summarizes the object and event selections from Section~\ref{sec:objectdef} and Section~\ref{sec:events}.
 
{\renewcommand{\arraystretch}{1.5}
\begin{table}[h!]
\centering
\caption{A summary of the object and event selection criteria at particle level and detector level.}
\begin{tabular}{|l|c|c|}
\hline
& Particle level & Detector level\\
\hline
\hline
Pileup & -- & Identify primary vertex \\
\hline
Jet algorithm& \multicolumn{2}{c|}{Anti-$k_t$, $R=0.4$} \\
\hline
Jet requirements & \multicolumn{2}{c|}{$|\eta| < 2.1$} \\
\hline
Jet constituents &  \begin{tabular}{@{}c@{}} Particles with $c\tau> 10$ mm prior to \\  detector interactions excluding $\mu$ and $\nu$\end{tabular} & Calorimeter energy clusters \\
\hline
Measurement inputs &  \begin{tabular}{@{}c@{}} Charged jet constituents, \\ $\pT>500$~\MeV\ and $|\eta|<2.5$ \end{tabular}&  \begin{tabular}{@{}c@{}} Ghost-associated tracks, \\ $\pT>500$~\MeV\ and $|\eta|<2.5$ \end{tabular} \\
\hline
Event selection & \multicolumn{2}{c|}{At least two jets, with the leading two satisfying $\pT^\text{lead}/\pT^\text{sublead} < 1.5$} \\
\hline
Jet selection &  \multicolumn{2}{c|}{Leading two, separated by $\eta$ (more forward/central)}\\
\hline
\end{tabular}
\label{tab:fiducial}
\end{table}
}
 
\clearpage
 
\section{Unfolding}
\label{sec:unfold}
 
The data are corrected for resolution and acceptance effects, and the fiducial phase space of the measurement is described by the particle-level object and event selection in Section~\ref{sec:obj}.   Equation~(\ref{eq:unfold}) symbolically summarizes the unfolding procedure for a binned distribution $x$:
 
\begin{align}
\label{eq:unfold}
x_\text{unfolded,$i$}=\frac{1}{n_\text{jets, unfolded}}\sum_{j=1}^{N_\text{total}}\theta_{ij} x_\text{detected,$j$}\left(\frac{1-\epsilon_\text{reco not true,$j$}}{1-\epsilon_\text{true not reco,$i$}}\right),
\end{align}
 
\noindent where $n_\text{jets, unfolded}$ is the unfolded number of forward or central jets (depending on the bin), determined by the number of entries in the $n_\text{ch}$ unfolding (as there is one entry per jet).  The symbols $\theta$ and $\epsilon$ represent the unfolding matrix and correction factors, described in more detail below.
 
The jet substructure observables are simultaneously unfolded with the jet $\pT$ and for the more forward and the more central jets at the same time.  For an observable with $n_\text{bins}$ bins in a given $\pT$ bin, this results in a total of $N_\text{total}=2\times (n_\text{bins})\times (\pT\text{ bins})$ bins.  All of these bins are concatenated to form a one-dimensional input.  To begin the unfolding, the data are corrected for the fraction of events that pass the detector-level selection but not the particle-level selection, $\epsilon_\text{reco not true}$.  This also corrects for non-dijet events, but their rate is negligible.  Then, an iterative Bayesian (IB) unfolding technique~\cite{DAgostini:1994zf} is used as a regularized matrix inversion to correct for the detector resolution in events that pass both the detector-level and particle-level selections.  The IB method is implemented in the RooUnfold framework~\cite{Adye:2011gm} with the unfolding matrix $\theta$ and one iteration, is chosen to minimize the total uncertainty.  After the application of the response matrix, a final correction is applied to account for the fraction of events that pass the particle-level but not detector-level selection, $\epsilon_\text{true not reco}$.  The resulting unfolded measurement is reorganized into individual distributions with $n_\text{bins}$ per $\pT$ bin for each of the more forward and more central jet.  The jet $\pT$ is also unfolded in parallel and each $\pT$ bin of the jet substructure observable is normalized by the number of measured jets in that bin.  For $n_\text{ch}$, this renders the distributions normalized to unity per jet $\pT$ bin; for the other observables, the normalization in each $\pT$ bin is (up to acceptance effects) $\langle n_\text{ch}\rangle$, as discussed in Section~\ref{sec:obs}.
 
To illustrate the jet-$\pT$ dependence of the measured observables, the evolution with the jet $\pT$ of various moments ($\kappa$) is computed using Eq.~(\ref{eq:unfoldmean}):
 
\begin{align}
\label{eq:unfoldmean}
\langle x^\kappa \rangle_\text{unfolded}(\text{$\pT$ bin $j$})&=c_\text{binning,$j$}(\kappa)\frac{\sum_{i=1}^{n_\text{bins}}x_\text{unfolded,$i$}\times (\text{bin center $i$})^\kappa }{ \sum_{i=1}^{n_\text{bins}}x_\text{unfolded,$i$}},
\end{align}
 
\noindent where the sum is over all $i$ that correspond to $\pT$ bin $j$.  Since the bin center is used to calculate the average, a correction $c_\text{binning}$ is applied to account for the difference between the bin center and the mean of the distribution within the bin.  This correction is calculated using \PYTHIA, and is computed by reweighting \PYTHIA so that it agrees with the unfolded distribution.  For $\zeta$, $\pTrel$, and $r$, Eq.~(\ref{eq:unfoldmean}) represents the $\kappa$ moment for individual particles.  For $\zeta$, the jet-based moments are also computed: $\langle \sum_\text{$i\in\text{jet}$} \zeta_i^\kappa\rangle$.  For these jet-based moments, Eq.~(\ref{eq:unfoldmean}) is modified by removing the denominator $\sum_{i=1}^{n_\text{bins}}x_\text{unfolded,$i$}$. By construction, the $\kappa=0$ jet-based moment of $\zeta$ is the $\kappa=1$ moment of $n_\text{ch}$.  The binning correction factor is mostly near unity, deviating by less than 1\% for $n_\text{ch}$ and up to about 10\% for the other observables.
 
Figure~\ref{fig:2dresponsematrix} shows the response matrix normalized per particle-level bin.  As stated above, the observable bins are concatenated with the jet $\pT$ and for both the more forward and more central jets to form a one-dimensional distribution that is unfolded.  A diagonal stripe represents events where the detector-level jet $\pT$ is the same as the particle-level value; off-diagonal components represent jet $\pT$ migrations.  Within a jet $\pT$ bin, there is a small dependence on $\zeta$, with a worse resolution at high $\zeta$ due to the deteriorating momentum resolution at high track $\pT$.  The diagonal strips in the upper left and lower right quadrants correspond to events where the more forward particle-level jet is the more central detector-level jet and vice versa.  This migration happens in about 1\% of events.   Within a given jet $\pT$ bin, the migrations to neighboring $\zeta$ bins are small.  Except at high $\zeta$ and high jet $\pT$ where the migrations can reach 50\%, the off-diagonal components of the response matrix are about 10\%.

\begin{figure}[htbp]
\centering
\includegraphics[width=0.75\textwidth]{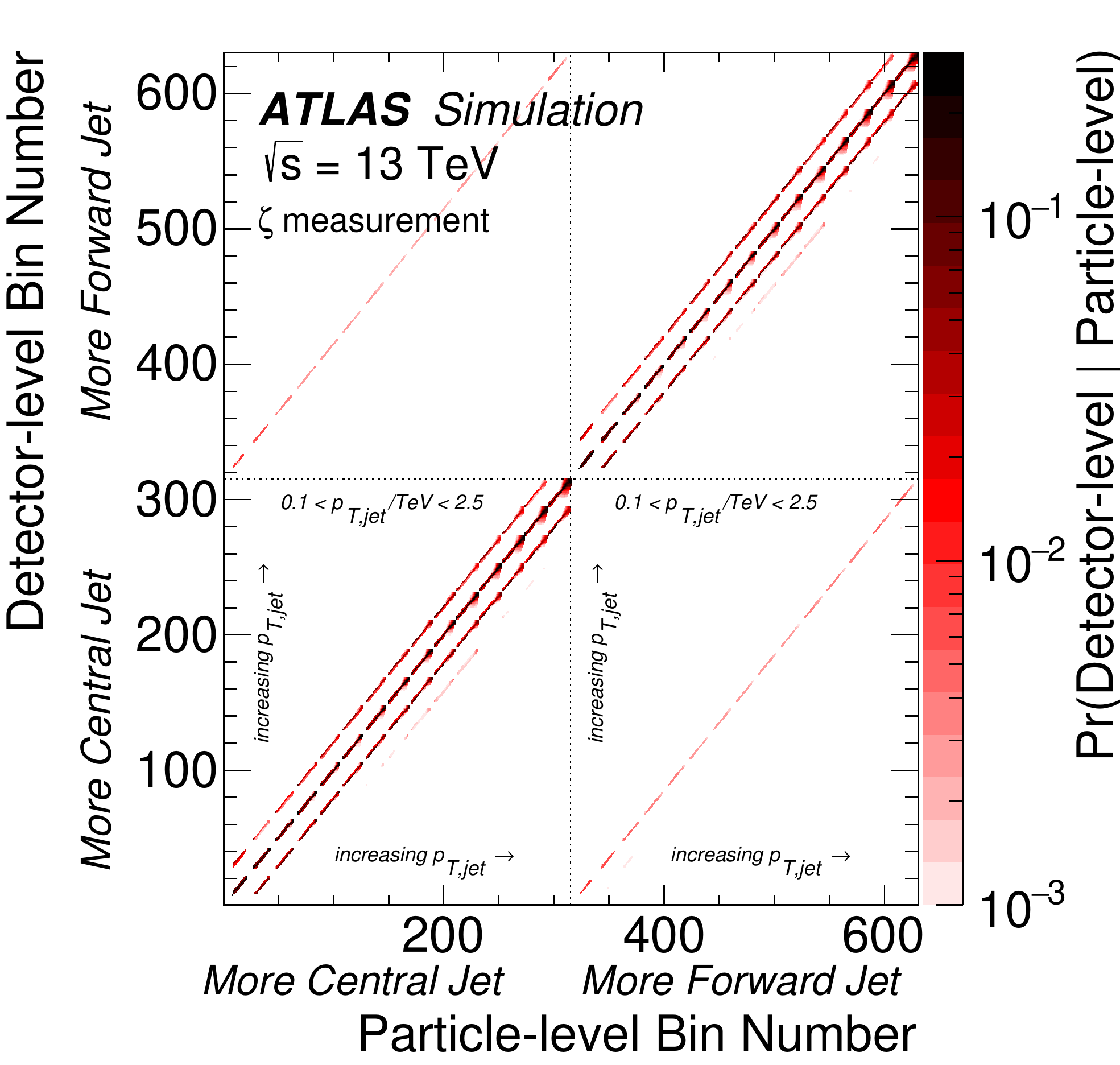}
\caption{The probability distribution for the detector-level combined transverse momentum fraction $\zeta$ and jet transverse momentum $\pT$ distribution normalized in bins of the particle-level variable using the \PYTHIA simulation. The $\zeta$ distribution is concatenated with the jet $\pT$ so that every 21 bins is a different jet-$\pT$ bin. The first 315 bins represent the more central of the two jets and the second 315 bins correspond to the more forward jet.  The $z$-axis is truncated at $10^{-3}$ for visualization only, to aid readability. }
\label{fig:2dresponsematrix}
\end{figure}
 
\clearpage
 
\section{Uncertainties}
\label{sec:systs}
 
Systematic and statistical uncertainties are assessed for each step of the analysis, including the acceptance correction factors, response matrix, and unfolding method.  For each uncertainty, some component of the analysis chain is varied and then the entire unfolding procedure is repeated.  Data and simulation statistical uncertainties are determined from pseudo-experiments using the bootstrap method~\cite{efron1979}.  The details of the experimental systematic uncertainties related to track and jet reconstruction are given in Section~\ref{sec:systs:exp} and the uncertainties in the unfolding method and fragmentation modeling are described in Section~\ref{sec:systs:theory}.  An additional source of uncertainty arising from binning effects is evaluated when computing the average value of an observable as a function of jet $\pT$.   The average values are determined using the bin centers, so the correction described in Section~\ref{sec:unfold} relies on the simulation for the distribution within a given bin.  An uncertainty in the binning correction is estimated by comparing the correction factors derived from \PYTHIA with those from \SHERPA, where both simulations are reweighted to match the unfolded data distribution.
 
Figure~\ref{fig:systoverview} provides an overview of the systematic uncertainties for a selection of observables, using the average value versus $\pT$ for illustration.  The uncertainty in the rate of fake and secondary tracks is the leading experimental reconstruction uncertainty for $\langle n_\text{ch}\rangle$ and $\langle r\rangle$ except at low jet $\pT$ where the uncertainties from the inclusive tracking efficiency and the unfolding procedure are larger.  The jet energy uncertainties are the most important for $\zeta$, with the tracking uncertainties matching in size in the highest jet-$\pT$ bins.  The tracking and jet energy uncertainties are about the same size for $\langle \pTrel\rangle$.  Fragmentation modeling uncertainties are large for $\langle n_\text{ch}\rangle$ at low jet $\pT$ and for $\langle \zeta\rangle$ at high jet $\pT$.  While the size of the binning correction uncertainty is less than 2\% for $\langle \pTrel\rangle$ and $\langle r\rangle$, it is still the dominant uncertainty for these observables.  Further details about each uncertainty source are given below and the full covariance matrices, including all correlation information, are made available in Ref.~\cite{hepdata}.
 
\begin{figure}[htbp]
\centering
\includegraphics[width=0.48\textwidth]{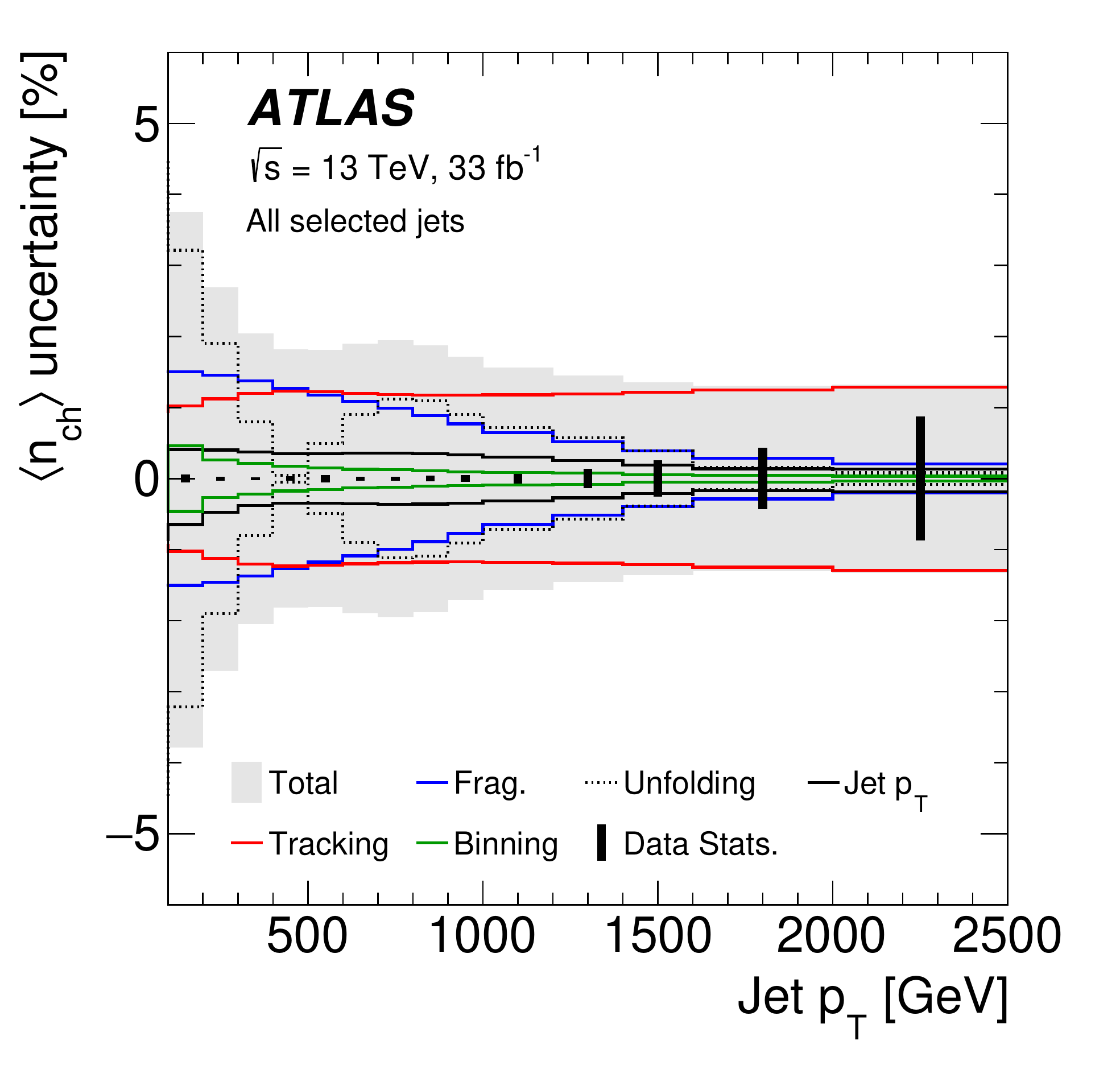}
\includegraphics[width=0.48\textwidth]{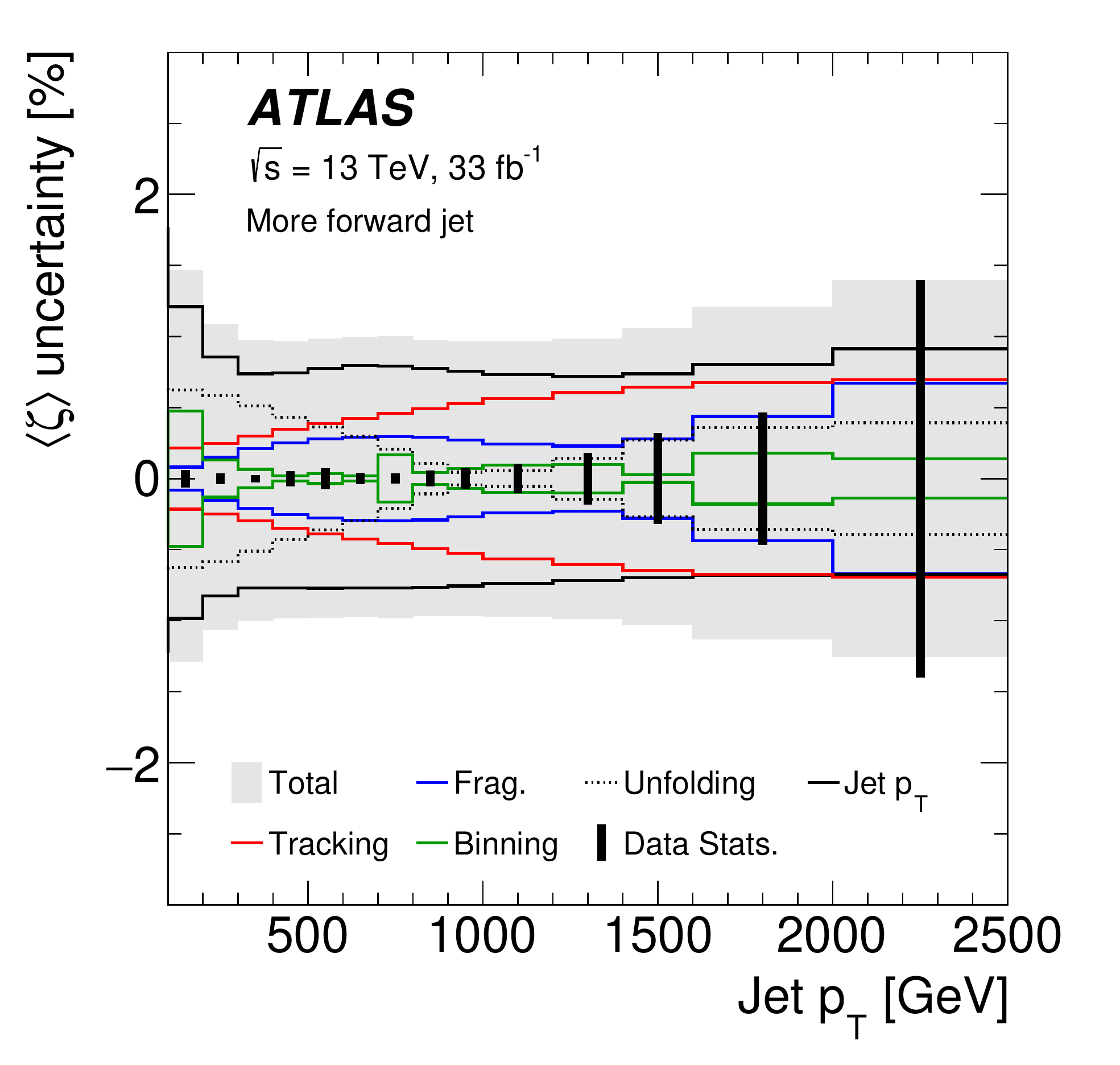}
\includegraphics[width=0.48\textwidth]{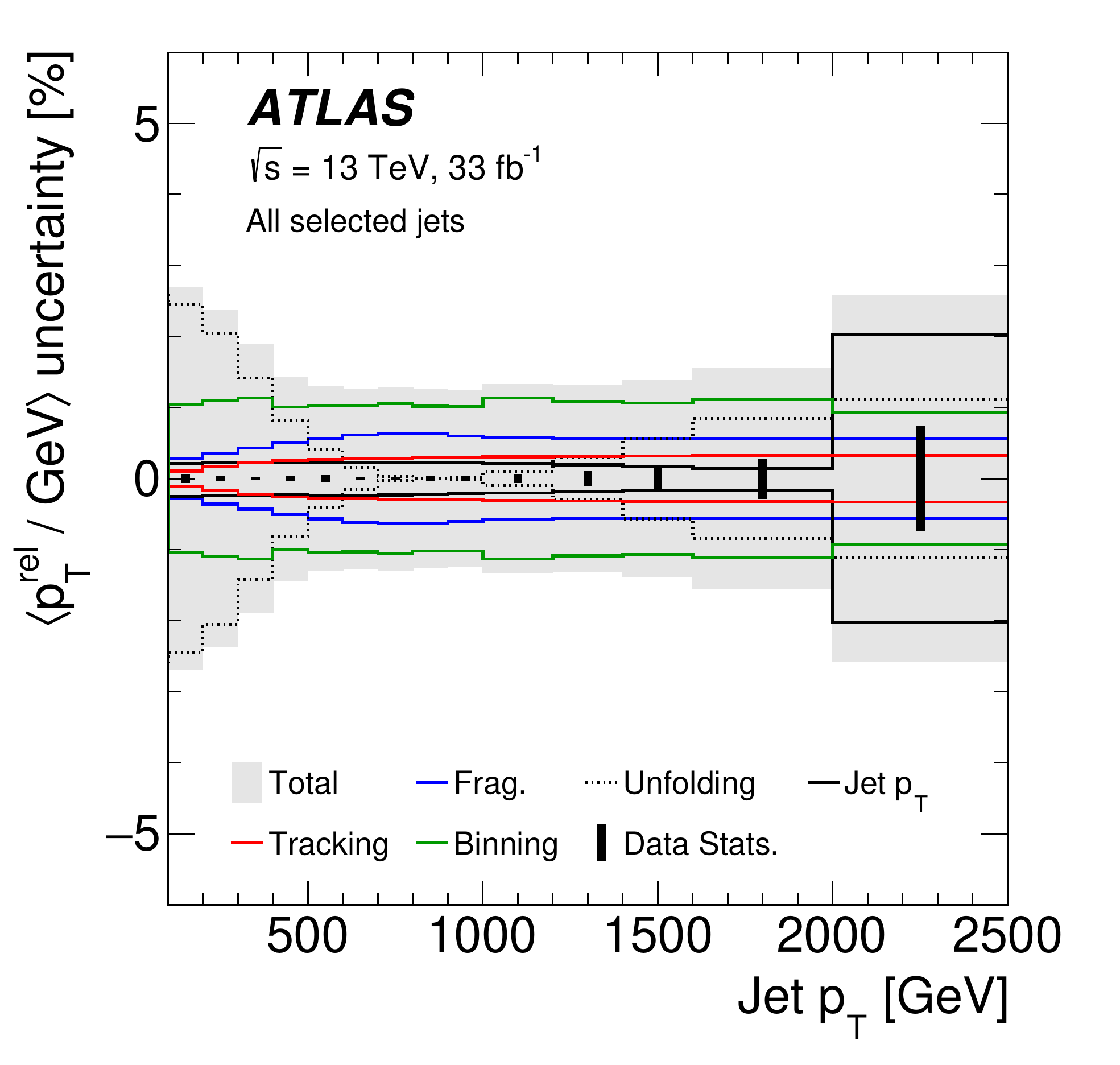}
\includegraphics[width=0.48\textwidth]{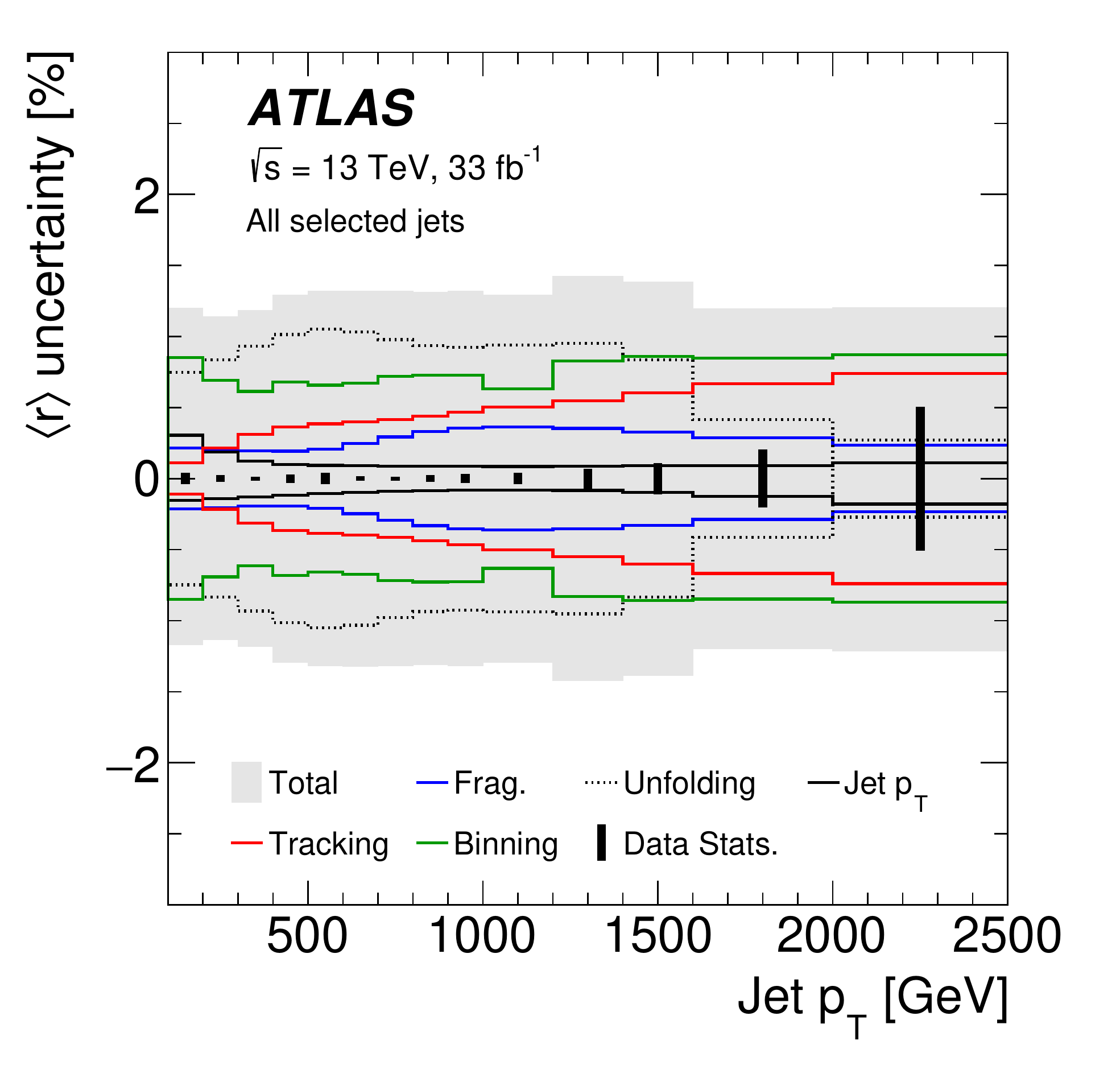}
\caption{An overview of the statistical and systematic uncertainties for the average value of (top left) charged-particle multiplicity $n_\text{ch}$, (top right) transverse momentum fraction $\zeta$, (bottom left) transverse momentum $\pTrel$, and (bottom right) radial profile as a function of jet transverse momentum $\pT$. The MC statistical uncertainties are negligible and are not shown.  When the uncertainties go through zero (as for the unfolding uncertainty in the top left), the signed uncertainty has changed from positive to negative or vice versa.   Most of the lines are the sum in quadrature of individual sources of uncertainty in each category, such as the various sources of tracking uncertainties as described in Sec.~\ref{sec:systs:exp}.}
\label{fig:systoverview}
\end{figure}
 
\subsection{Track and jet reconstruction}
\label{sec:systs:exp}
 
Except for $\zeta$, the jet energy is only used to determine the $\pT$ bin.  Since the fragmentation properties vary slowly with jet $\pT$, the resulting impact of jet energy scale and resolution uncertainties on the analysis is often less important than other sources of uncertainty.  Nonetheless, the impact of a 19-parameter decomposition of the jet energy scale uncertainty was evaluated~\cite{Aaboud:2017jcu}. Six of these 19 components are due to in situ constraints on the jet energy scale from various multi-object balance studies, such as $Z$+jets.  Additional sources of uncertainty are related to pileup, jet flavor, and extrapolations to high $\pT$.   The total uncertainty in the jet energy scale is about 1\% for jets with $\pT$ between 100 and 1000~\GeV\ and the impact on this measurement is much less than 1\% except at high $\zeta$, where it can reach as high as 2\%.  The impact of the jet energy resolution is determined from an ensemble of event samples with jet energies smeared within the uncertainty.
 
The most important experimental uncertainties are related to track reconstruction and cover the track reconstruction efficiency, the rate of fake and secondary tracks, the momentum scale, and density effects from pixel and strip cluster merging.   In the \PYTHIA simulation, approximately 60\% of the charged particles / tracks inside jets are charged pions that are well matched,\footnote{Reconstructed tracks are matched to charged particles by examining the pattern of sensors where energy was deposited.  If over 50\% of the weighted number of measurements on a track are due to one charged particle, it is declared matched to the track.  The weights are chosen to reflect the amount of information present in each detector and are ten for the pixel detector, five for the strip detector, and one for the straw tube tracker.} 10\% are well-matched kaons, 5\% are well-matched protons, 15\% are charged particles that are not matched to reconstructed tracks (inefficiency), 5\% are secondaries (split equally between photon conversions and nuclear interactions), 1\% are not well-matched tracks (fake tracks), and about $\mathcal{O}(0.1\%)$ are pileup tracks wrongly matched to the primary hard-scattering vertex.  The pileup contribution decreases with jet $\pT$ and momentum fraction, but increases with jet cone size (reaching 1\% at $\Delta$ = 0.4).  In contrast, the fake-track rate increases slightly with jet $\pT$ and has a contribution at high-momentum fraction of a few percent from kinked tracks reconstructed with a very high $\pT$.  The reconstruction inefficiency grows with jet $\pT$, and is peaked at both low and high radial distance from the center of the jet and is reduced at high momentum fraction.  This is because tracks with a larger radial distance from the jet axis tend to have lower $\pT$ (larger material effects and thus lower efficiency), while tracks in the core of the jet suffer from an inefficiency in the pattern recognition in the dense environment.
 
The uncertainty in the inclusive track reconstruction efficiency is dominated by the uncertainty in the amount of material in the inner detector.  Variations in the amount of material that are consistent with detector construction knowledge and measurements from secondary vertices~\cite{Aaboud:2017pjd} result in an uncertainty of 0.5\% for $|\eta| < 0.1$, which grows to 2.7\% for $2.3<|\eta|<2.5$.  This uncertainty is applied in the simulation by randomly removing tracks with a $\pT$- and $|\eta|$-dependent probability.   This uncertainty dominates the $n_\text{ch}$ measurement for jet $\pT\lesssim 1$~\TeV.
 
Since the ATLAS pixel detector measures the charge collected from ionization, it is possible to constrain the inefficiency from density effects by looking for single tracks with pixel charge consistent with two minimum-ionizing particles~\cite{Aaboud:2017all}.  The resulting uncertainty is about 0.4\% for tracks with $\Delta R < 0.1$ and is validated with additional studies related to the charged-to-neutral ratio in the jet as well as the geometric orientation of pixel clusters~\cite{myATL-PHYS-PUB-2017-016}.  This uncertainty is most important for the radial energy measurement at small radii from the jet axis and for the $n_\text{ch}$ measurement in the highest jet-$\pT$ bins.
 
The rate of fake tracks is studied inside jets by inverting some of the track quality criteria such as the fit $\chi^2/\text{NDF}$ and is found to agree between data and simulation at the 30\% level~\cite{myATL-PHYS-PUB-2017-016}.  A related source of uncertainty is due to the rate of secondary tracks.  These tracks originate from real charged particles, but are the result of interations in detector material and not direct fragmentation processes.  The rate of secondaries is estimated by fitting the track $d_0$ distribution and is found to agree with simulation within about 30\%.   These rates are then varied to determine an uncertainty in the measurement.  The fake-track rate is the leading source of uncertainty for $n_\text{ch}$ when $\pT\gtrsim 1$~\TeV\ and when $\zeta\sim 1$ or $r \lesssim 0.05$ for all jet-$\pT$ bins.  Uncertainties related to the modeling of pileup have a negligible impact.
 
The leading source of uncertainty in the track parameters is in the $q/\pT$ ($q$ is the electric charge) from a potential sagitta distortion due to detector misalignment weak modes~\cite{ATL-PHYS-PUB-2015-051}.  This bias is corrected and the uncertainty in the correction is about 0.1/\TeV\ except at $\phi\approx 0$ and $|\eta|\sim 2.5$ where the correction can reach 1/\TeV.  The impact on the measurement is smaller than the other tracking uncertainties.
 
\subsection{Unfolding method and fragmentation modeling}
\label{sec:systs:theory}
 
An uncertainty resulting from the unfolding method described in Section~\ref{sec:unfold} is determined by unfolding the prediction from a reweighted simulation with the nominal procedure.  The reweighted simulation is constructed by modifying the nominal \PYTHIA~8 particle-level spectrum so that the simulated detector-level spectrum, from propagating the reweighted particle-level spectrum through the response matrix, has significantly improved agreement with the data.  The modified detector-level distribution is unfolded with the nominal response matrix and the difference between this and the reweighted particle-level spectrum is an indication of the bias due to the unfolding method (in particular, the choice of prior)~\cite{Malaescu:2009dm}.  The weights are chosen by comparing the \PYTHIA~8 particle-level spectrum with the unfolded data.  After applying the reweighting, the $\chi^2/\text{NDF}$ calculated using only the statistical uncertainties improves significantly in each jet $p_\text{T}$ bin.  The resulting systematic uncertainties are generally much smaller than the detector-level differences between the data and simulation, as desired.
 
The unfolded result depends on the modeling of jet fragmentation through the prior, the response matrix, and the correction factors.  Variations in the prior are already accounted for in the data-driven non-closure uncertainty described above.  The other contributions are evaluated by comparing the result using \PYTHIA~8 with the result using the alternative \HERWIGpp sample described in Section~\ref{sec:analysis}.  A similar uncertainty is obtained when using \HERWIGpp or \SHERPA as the alternative model.  This comparison is decomposed into components corresponding to varying only the response matrix or only the initial/final correction factors, $\epsilon_\text{reco not true}$ and $\epsilon_\text{true not reco}$ in Eq.~(\ref{eq:unfold}).  All three components are added in quadrature to determine the total uncertainty due to fragmentation modeling.   Even though these sources of uncertainty are correlated, they were treated as independent because the level of correlation is unknown given that there are only two alternative models.  The resulting uncertainty is much smaller than the difference between \PYTHIA~8 and \HERWIGpp at particle level.  For $n_\text{ch}$, the response matrix is the dominant contribution to this uncertainty, except in the first jet-$\pT$ bin where the correction factors and their uncertainty are also important.   For the per-particle observables ($\zeta$, $r$, $\pTrel$), the correction factors dominate the uncertainty because acceptance effects are much more important.
 
\section{Results}
\label{sec:result}
 
The unfolded data are presented in two ways.  Section~\ref{sec:results:inclusive} focuses on the inclusive spectra for both jets together, while Section~\ref{sec:results:qg} uses the differences between forward and central jets to determine the unique features of quark-initiated and gluon-initiated jets, some of which can be compared with perturbative QCD calculations.  These sections show a selection of jet $\pT$ bins; a complete set of results can be found in Ref.~\cite{hepdata}.
 
\subsection{Inclusive distributions}
\label{sec:results:inclusive}
 
The unfolded averages of the measured observables are presented as a function of the jet $\pT$ in Figure~\ref{fig:result:inclusive:averagevpT_fc} for the more forward and more central jets separately and then combined in Figure~\ref{fig:result:inclusive:averagevpT}.   All other figures in this section combine measurements of both jets.  The more central jets show properties that are more gluon-like than the more forward jets: they have a larger charged-particle multiplicity and a softer momentum-fraction spectrum.  The modeling of the all-jet spectra is very similar to that of the more forward/backward jets and is described in detail for the all-jet spectra only.
 
As the jet $\pT$ increases, the average charged-particle multiplicity increases, the average momentum fraction decreases, the average $\pTrel$ increases, and the average multiplicity-weighted radius decreases.  Charged-particle multiplicity increases from about 10 at jet \pT\ of 100~\GeV\ to just over 20 at 2.5~\TeV.  In most cases, \PYTHIA~8 and \SHERPA bracket the data, and are accurate to better than 10\%; \HERWIGpp is often between these two and closer to the data. As the distribution of $n_\text{ch}$ is almost Poissonian, nearly all of the information about the distribution is encoded in the mean value.  In contrast, the distribution of $\zeta$ is more complicated.\footnote{The distribution is nearly Gaussian in $\log\zeta$, so it is well specified by two parameters instead of one~\cite{pinkbook}.} The average momentum fraction is about 5\% at jet \pT\ of 100~\GeV\ and decreases to about 2.5\% at 2.5~\TeV\ (the most probable value, shown below, is lower).   The distributions of $\pTrel$ and the radial profiles fall steeply (nearly exponentially) away from zero and the average values in Figure~\ref{fig:result:inclusive:averagevpT} give a sense of how fast they fall (exponential distributions are uniquely specified by their mean).  The average $\pTrel$ at $\pTjet=100$~\GeV\ is about 0.35~\GeV\ and increases to about 0.55~\GeV\ at $\pTjet=2.5$~\TeV.  If the angular distribution about the jet axis is independent of \pT, the average value of $\pTrel$ should be proprotional to $\langle\zeta\rangle(\pTjet)\times \pTjet$.  This would suggest an increase by a factor of $(2.5\%/5\%)\times(2500/100)\sim 12.5$ across the measured range; instead it only increases by a factor of about 1.5.  This means that the angular distribution is not independent of $\pT$ and in particular, the jets become more collimated.  This is also consistent with direct measurement of the radial profile, where the average value drops from about 0.06 at $\pTjet=100$~\GeV\ to about 0.03 at $\pTjet=2.5$~\TeV.  While \PYTHIA~8, \SHERPA, and \HERWIGpp agree well with the data for $\pTrel$, \SHERPA provides a poorer model of the average radial profile as a function of the jet $\pT$.
 
As noted above, the distribution of $\zeta$ cannot be described simply by its average value, in contrast to $n_\text{ch}$, $\pTrel$ and $r$, which are nearly Poisson or exponentially distributed.   Therefore, it is useful to summarize the $\pTjet$ dependence of other aspects of the $\zeta$ distribution.  Figure~\ref{fig:result:inclusive:fractions} shows partial integrals of the $\zeta$ distribution and Figure~\ref{fig:result:inclusive:moments} shows the average values of $\zeta^{1/2}$, $\zeta^2$, $\sum_{i\in\text{jet}}\zeta^{1/2}$ and $\sum_{i\in\text{jet}}\zeta^{2}$.  Figure~\ref{fig:result:inclusive:fractions} illustrates how the average fraction of charged particles with a given momentum fraction evolves with $\pTjet$.   There is no correction for binning effects, as the measured $\zeta$ distribution has bin edges which nearly align with $0.1\%$, 1\%, and 10\%.   In particular, the $\zeta$ bins are $1/1.5^n$, for $n=0,\ldots,21$, and the fractions in Figure~\ref{fig:result:inclusive:fractions} are estimated as $0.1\%\approx 1/1.5^{17}$, $1\%\approx 1/1.5^{11}$, and $10\%\approx 1/1.5^5$.  The fraction of particles carrying 10\% or less of the momentum changes very little across the entire $\pTjet$ range and is also near unity ($>90\%$ for all $\pTjet$).  A strong $\pTjet$ dependence is introduced when the $\zeta$ threshold is lowered to 1\% and to 0.1\%. Since charged particles are required to have $\pT>500$~\MeV, only jets with $\pTjet>500$~\GeV\ can have particles with $\zeta<0.1\%$.  The fraction of particles with $\zeta < 1\%$ has a logarithmic increase while the fraction of particles with $\zeta < 0.1\%$ appears to increase faster than linearly with $\pTjet$.  Both of these general trends are reproduced by \PYTHIA~8, \SHERPA, and \HERWIGpp, although for example, \PYTHIA~8 disagrees with the exact value at low $\pTjet$ for $\zeta<1\%$ and all $\pTjet$ for $\zeta<0.1\%$.  For the $\zeta<0.1\%$ case, \SHERPA and \PYTHIA~8 bracket the data, with \SHERPA predicting more particles with a lower $\zeta$ fraction, while \HERWIGpp is much closer to the data.  The average values of $\sqrt{\zeta}$ and $\zeta^2$ for individual particles as a function of jet $\pTjet$ in the top panel of Figure~\ref{fig:result:inclusive:moments} show a decreasing trend that is qualitatively similar to the trend for the average $\zeta$ in Figure~\ref{fig:result:inclusive:averagevpT}.  For $\sqrt{\zeta}$, \PYTHIA~8/\HERWIGpp and \SHERPA bracket the data, although \PYTHIA~8 agrees with the data within the uncertainty.  \SHERPA predicts a significantly higher average $\zeta^2$ than is present in the data.  As with $\langle n_\text{ch}\rangle$, the average value of $\sum_{i\in\text{jet}}\zeta^{1/2}$ increases with the jet $\pT$, while the $\pT$ dependence of $\langle \sum_{i\in\text{jet}}\zeta^{2}\rangle$ is more complicated as it first decreases and then slowly increases with jet $\pT$.  This trend is well reproduced by \PYTHIA and \HERWIGpp, but not by \SHERPA.
 
To present more differential information, the full unfolded distributions for $n_\text{ch}$, $\zeta$, $\pTrel$, and $r$ are shown in Figures~\ref{fig:results:nch},~\ref{fig:results:zeta},~\ref{fig:results:ptrel}, and~\ref{fig:results:r}, respectively, for representative $\pTjet$ bins.  Many of the relevant trends are captured in the above discussion about the $\pTjet$ dependence of the moments.  However, finer information that may be useful for generator tuning is provided by the differential distributions.
 
\begin{figure}[htbp]
\centering
\includegraphics[width=0.48\textwidth]{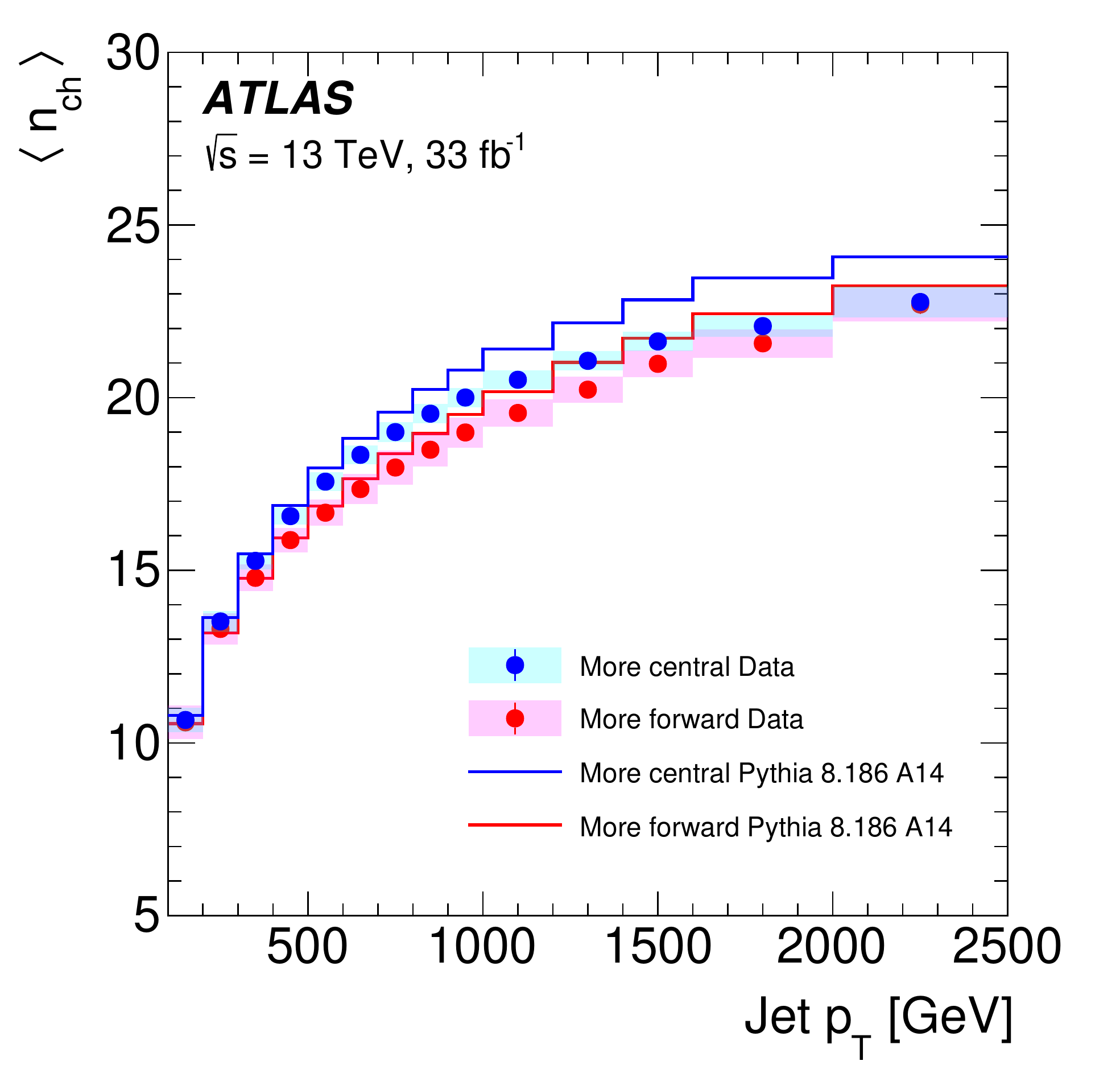}
\includegraphics[width=0.48\textwidth]{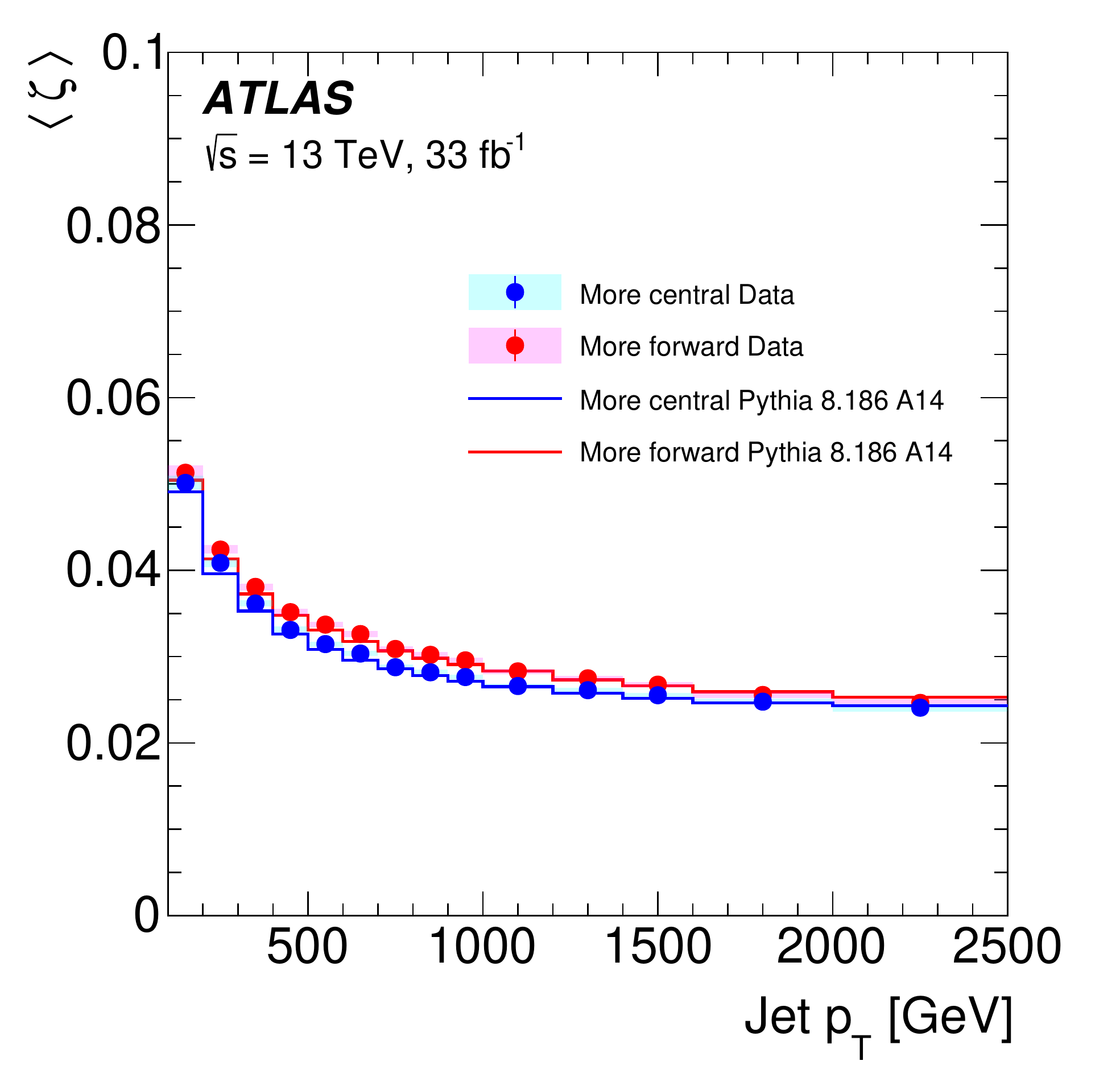}
\includegraphics[width=0.48\textwidth]{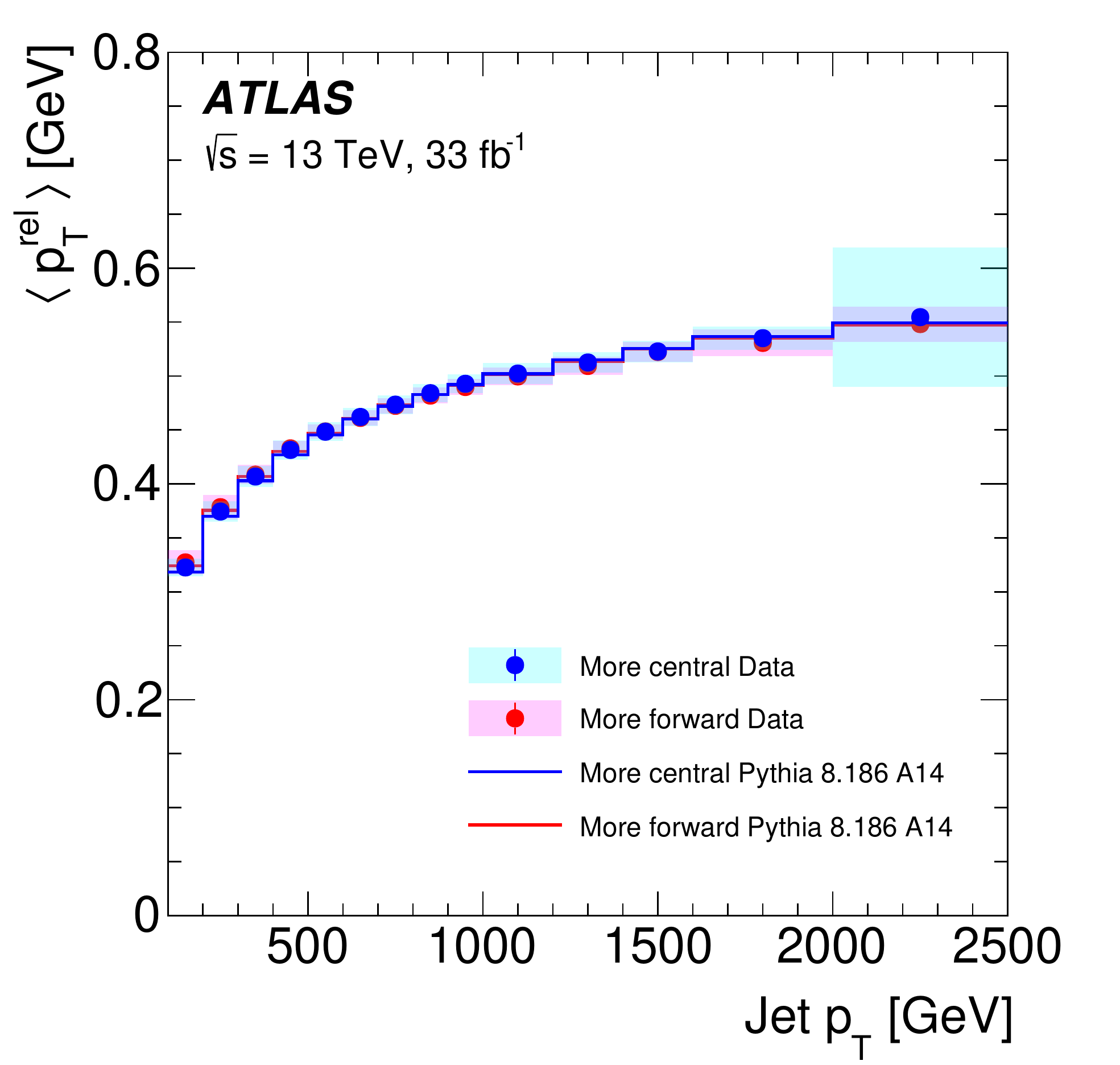}
\includegraphics[width=0.48\textwidth]{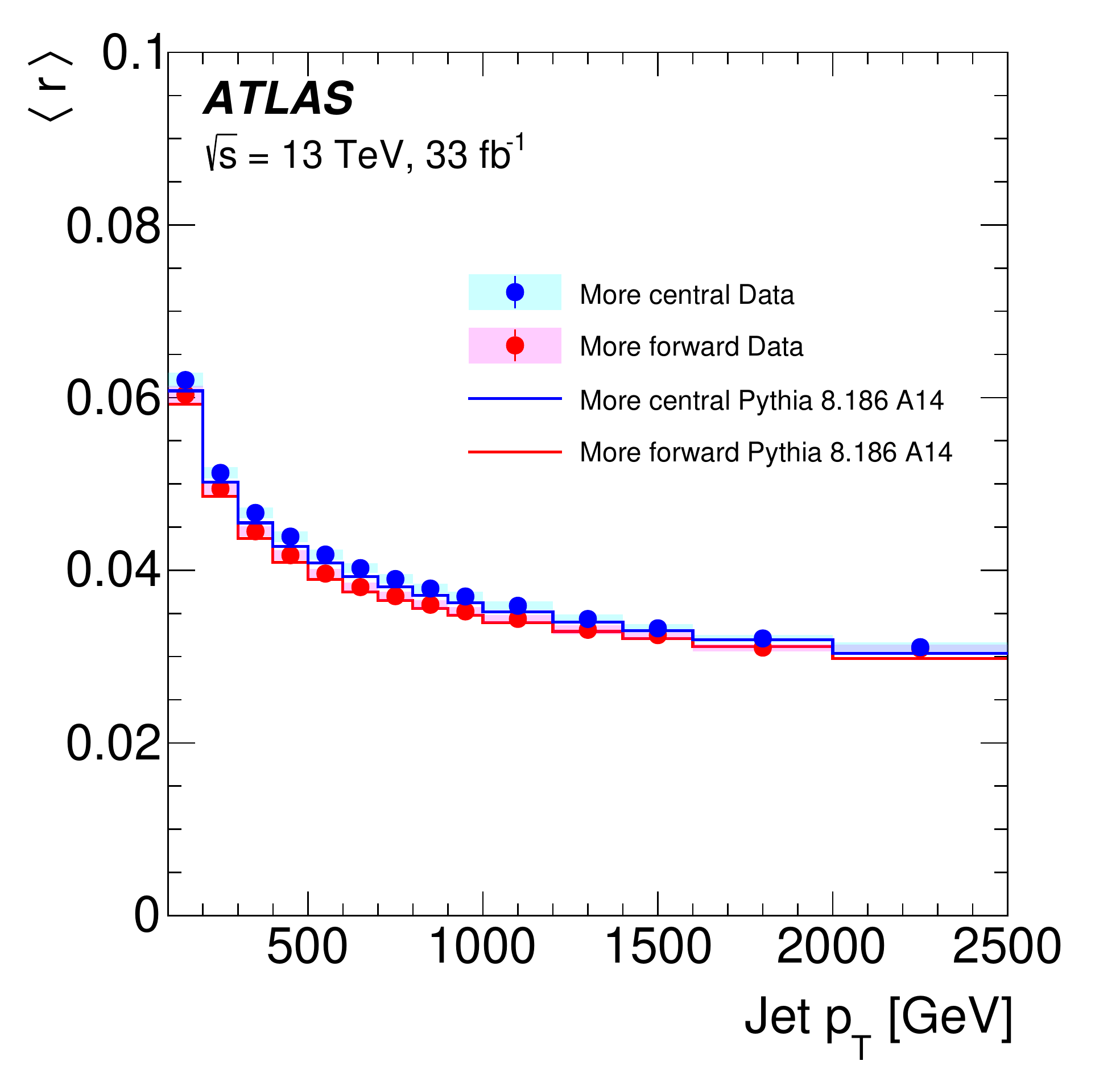}
\caption{The unfolded measured averages for the (top left) charged-particle multiplicity $n_\text{ch}$, (top right) transverse momentum fraction $\zeta$, (bottom left) transverse momentum $\pTrel$, and (bottom right) radial profile as a function of the jet transverse momentum $\pT$ for the more forward and more central of the two jets, separately. The uncertainty bands show the combined statistical and systematic uncertainty. }
\label{fig:result:inclusive:averagevpT_fc}
\end{figure}
 
\begin{figure}[htbp]
\centering
\includegraphics[width=0.48\textwidth]{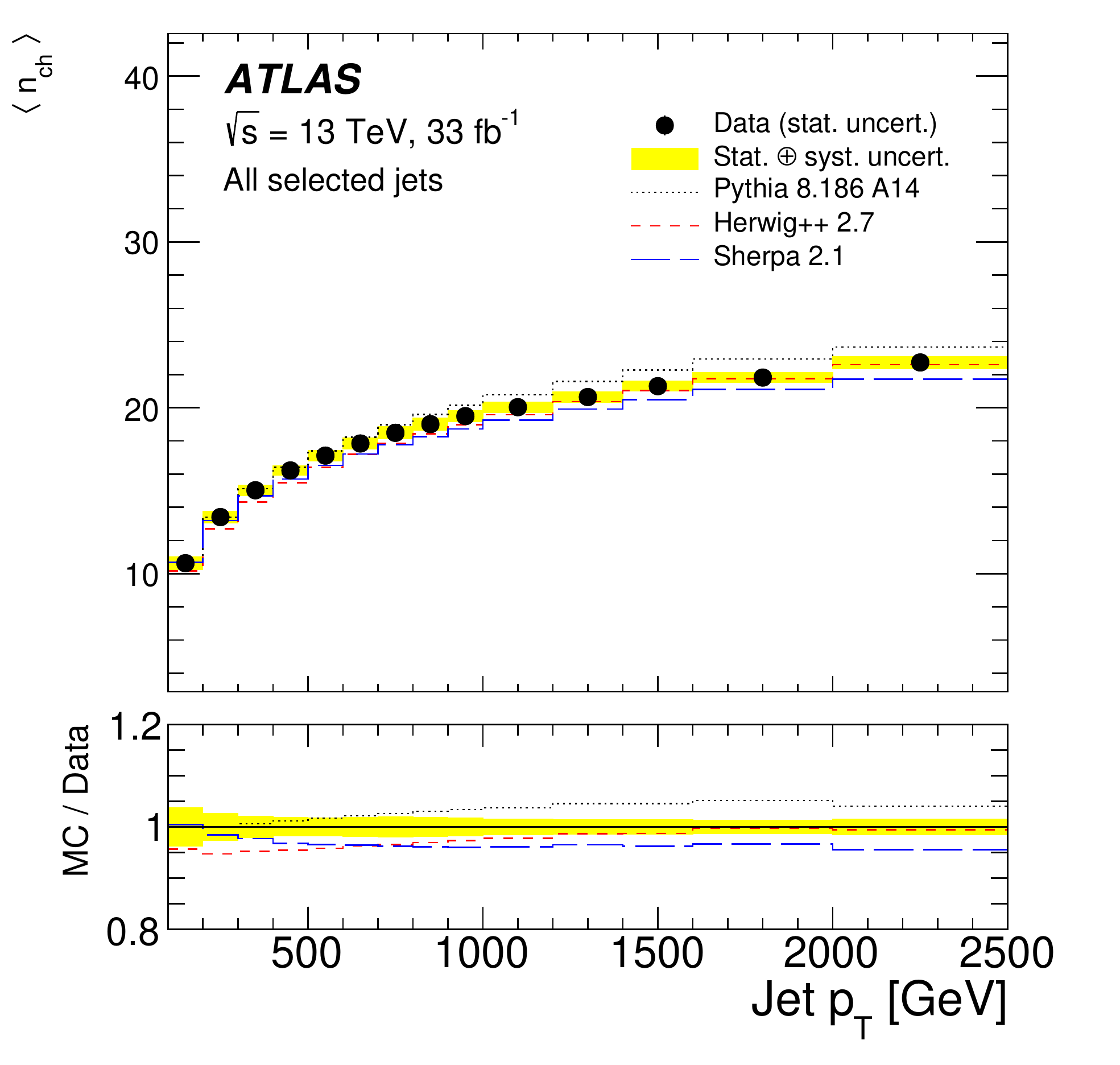}
\includegraphics[width=0.48\textwidth]{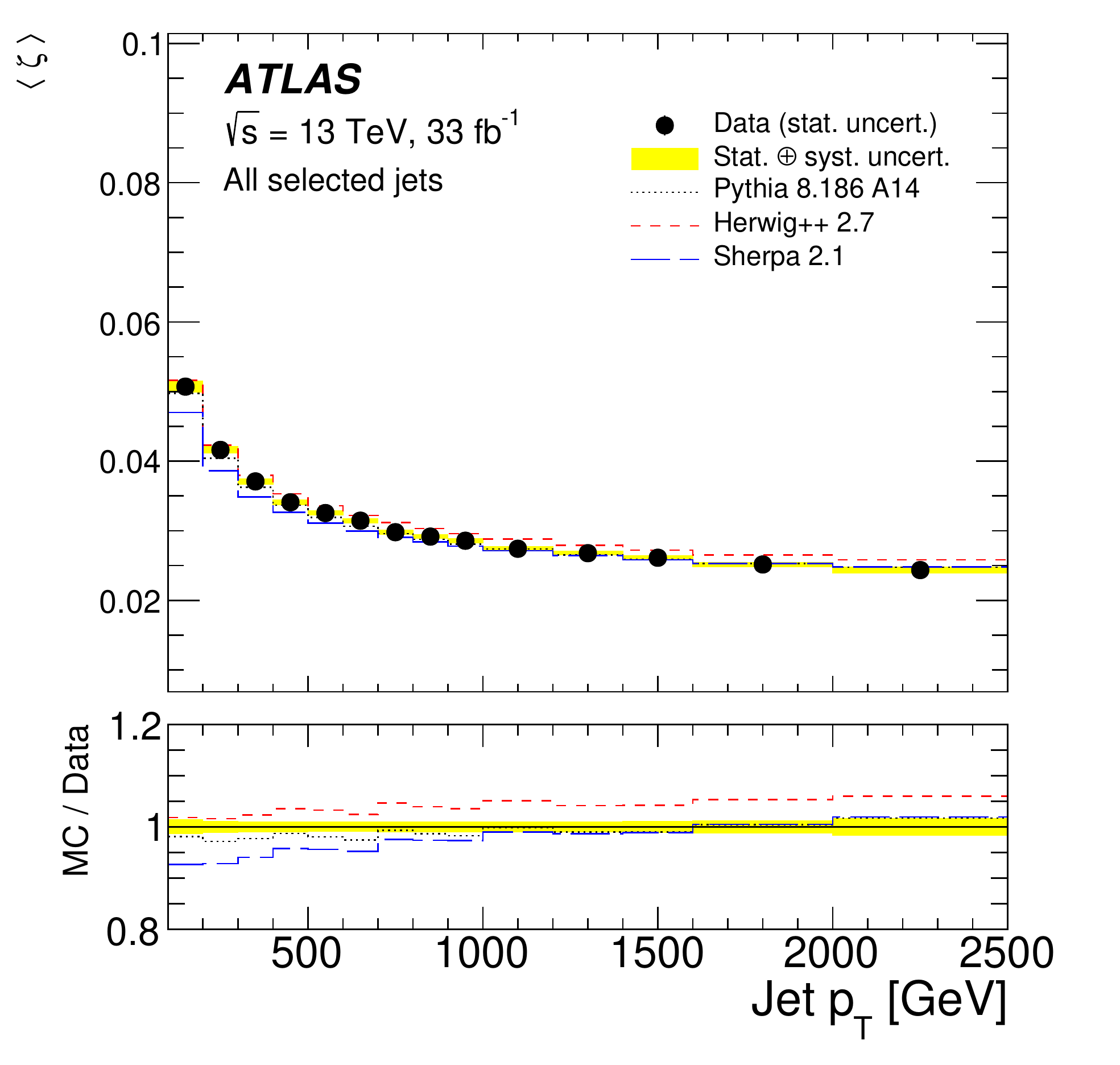}
\includegraphics[width=0.48\textwidth]{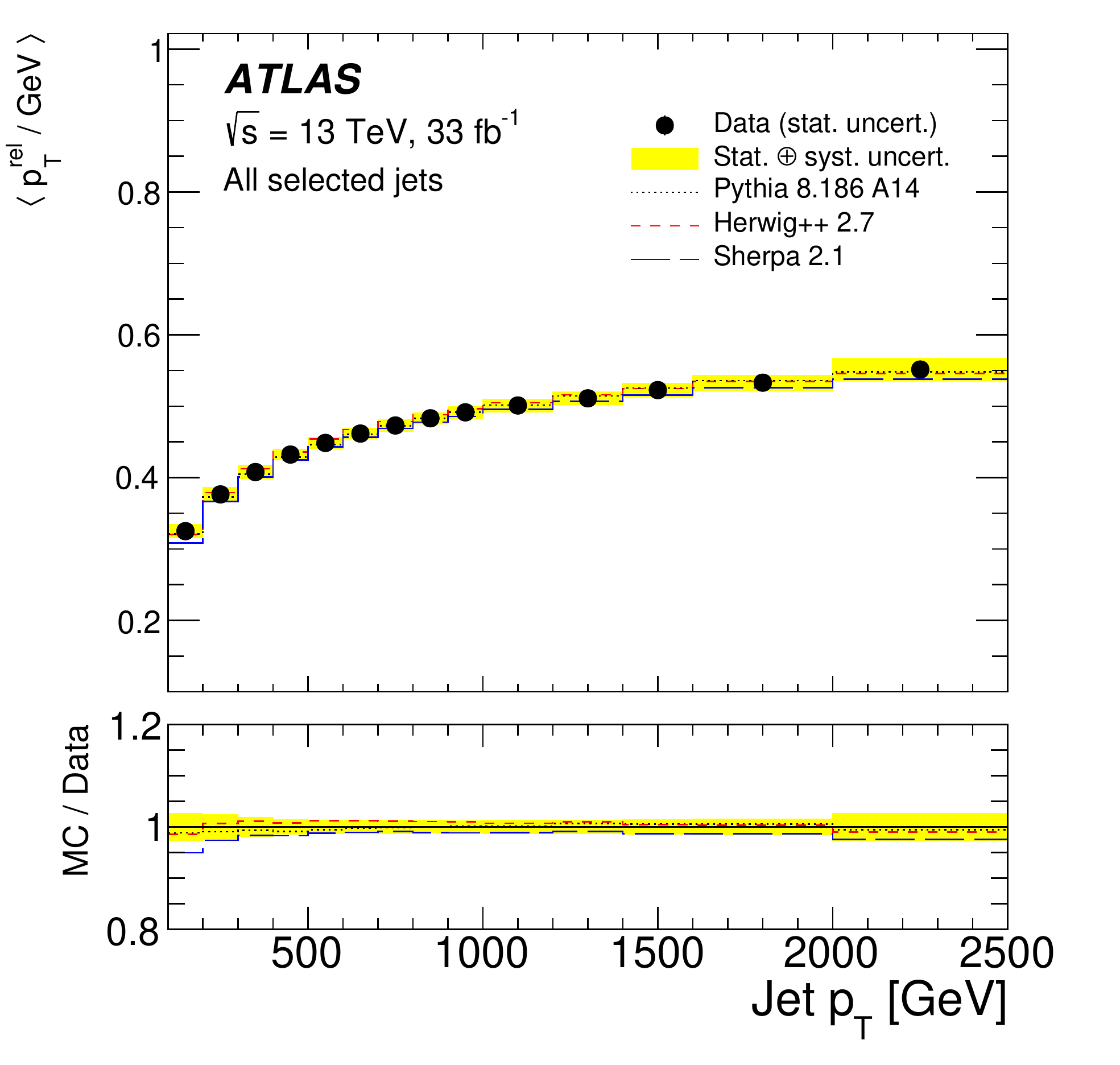}
\includegraphics[width=0.48\textwidth]{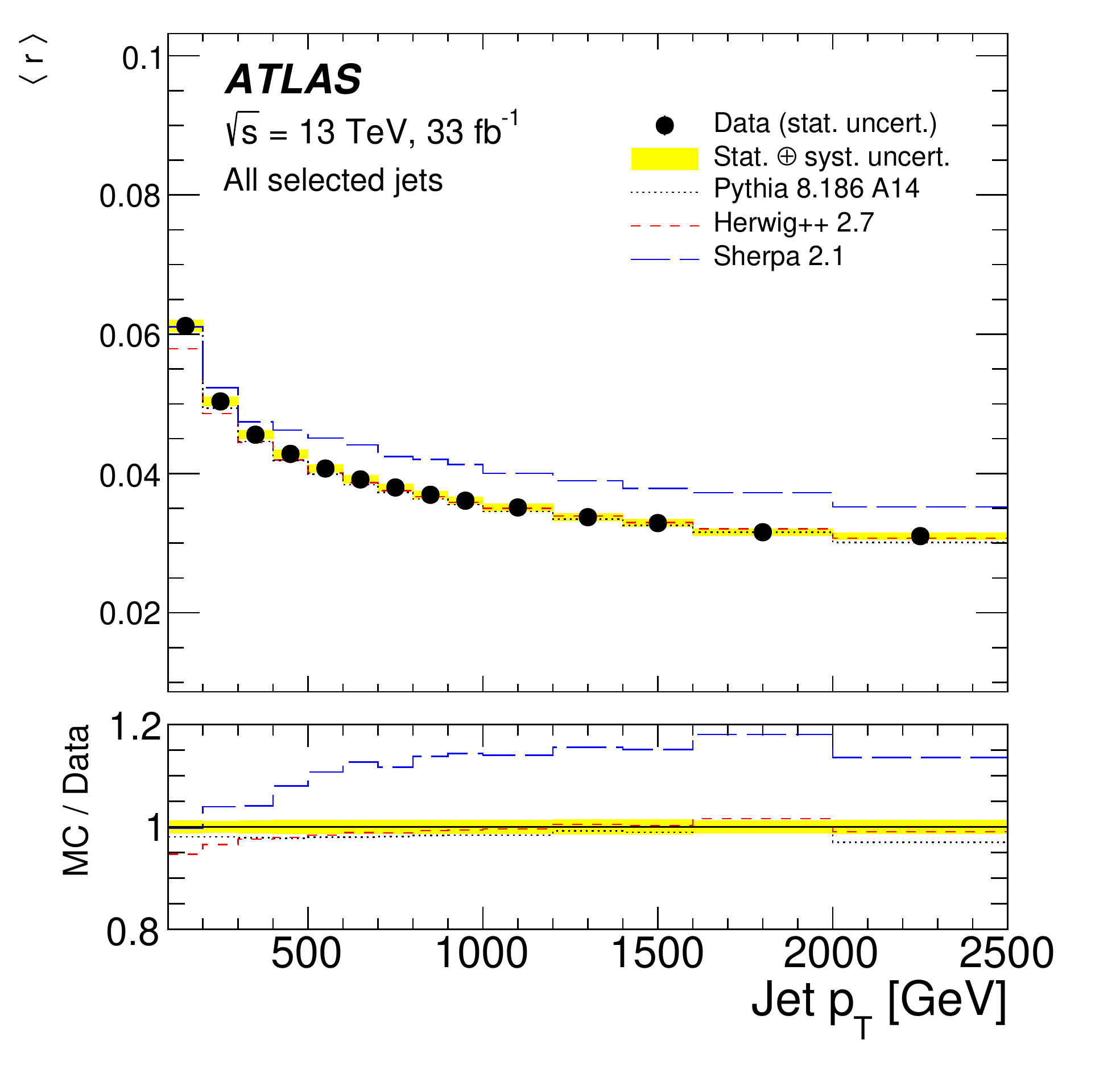}\\
\caption{The unfolded measured averages (top left) charged-particle multiplicity $n_\text{ch}$, (top right) transverse momentum fraction $\zeta$, (bottom left) transverse momentum $\pTrel$, and (bottom right) radial profile as a function of the jet transverse momentum $\pT$.  The lower panel shows the ratio of various MC predictions to the data, with the total uncertainty band centered on the data at unity.  }
\label{fig:result:inclusive:averagevpT}
\end{figure}
 
\begin{figure}[htbp]
\centering
\includegraphics[width=0.48\textwidth]{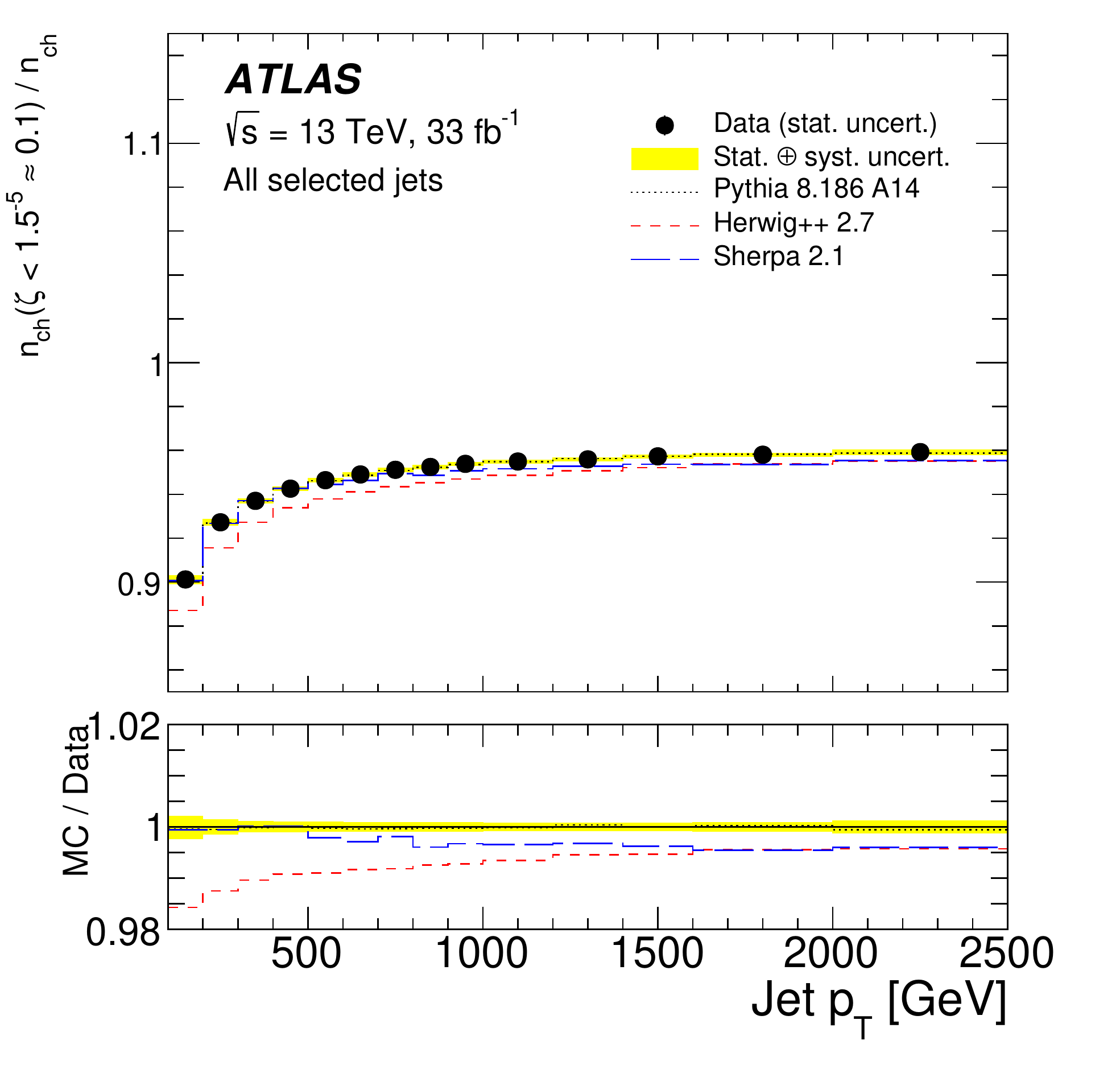}
\includegraphics[width=0.48\textwidth]{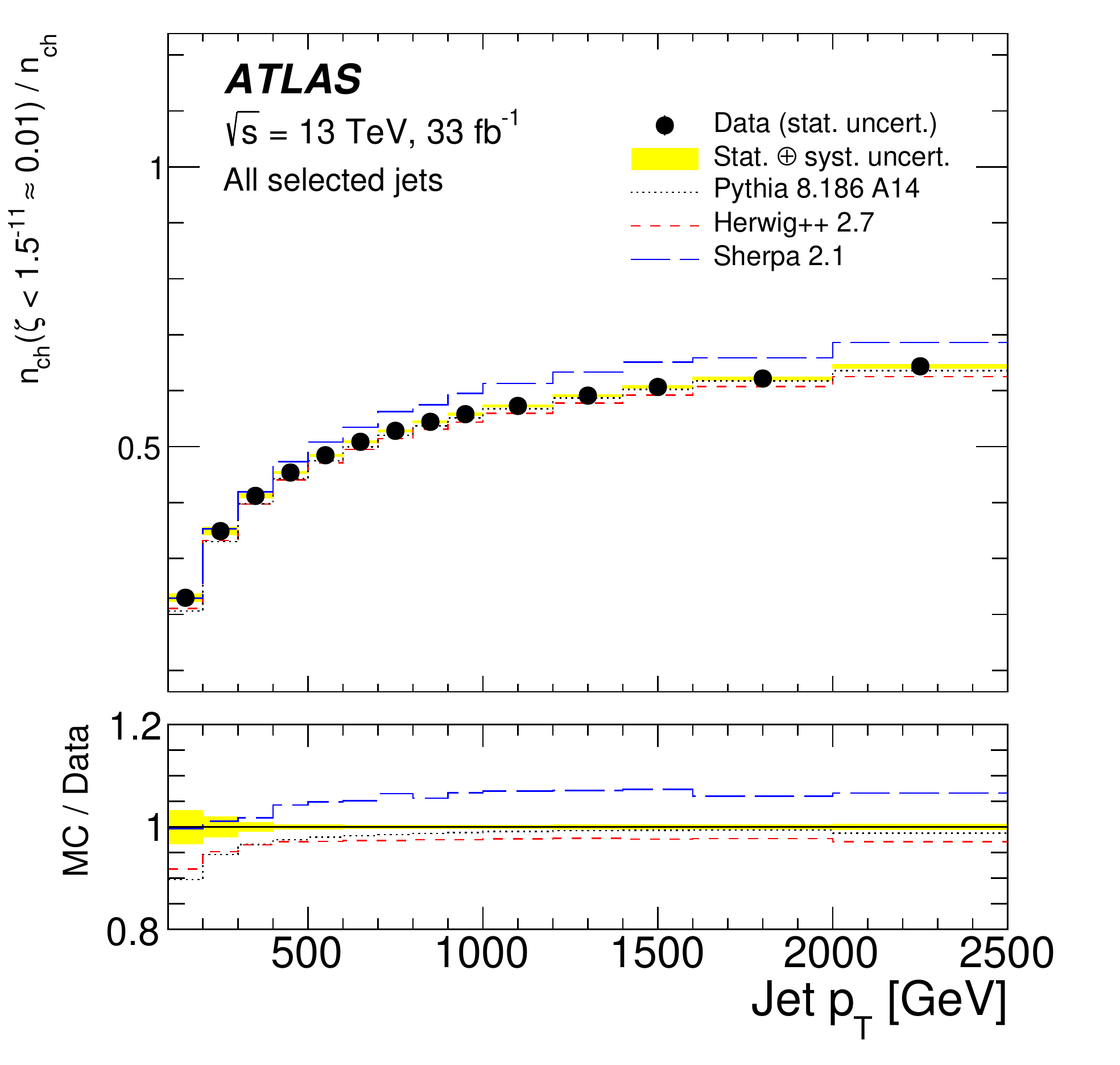}
\includegraphics[width=0.48\textwidth]{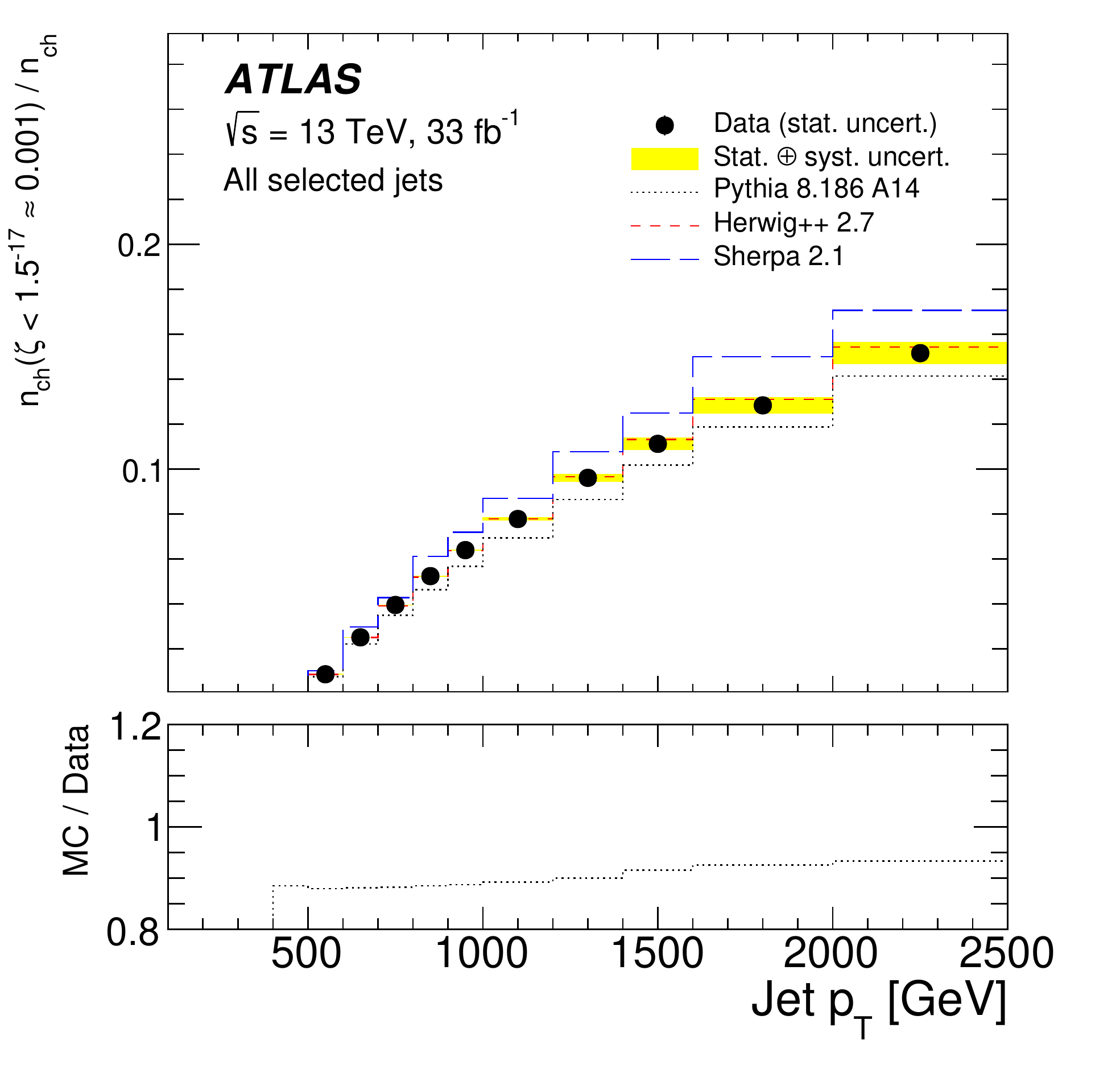}
\caption{The unfolded fraction of charged particles carrying a fraction $\zeta \lesssim $ 10\% (top left), 1\% (top right), and 0.1\% (bottom) of the jet transverse momentum $\pT$.  The exact fractions are given in the figures and correspond to powers of $1.5$ so that the values align precisely with bin edges and thus no binning correction is required.  The lower panels show the ratio of various MC predictions to the data, with the total uncertainty band centered on the data at unity. }
\label{fig:result:inclusive:fractions}
\end{figure}
 
\begin{figure}[htbp]
\centering
\includegraphics[width=0.48\textwidth]{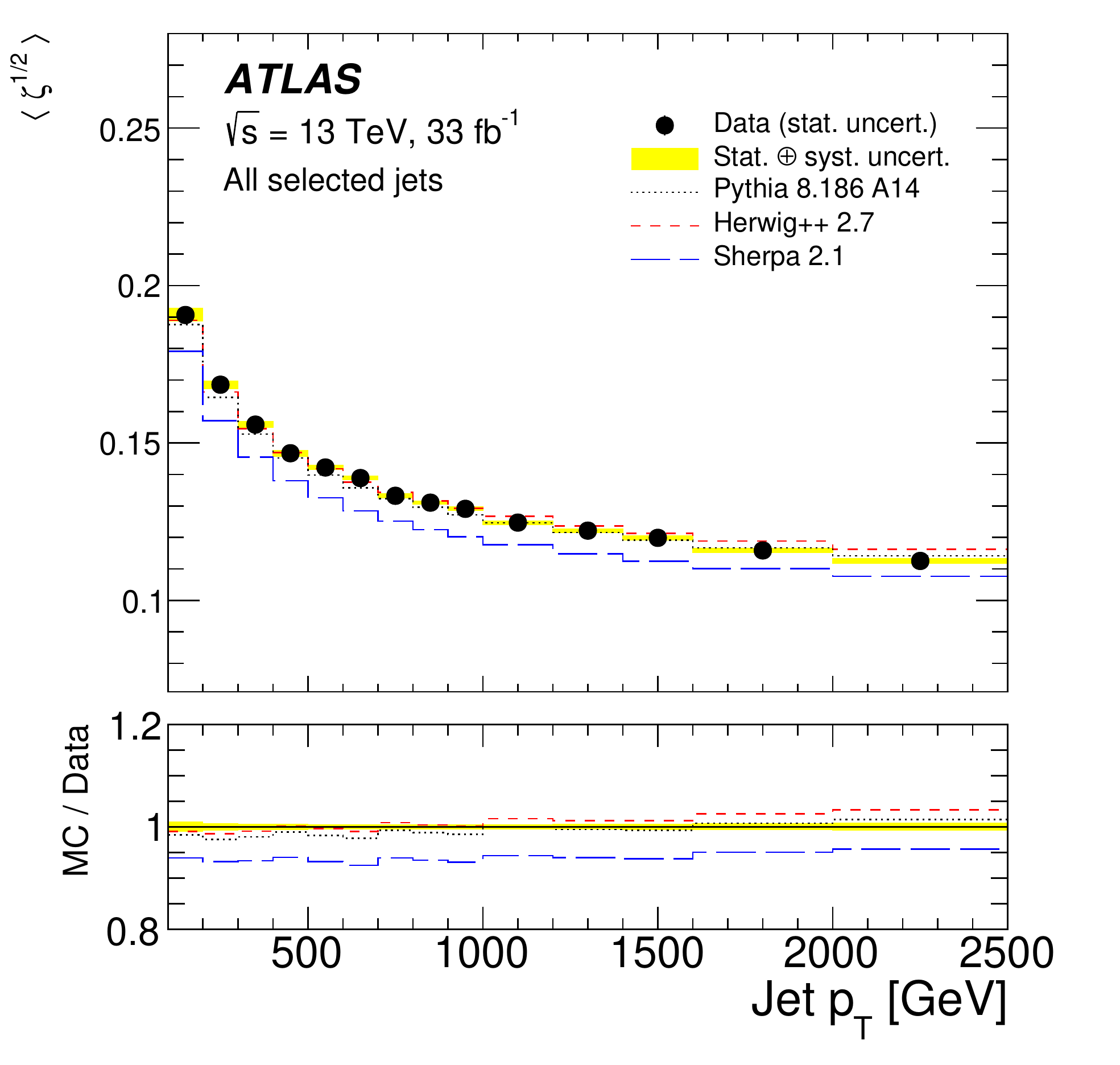}
\includegraphics[width=0.48\textwidth]{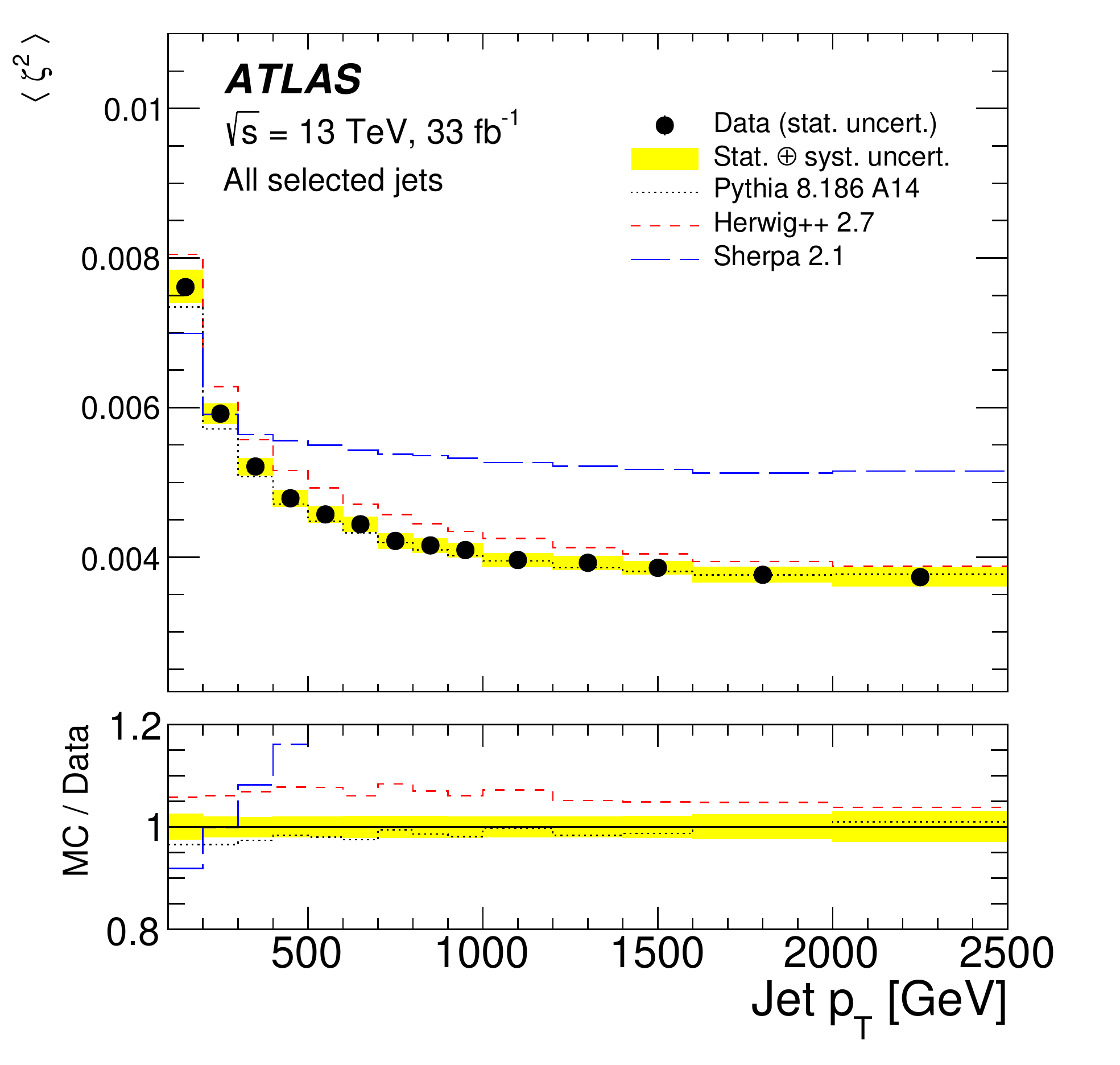}
\includegraphics[width=0.48\textwidth]{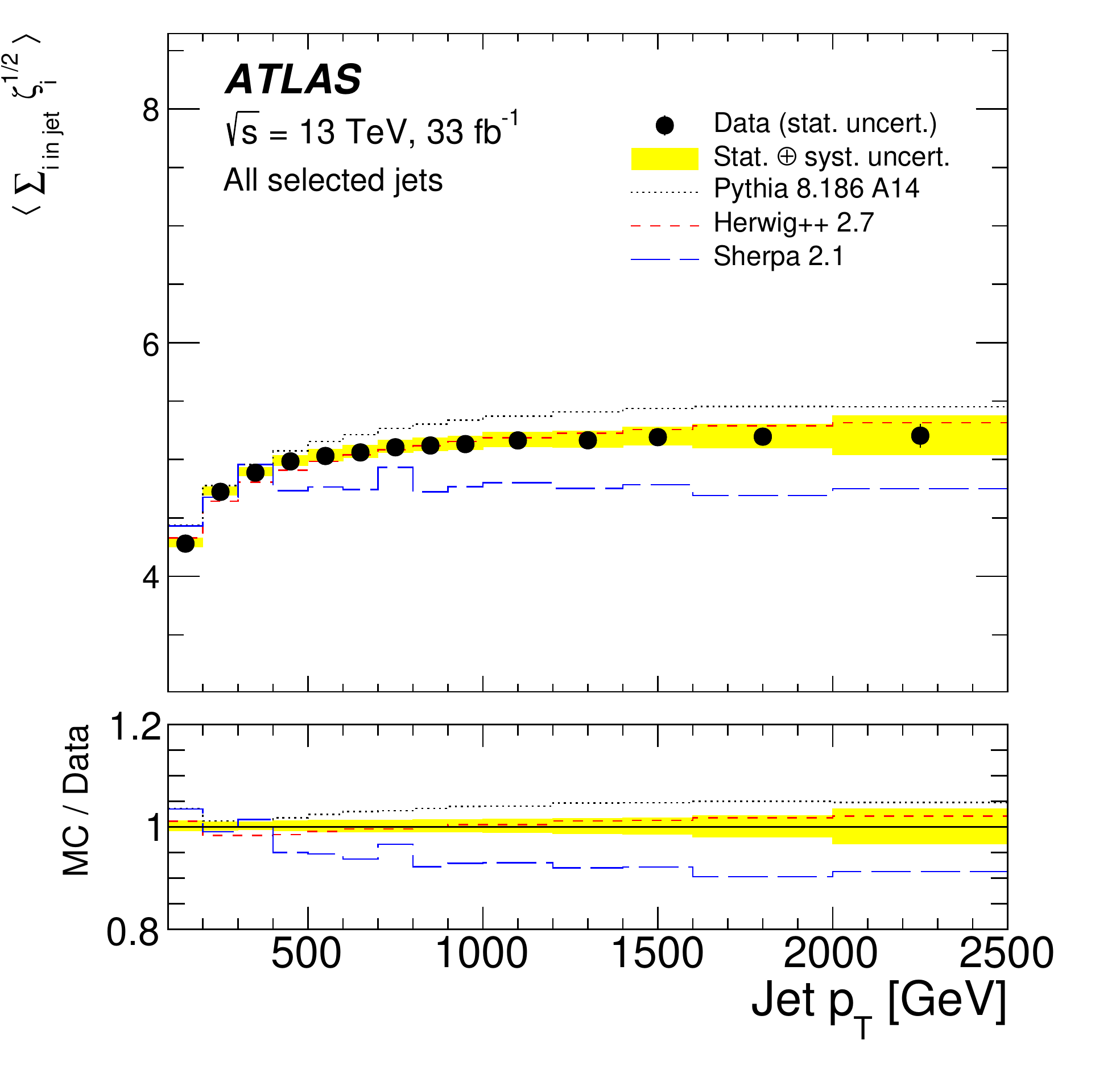}
\includegraphics[width=0.48\textwidth]{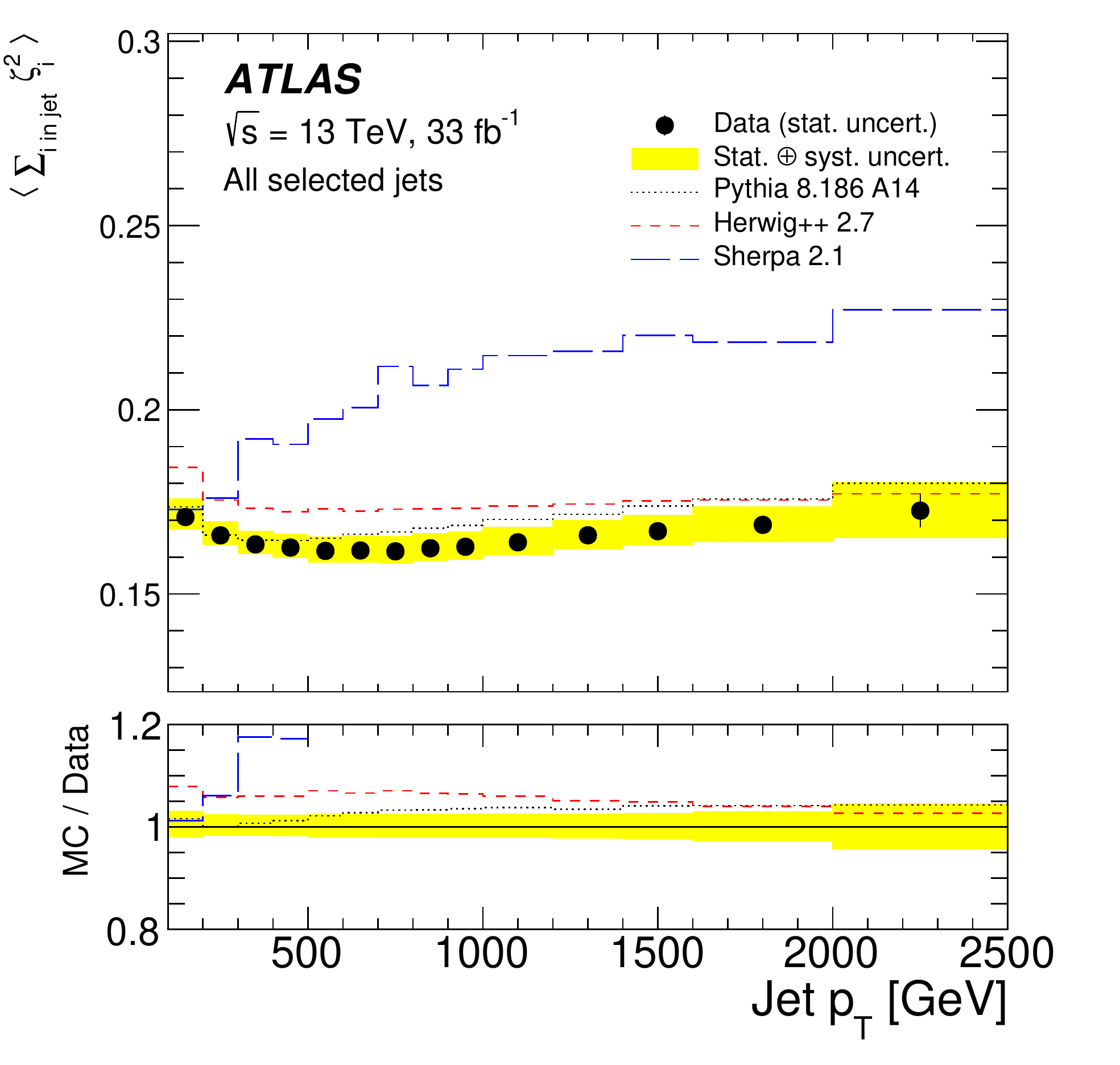}\\
\caption{Top: The unfolded (left) $1/2$ and (right) $2^\text{nd}$ moments of the transverse momentum fraction, $\zeta$, distribution as a function of the jet transverse momentum $\pT$.  Bottom: The unfolded moment of (left) $\sum_{i\in\text{jet}}\zeta^{1/2}$ and (right) $\sum_{i\in\text{jet}}\zeta^{2}$ as a function of the jet $\pT$.  Section~\ref{sec:obs} presents the definitions of both classes of observables.  The lower panels show the ratio of various MC predictions to the data, with the total uncertainty band centered on the data at unity.  The values are computed from the moments of the unfolded distributions in each jet-$\pT$ bin, with a small correction added to account for binning effects.}
\label{fig:result:inclusive:moments}
\end{figure}
 
\begin{figure}[htbp]
\centering
\includegraphics[width=0.48\textwidth]{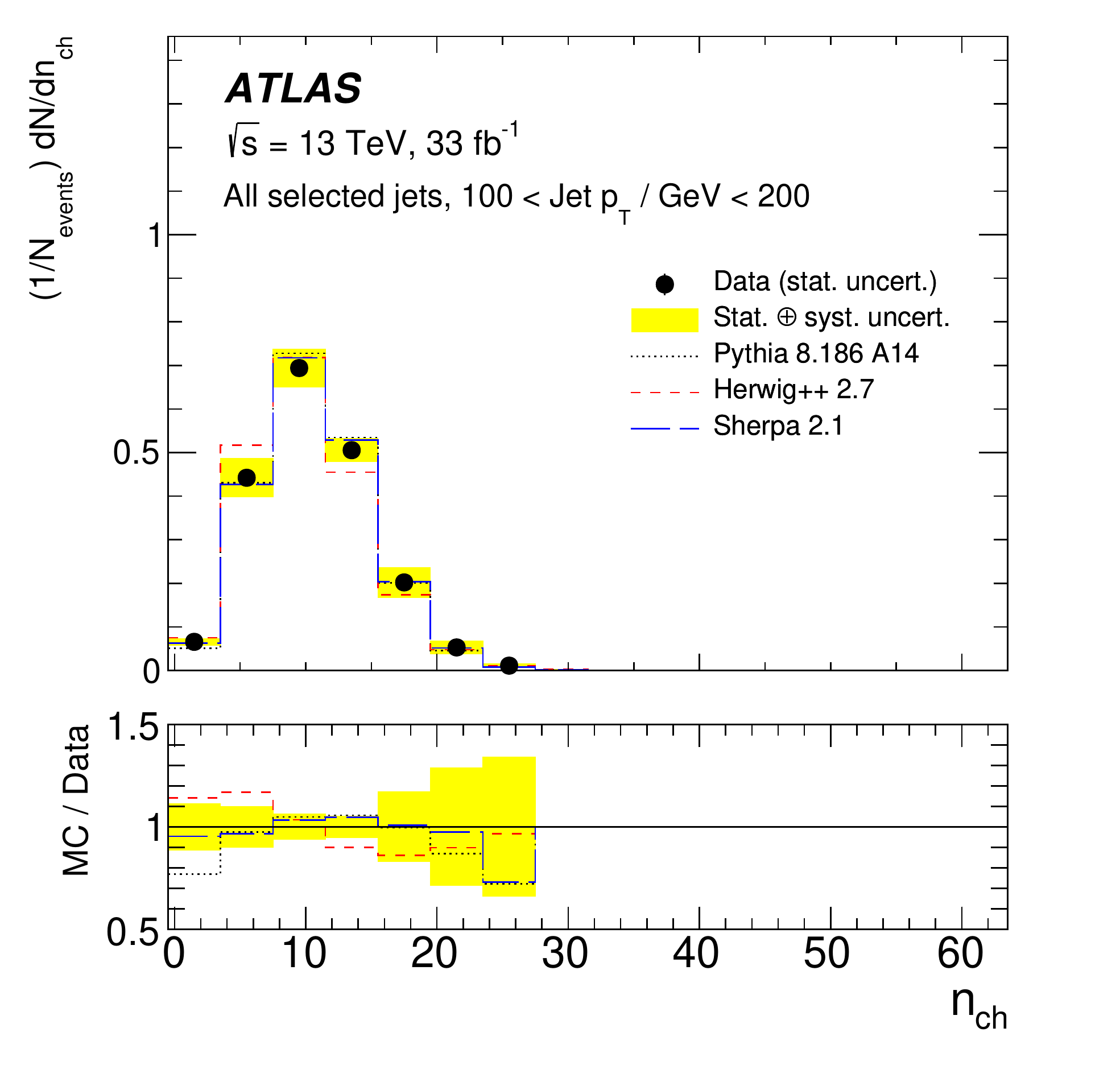}
\includegraphics[width=0.48\textwidth]{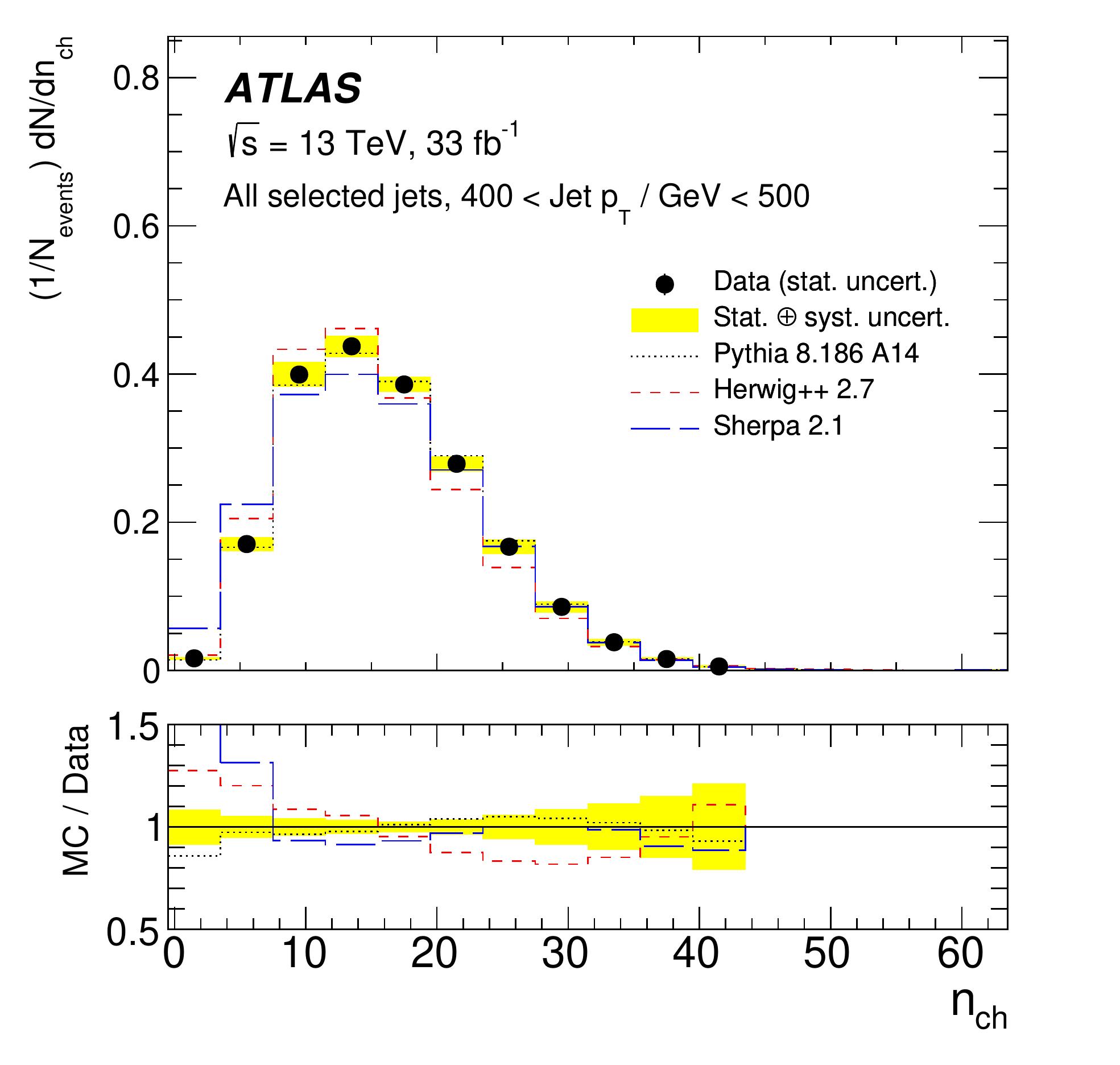}
\includegraphics[width=0.48\textwidth]{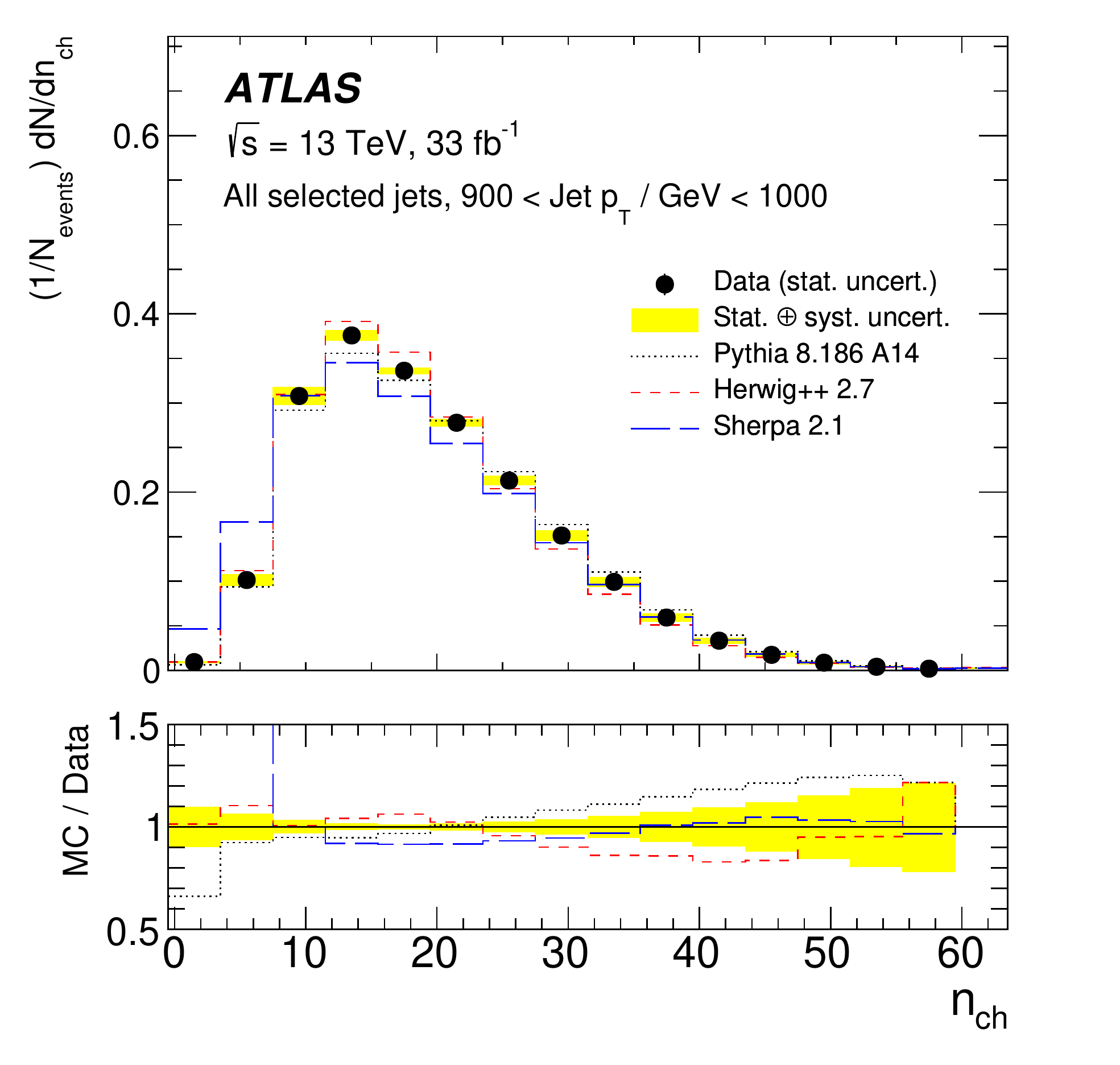}
\includegraphics[width=0.48\textwidth]{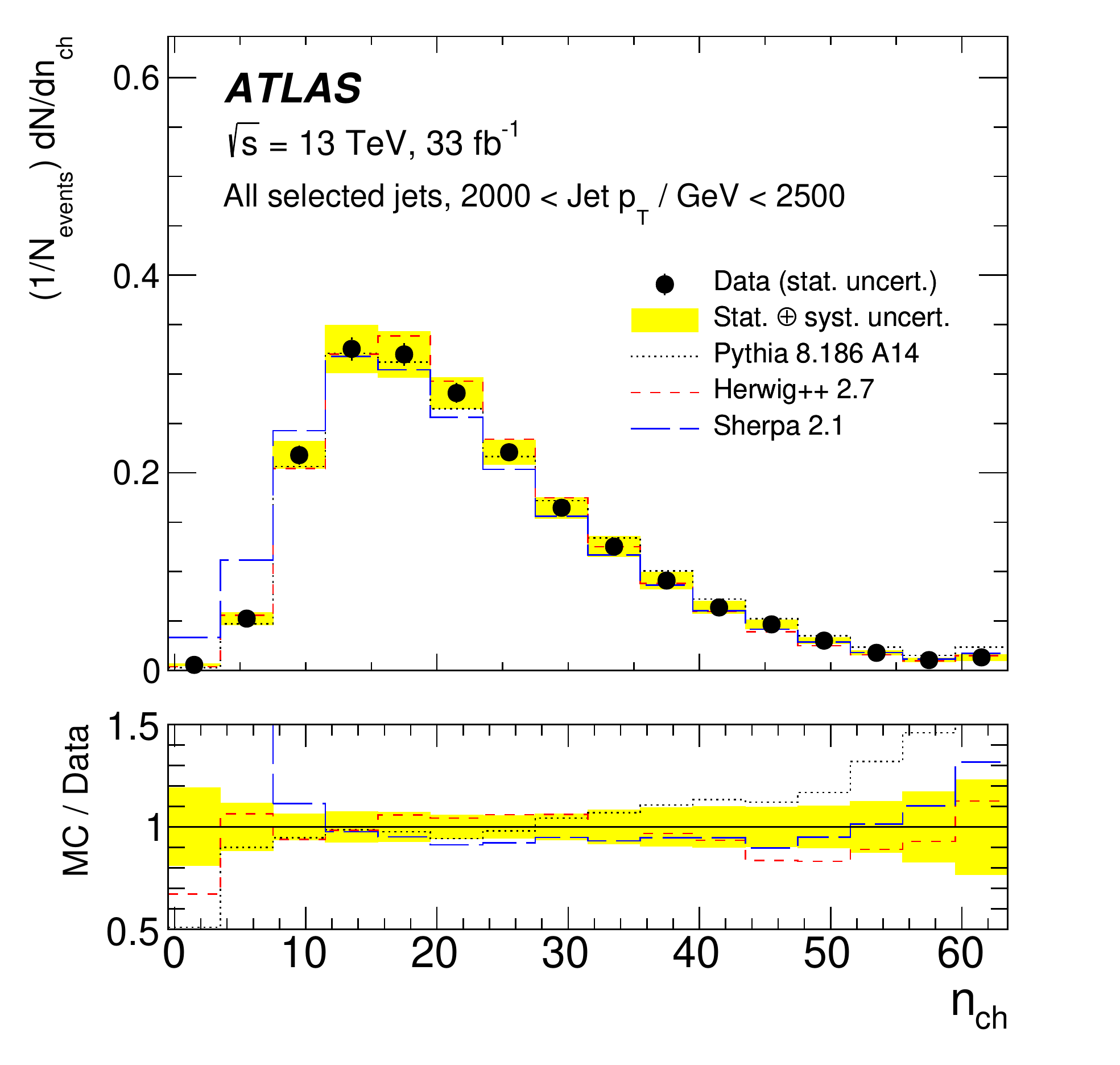}\\
\caption{The distribution of charged-particle multiplicity $n_\text{ch}$ in four bins of jet transverse momentum: (top left) 100~\GeV~$<\pTjet<200$~\GeV, (top right) 400~\GeV~$<\pTjet<500$~\GeV, (bottom left) 900~\GeV~$<\pTjet<1000$~\GeV, and (bottom right) 2000~\GeV~$<\pTjet<2500$~\GeV.   The lower panels show the ratio of various MC predictions to the data, with the total uncertainty band centered on the data at unity.  Additional $\pTjet$ bins can be found in Ref.~\cite{hepdata}.}
\label{fig:results:nch}
\end{figure}
 
\begin{figure}[htbp]
\centering
\includegraphics[width=0.48\textwidth]{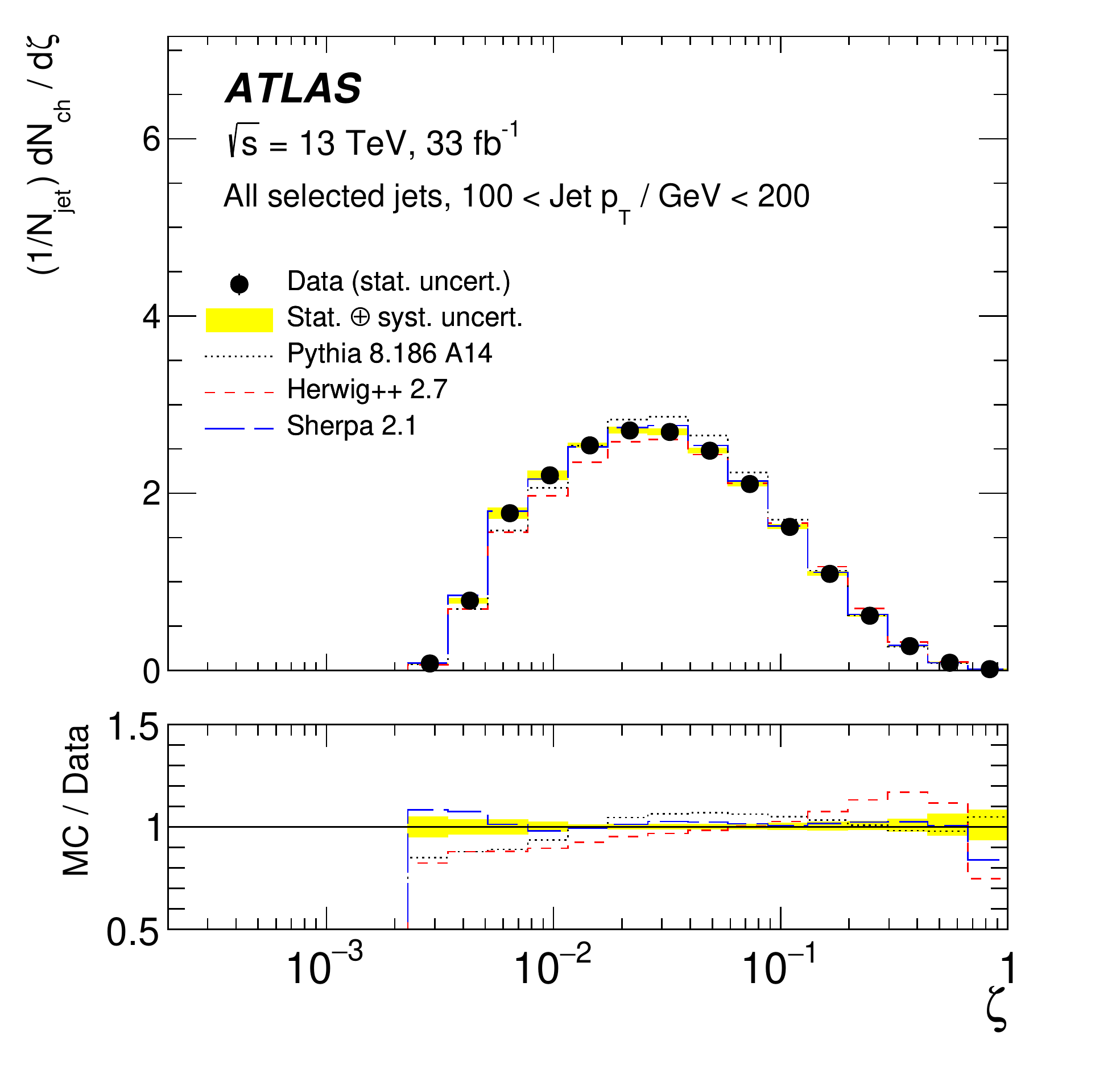}
\includegraphics[width=0.48\textwidth]{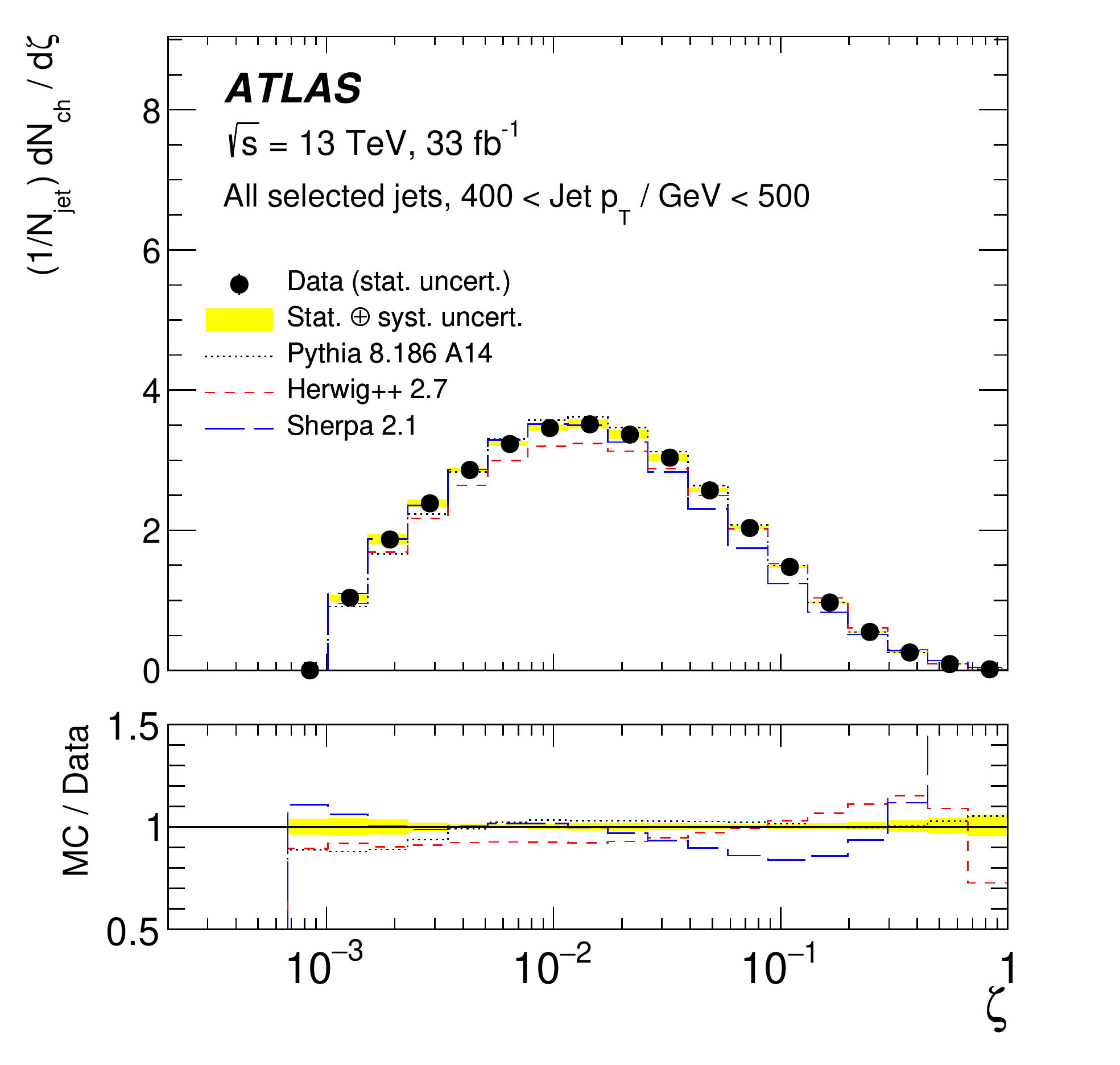}
\includegraphics[width=0.48\textwidth]{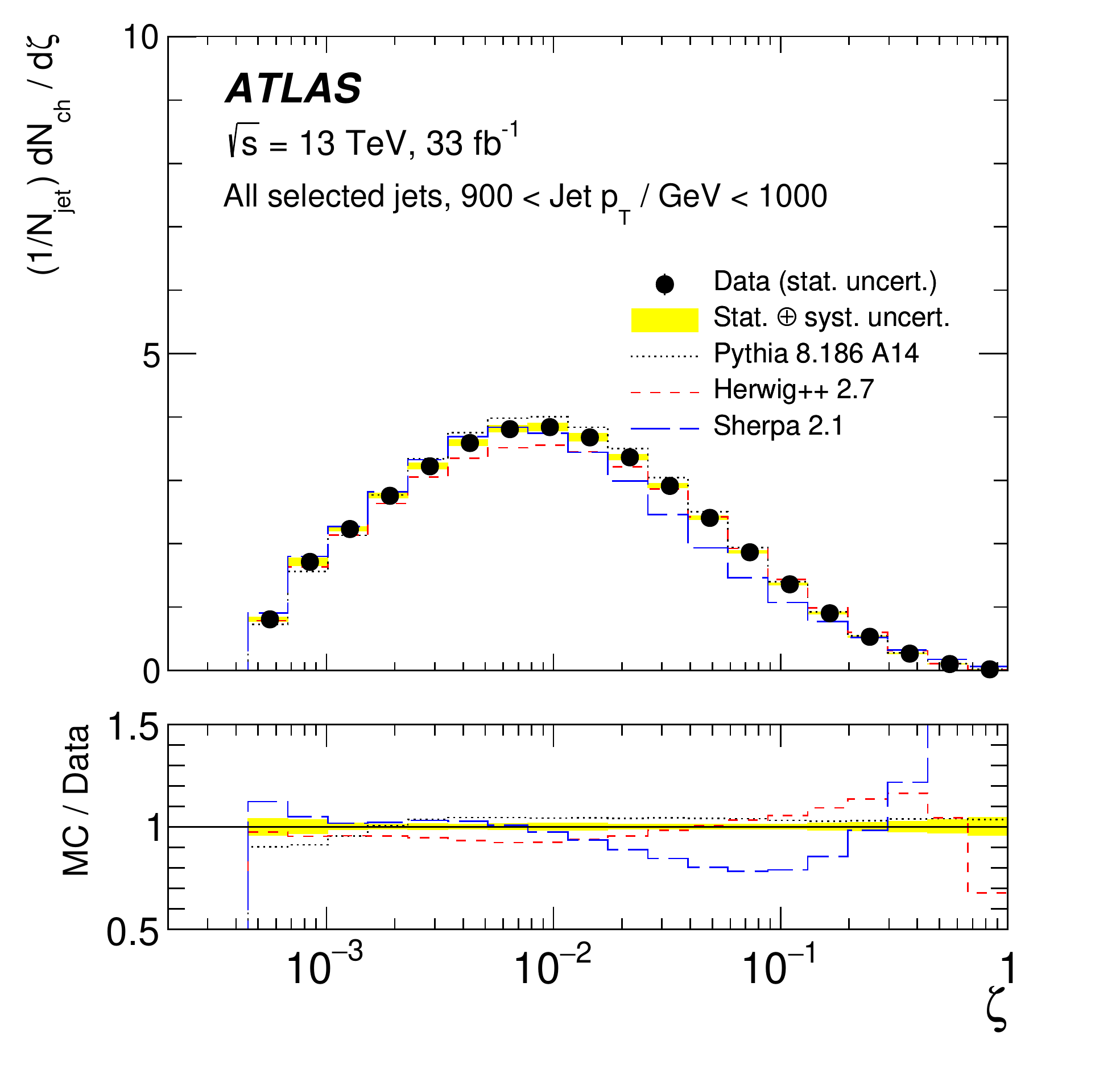}
\includegraphics[width=0.48\textwidth]{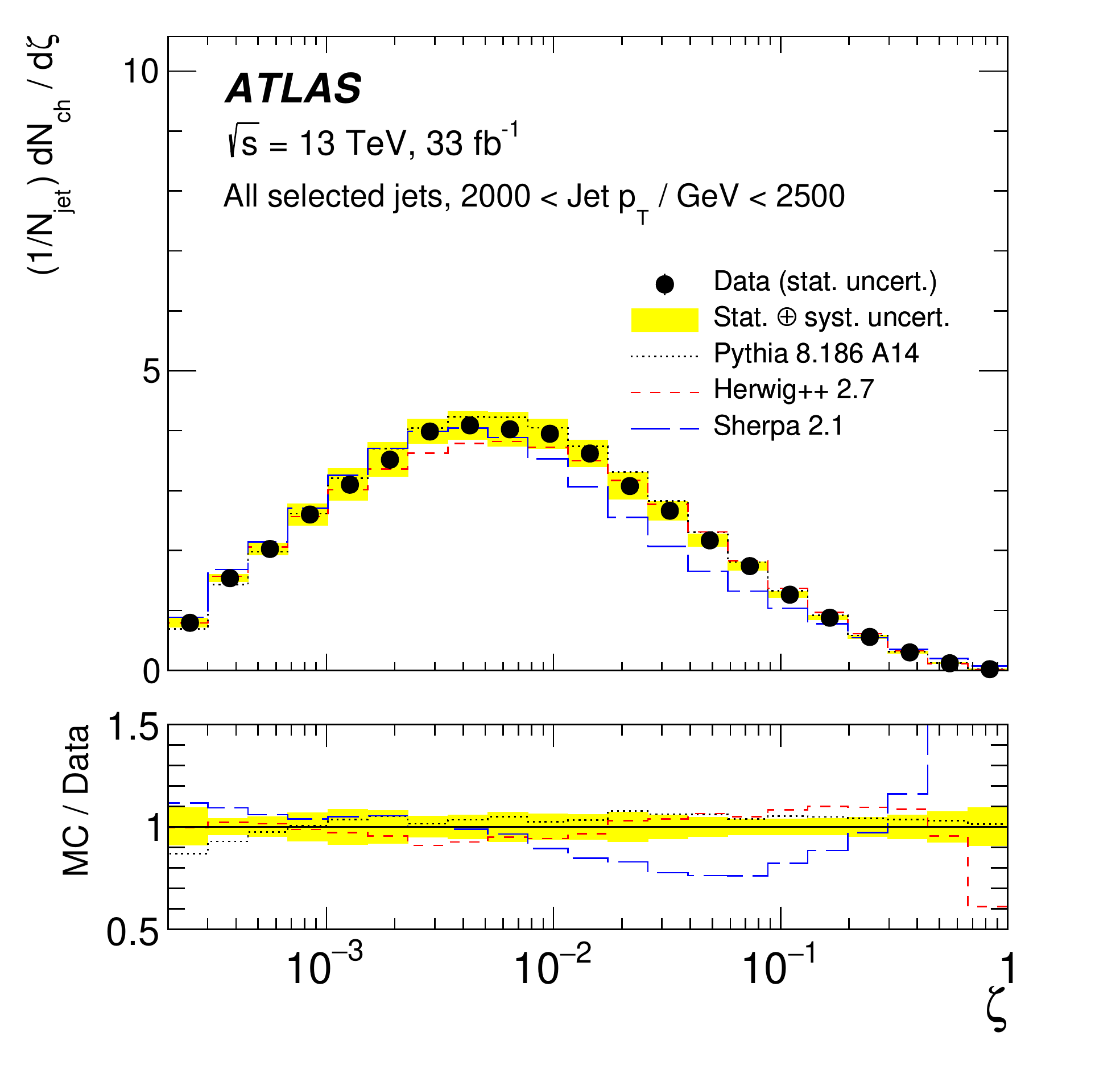}\\
\caption{The transverse momentum fraction $\zeta$ distribution in four bins of jet transverse momentum: (top left) 100~\GeV~$<\pTjet<200$~\GeV, (top right) 400~\GeV~$<\pTjet<500$~\GeV, (bottom left) 900~\GeV~$<\pTjet<1000$~\GeV, and (bottom right) 2000~\GeV~$<\pTjet<2500$~\GeV.   The lower panels show the ratio of various MC predictions to the data, with the total uncertainty band centered on the data at unity.   Additional $\pTjet$ bins can be found in Ref.~\cite{hepdata}.}
\label{fig:results:zeta}
\end{figure}
 
\begin{figure}[htbp]
\centering
\includegraphics[width=0.48\textwidth]{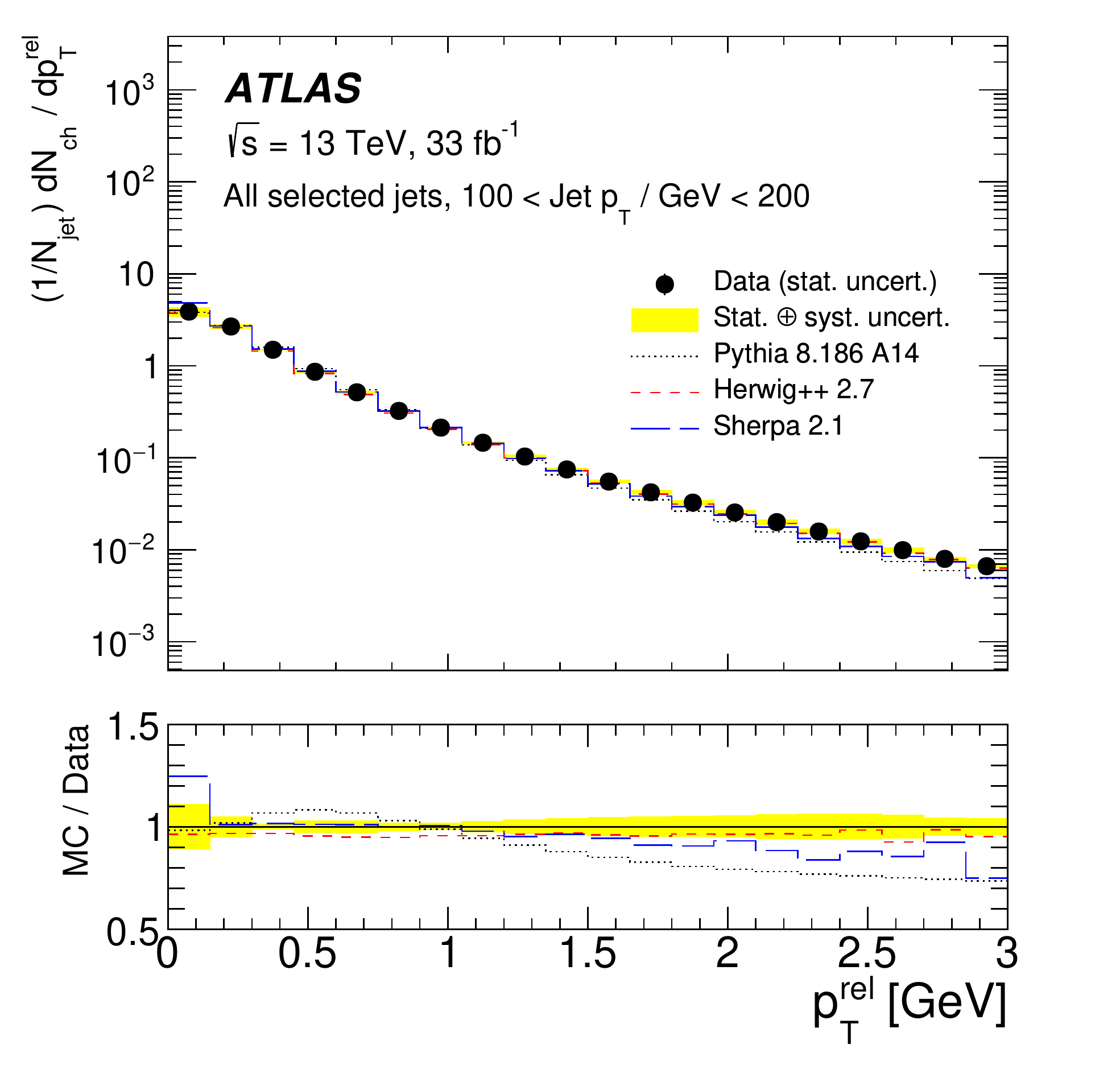}
\includegraphics[width=0.48\textwidth]{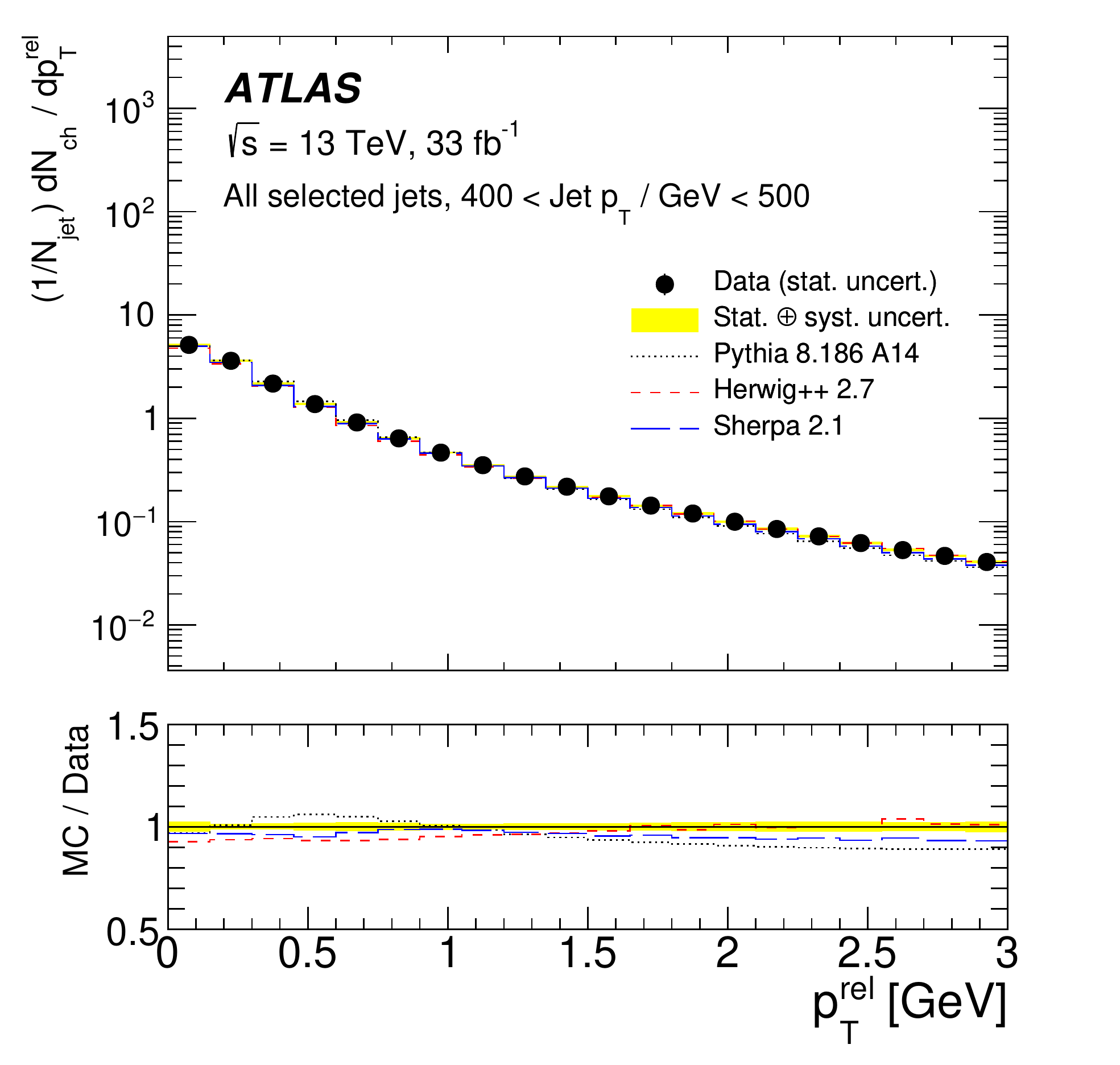}
\includegraphics[width=0.48\textwidth]{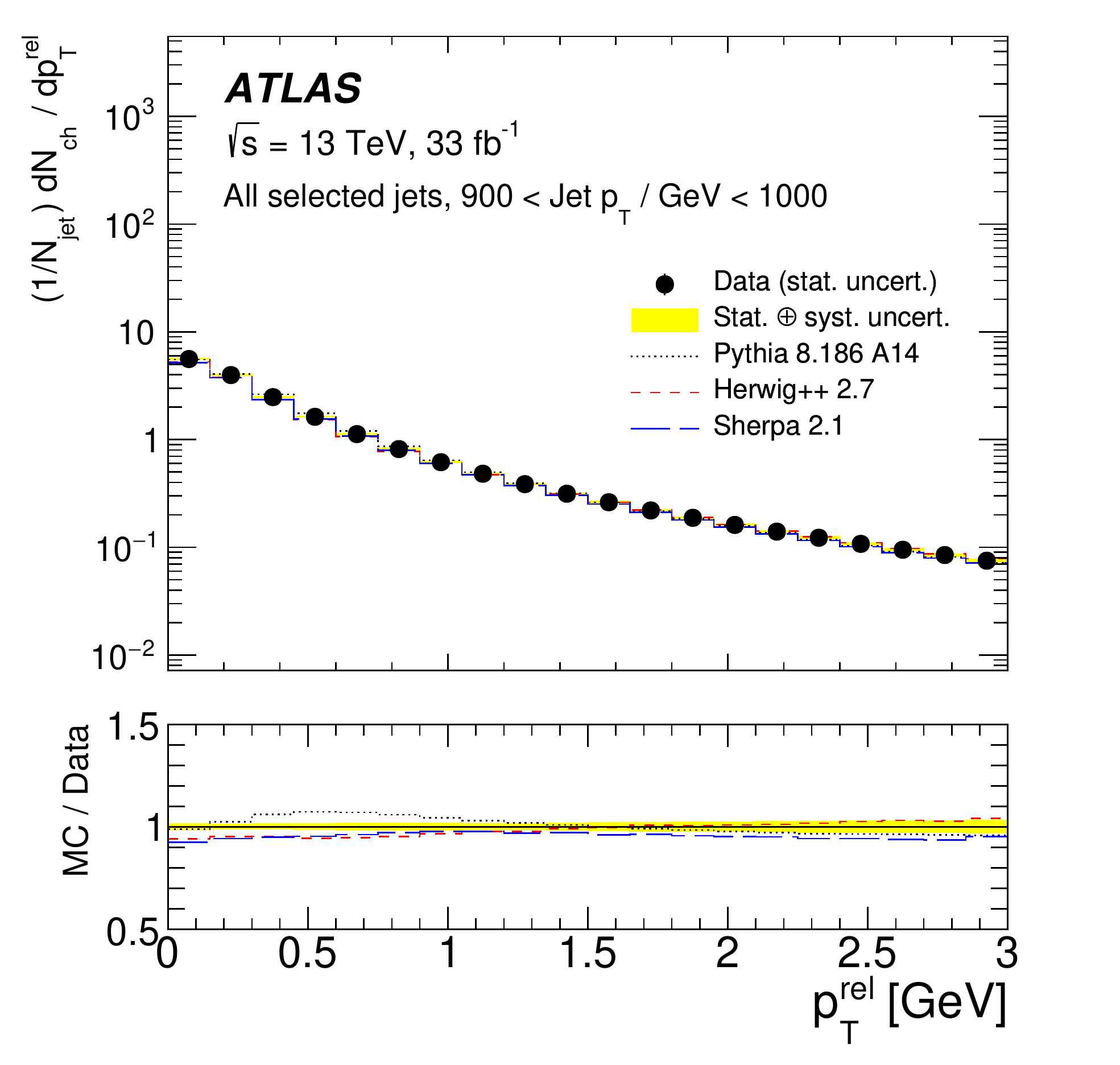}
\includegraphics[width=0.48\textwidth]{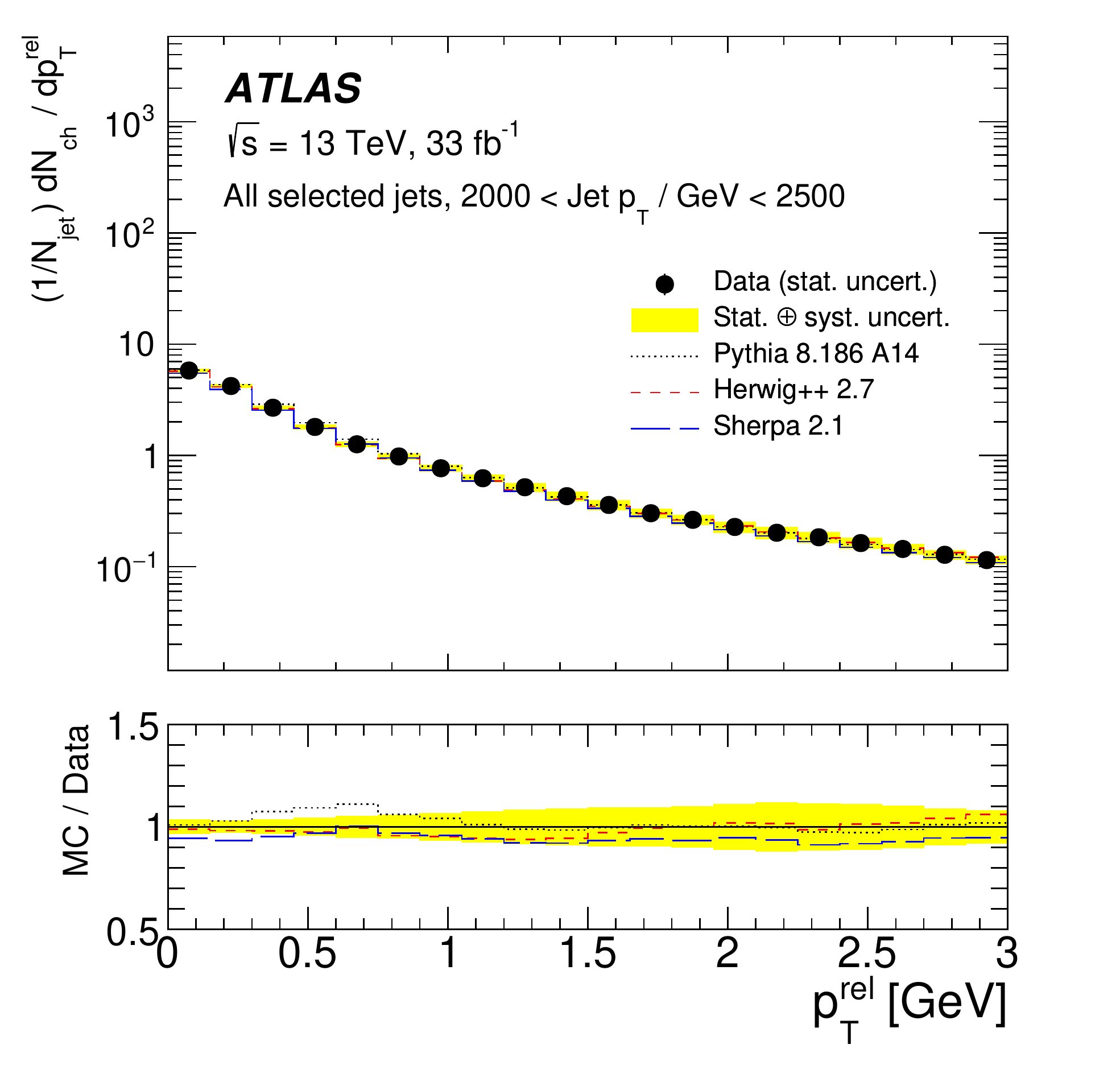}\\
\caption{The transverse momentum $\pTrel$ distribution in four bins of jet transverse momentum: (top left) 100~\GeV~$<\pTjet<200$~\GeV, (top right) 400~\GeV~$<\pTjet<500$~\GeV, (bottom left) 900~\GeV~$<\pTjet<1000$~\GeV, and (bottom right) 2000~\GeV~$<\pTjet<2500$~\GeV.   The lower panels show the ratio of various MC predictions to the data, with the total uncertainty band centered on the data at unity.   Additional $\pTjet$ bins can be found in Ref.~\cite{hepdata}.}
\label{fig:results:ptrel}
\end{figure}
 
\begin{figure}[htbp]
\centering
\includegraphics[width=0.48\textwidth]{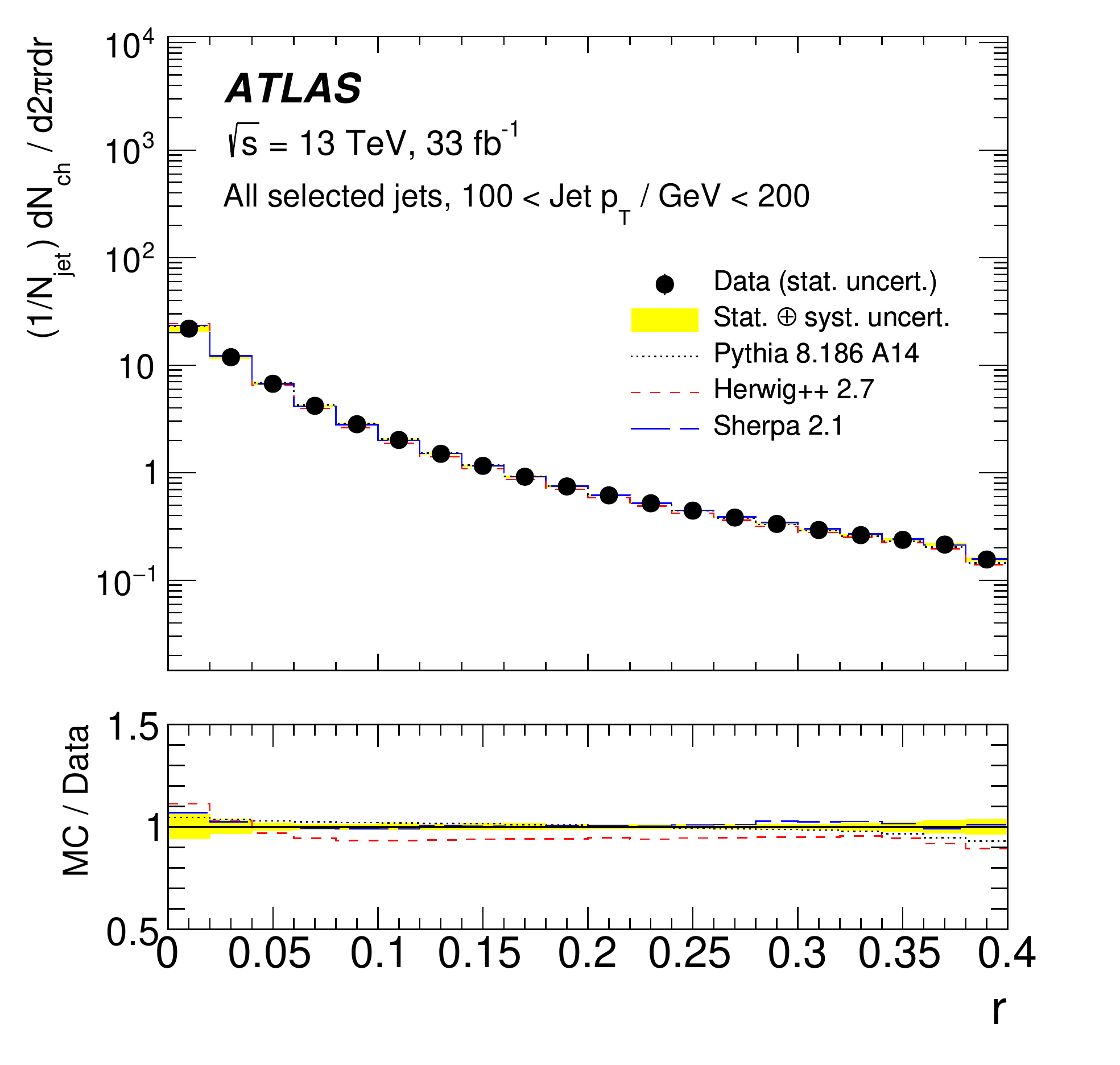}
\includegraphics[width=0.48\textwidth]{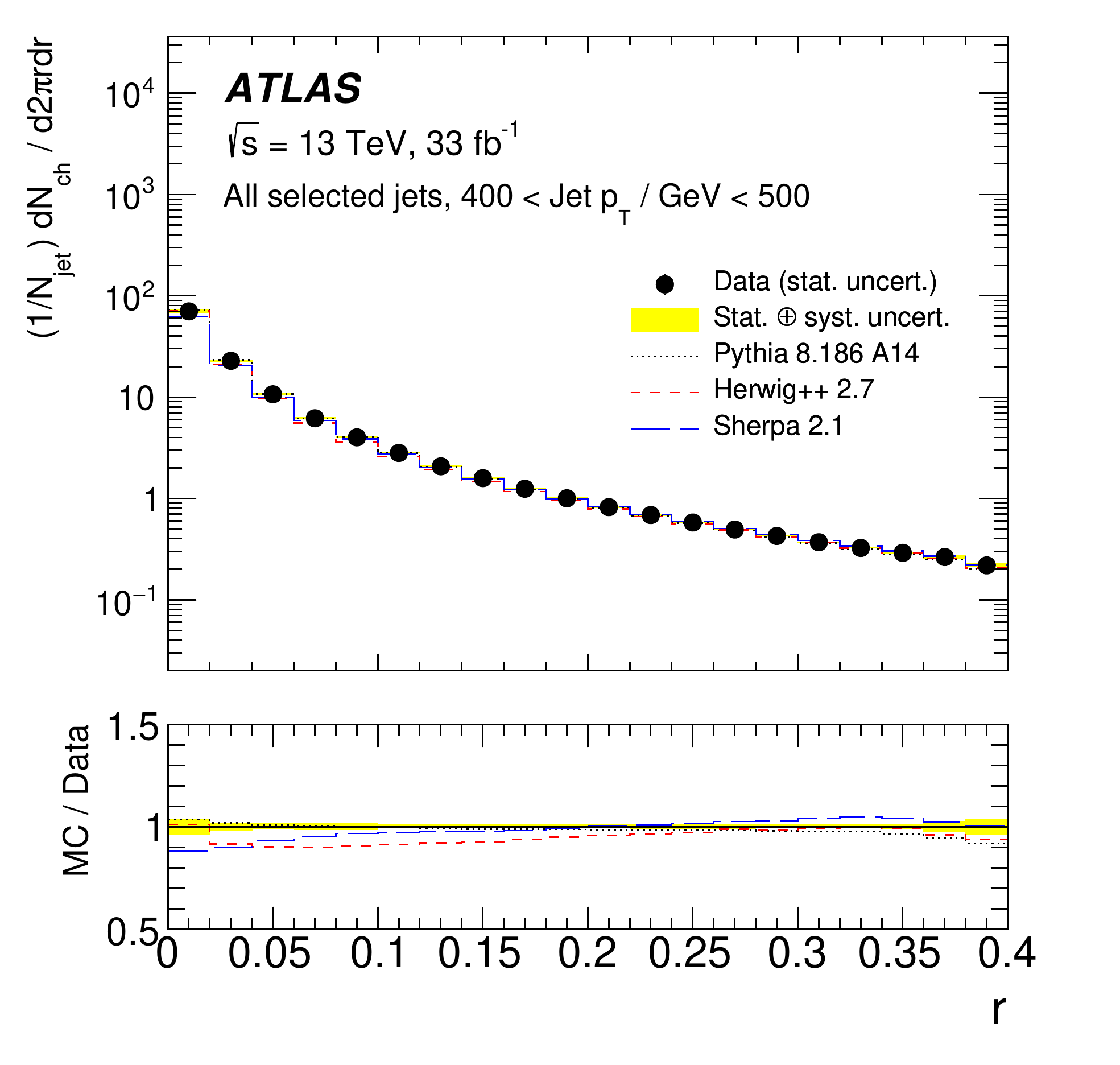}
\includegraphics[width=0.48\textwidth]{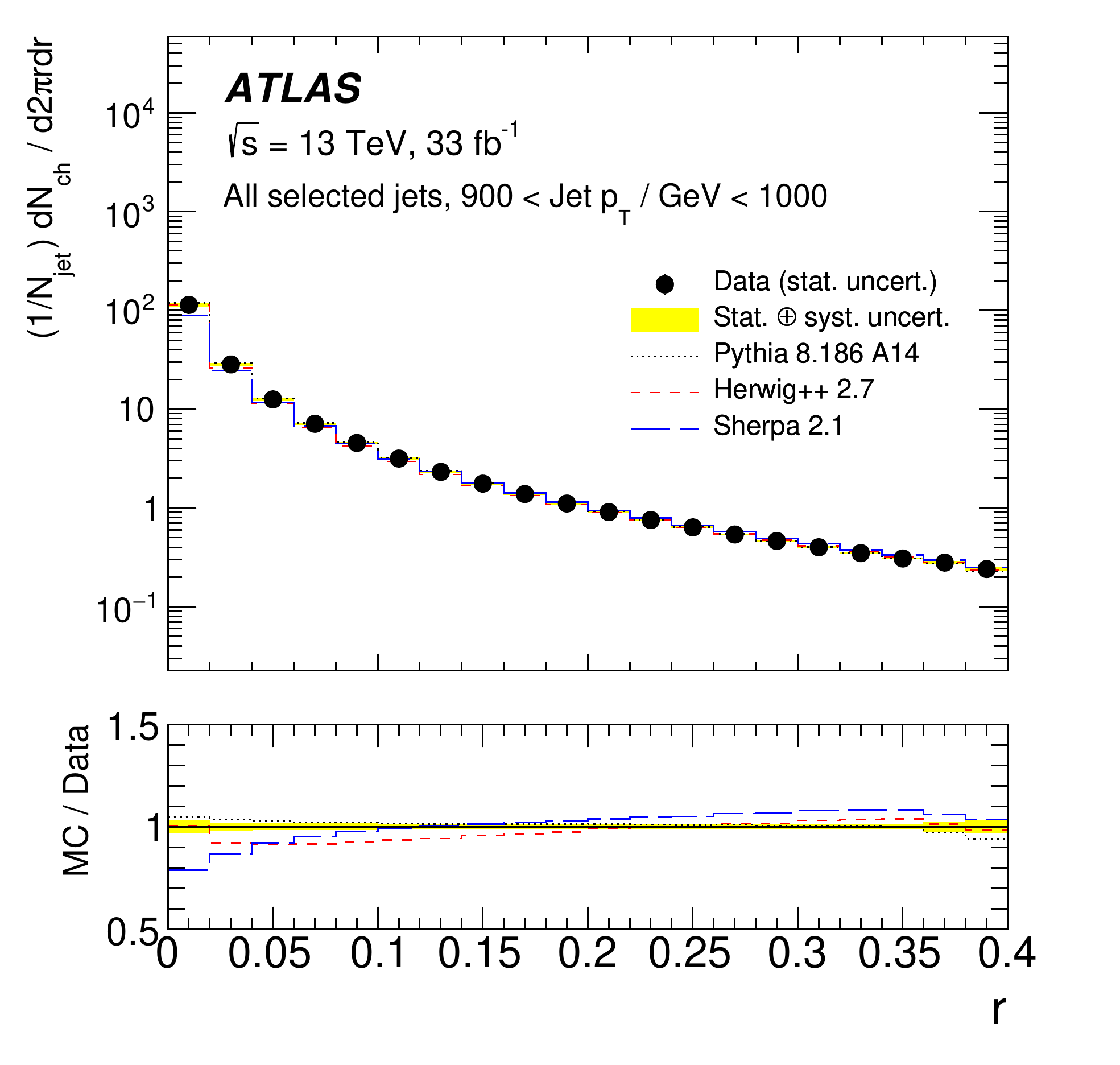}
\includegraphics[width=0.48\textwidth]{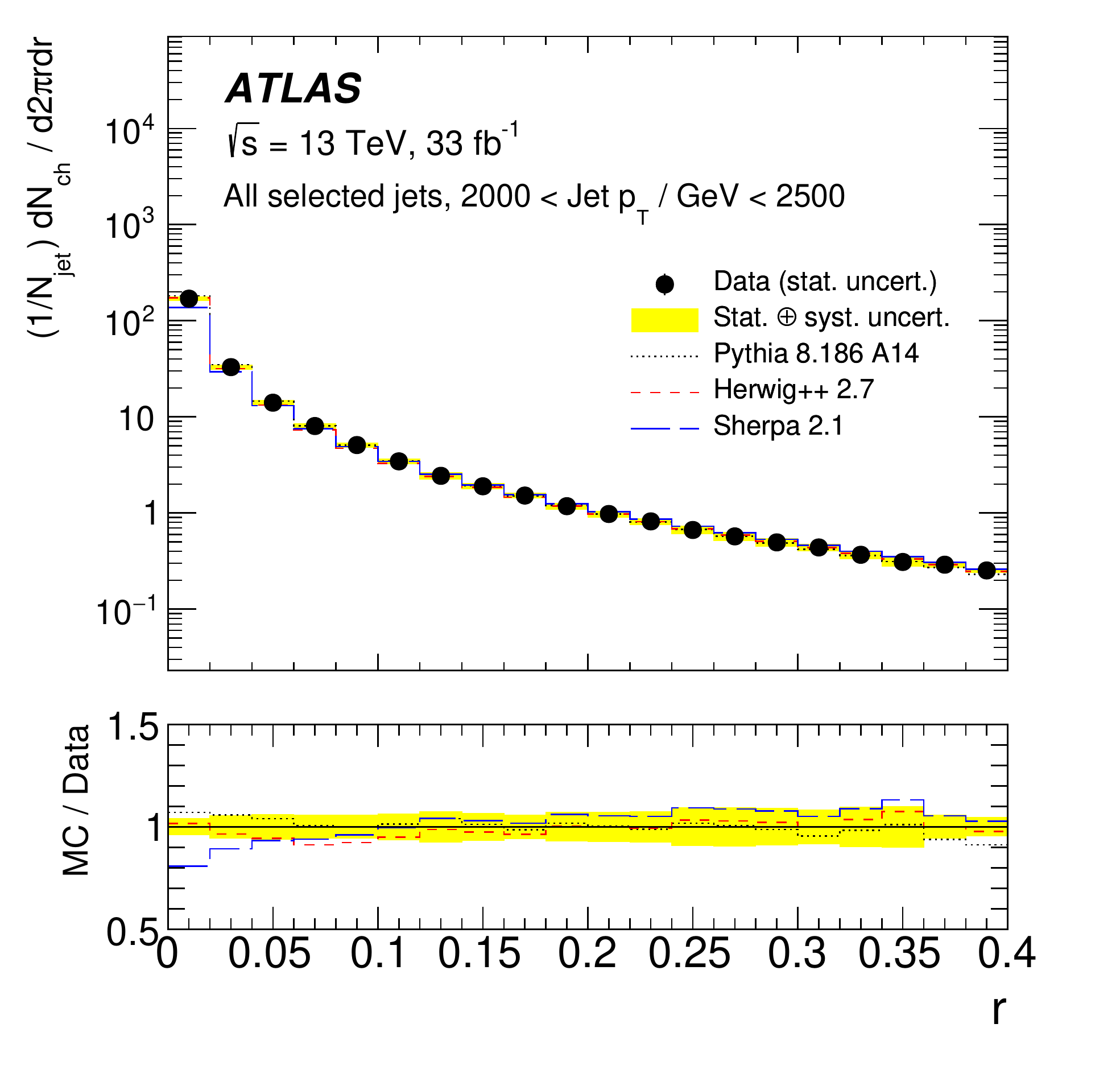}\\
\caption{Distribution of track radial profile in bins of the radial distance $r$ from the jet axis in four bins of jet transverse momentum: (top left) 100~\GeV~$<\pTjet<200$~\GeV, (top right) 400~\GeV~$<\pTjet<500$~\GeV, (bottom left) 900~\GeV~$<\pTjet<1000$~\GeV, and (bottom right) 2000~\GeV~$<\pTjet<2500$~\GeV.   The lower panels show the ratio of various MC predictions to the data, with the total uncertainty band centered on the data at unity.  Additional $\pTjet$ bins can be found in Ref.~\cite{hepdata}. }
\label{fig:results:r}
\end{figure}
 
\clearpage
 
\subsection{Quark and gluon distributions}
\label{sec:results:qg}
 
As discussed in Section~\ref{sec:events}, the more forward and the more central of the two selected jets can be separated to study differences between the radiation patterns within quark and gluon jets.  Using the fraction of quark jets $f_q$ in the two jet samples (forward $f$ and central $c$), one can extract the quark ($h_i^q$) and gluon ($h_i^g$) jet fragmentation properties separately by solving a system of equations per bin $i$ of an observable:
 
\begin{align}
\label{eq:mix}
h_i^f&=f_q^fh_i^q+(1-f_q^f)h_i^g,\\\label{eq:mix2}
h_i^c&=f_q^ch_i^q+(1-f_q^c)h_i^g,
\end{align}
 
\noindent where $f_q^x$ is the fraction of quark jets in sample $x$ (see Figure~\ref{fig:qg} for the gluon fraction) and the nominal fractions are taken from the default \PYTHIA simulation described in Section~\ref{sec:analysis}.  The flavor of a jet is defined as the type of the highest-energy parton from the event record (all partons prior to hadronization) matched to the jet via ghost association.  This definition is not unique because quark and gluon labels are not universal due to color connections with other partons in the event.\footnote{However, for isolated jets, the topology dependence is predicted to be much smaller than the difference between quark and gluon jets~\cite{Bright-Thonney:2018mxq}. }  In addition to the uncertainty in $h_i^f$ and $h_i^c$ from the unfolding method, uncertainties in the extracted $h_i^g$ and $h_i^q$ distributions arise from the PDF choice, from the matrix elements, from the fragmentation model (due to flavor changing), and from the method non-closure.  The determination of the uncertainty from the choice of PDF uses the NNPDF uncertainty set (NNPDF 2.3 at LO in QCD and QED with $\alphas(m_Z)=0.119$) and the matrix-element uncertainty is estimated by comparing the nominal fractions from \PYTHIA with those from \HERWIG.\footnote{These two generators also use different PDF sets, so this uncertainty is double-counted in the overall uncertainty.}  The non-closure uncertainty is due to the small (sub-percent level) differences between forward and central quark jets, as well as forward and central gluon jets, resulting from an $\eta$ dependence in the jet fragmentation at a fixed jet $\pT$~\cite{Bright-Thonney:2018mxq}.  When presenting the average properties in bins of jet $\pT$, the binning correction described in Section~\ref{sec:unfold} is also applied and the corresponding uncertainty contributes to the total uncertainty (though it is smaller than other sources of uncertainty).
 
The matrix-element uncertainty dominates the total uncertainty in the extraction procedure, resulting in an uncertainty that is about 1\% at high jet $\pT$ and about 5\% at low to moderate jet $\pT$ for quark jets, with the inverse trend for gluon jets (low uncertainty at low jet $\pT$ and large uncertainty at high jet $\pT$).  The extractions presented here use leading-order matrix elements and leading-logarithm parton shower programs; higher-order effects that modify the fractions $f$ are not included in this leading-order extraction.  Figure~\ref{fig:extractedMCntrackdata} shows the extracted quark and gluon distributions for jets with $1000$~\GeV\ $<\pTjet< 1200$~\GeV.   To reinforce the simulation dependence of these extractions, the data distributions are referred to as `extracted quark-like' and `extracted gluon-like'.
 
\clearpage
 
A key challenge with the extraction based on Eqs.~(\ref{eq:mix}) and~(\ref{eq:mix2}) is that it strongly depends on simulation for the fractions $f_q$ and $f_g$.  A new approach that does not require the input of any fractions is topic modeling~\cite{Metodiev:2018ftz,Komiske:2018vkc}, which holds great promise for learning about quark- and gluon-like jets with less input from theory.   In this approach, one can extract distributions of `topics' $T_1$ and $T_2$:
 
\begin{align*}
h^{T_1}_i &= \frac{h^f_i-\left(\text{min}_j \{h^f_j/h^c_j\}\right)\times h^c_i}{1-\text{min}_j h^f_j/h^c_j}, \\
h^{T_2}_i &= \frac{h^c_i-\left(\text{min}_j \{h^c_j/h^f_j\}\right)\times h^f_i}{1-\text{min}_j h^c_j/h^f_j} .
\end{align*}
 
\noindent In the limit that $\text{min}_j \{h^g_j/h_j^q\}=\text{min}_j \{h^q_j/h^g_j\}=0$, $h^{T_1}=h^q$ and $h^{T_2}=h^g$.  When this is not exactly the case, the topics are universal but not pure combinations of quarks and gluons. The extracted topics using $n_\text{ch}$ in two jet $\pT$ bins are shown in Figure~\ref{fig:extractedMCntrackdatatopics}.  The very low $n_\text{ch}$ region is dominated by quarks and the very high $n_\text{ch}$ region is dominated by gluons and therefore $n_\text{ch}$ nearly has the property that $\text{min}_j \{h^g_j/h_j^q\}\approx \text{min}_j \{h^q_j/h^g_j\}\approx 0$.  Therefore, the first topic is well aligned with quarks and the second topic is more gluon-like.  This alignment is better for quarks than for gluons, but the second topic does converge to the gluon distribution at high jet $\pT$.  Other observables aside from $n_\text{ch}$ are not considered for topic modeling because there are no bins where $h^g_j/h_j^q=0$ or $h^q_j/h^g_j=0$ is approximately true and therefore the topics do not align with quark- and gluon-like quantities.
 
\begin{figure}[htbp]
\centering
\includegraphics[width=0.5\textwidth]{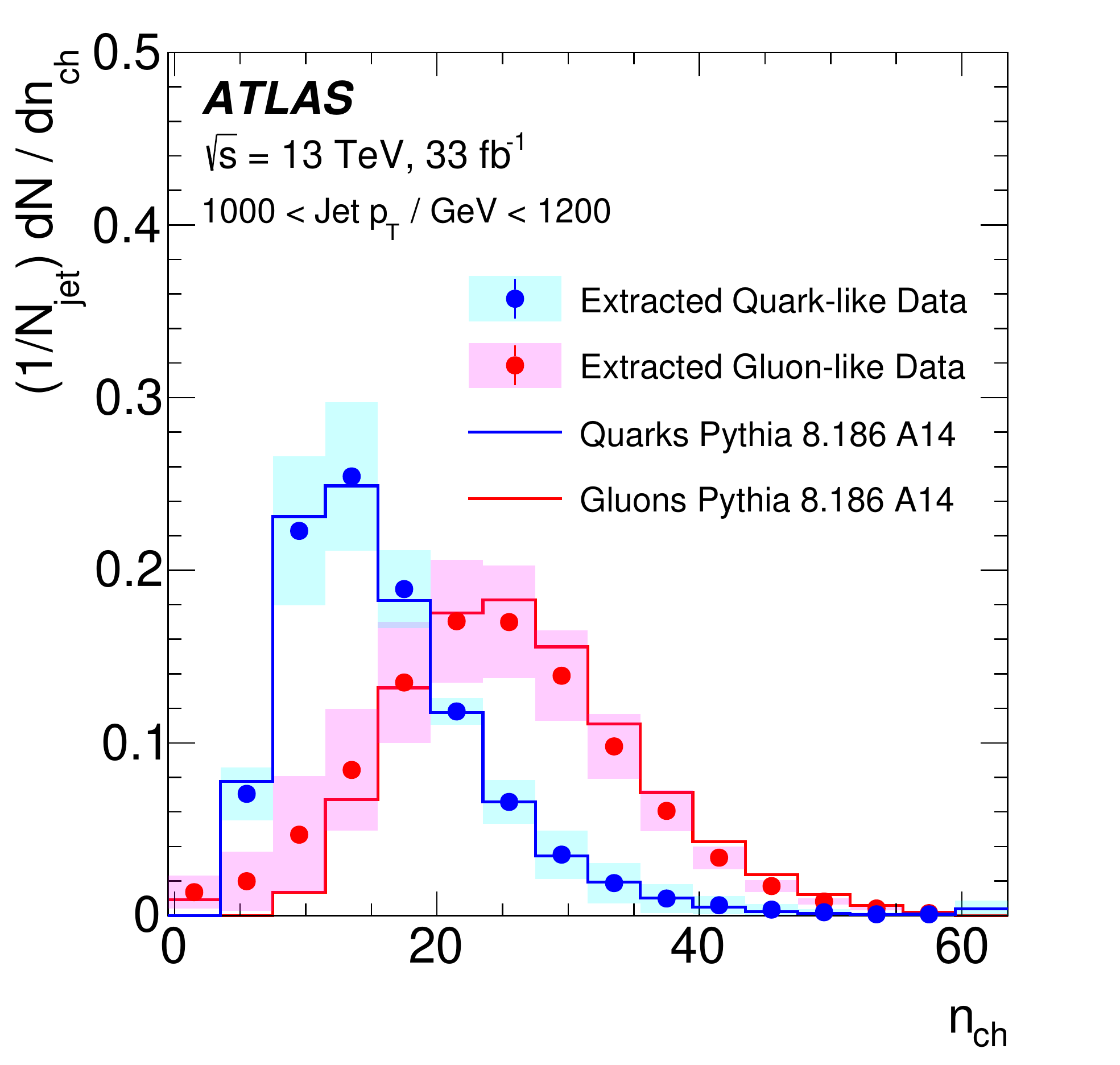}\includegraphics[width=0.5\textwidth]{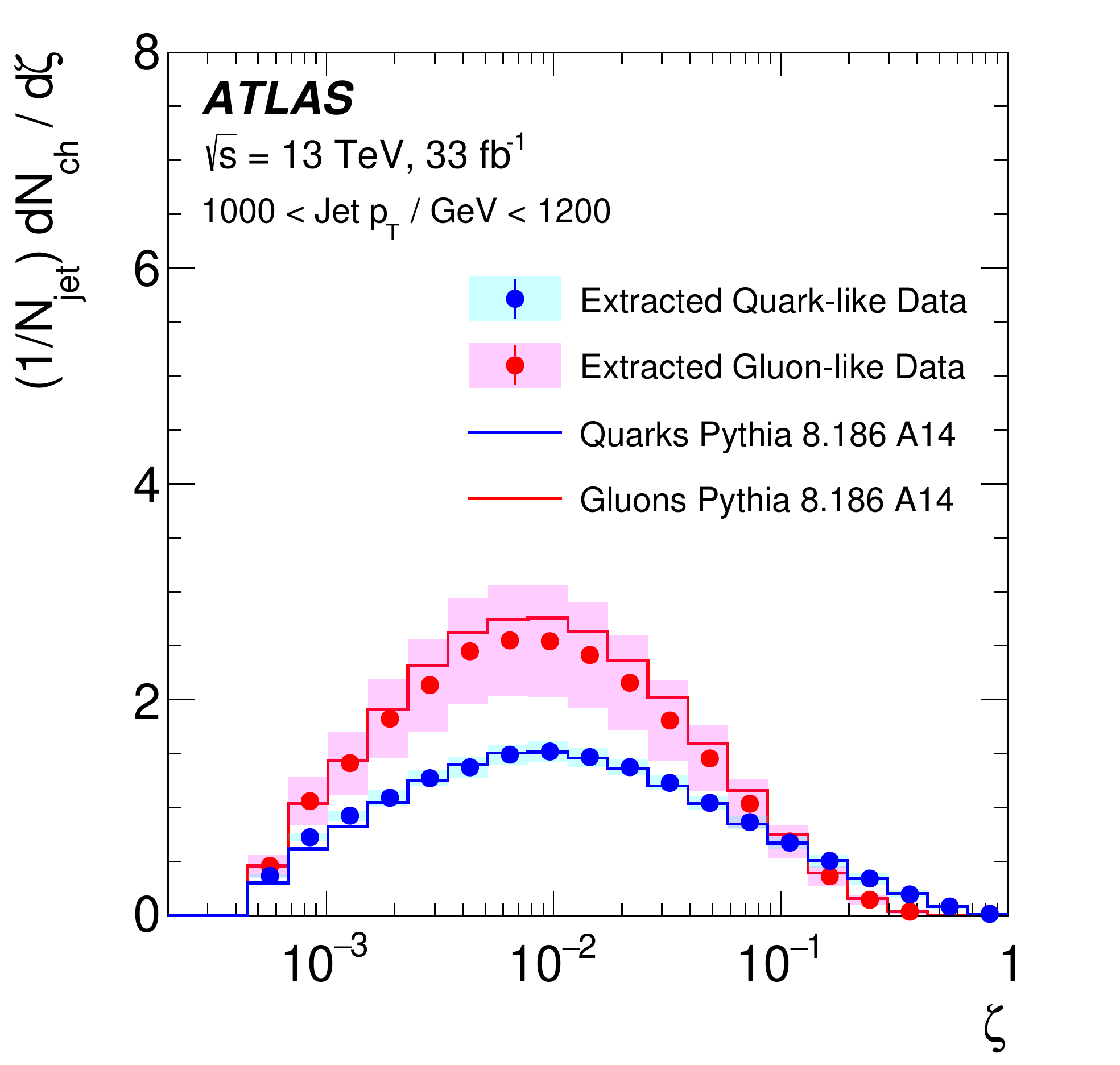}\\
\includegraphics[width=0.5\textwidth]{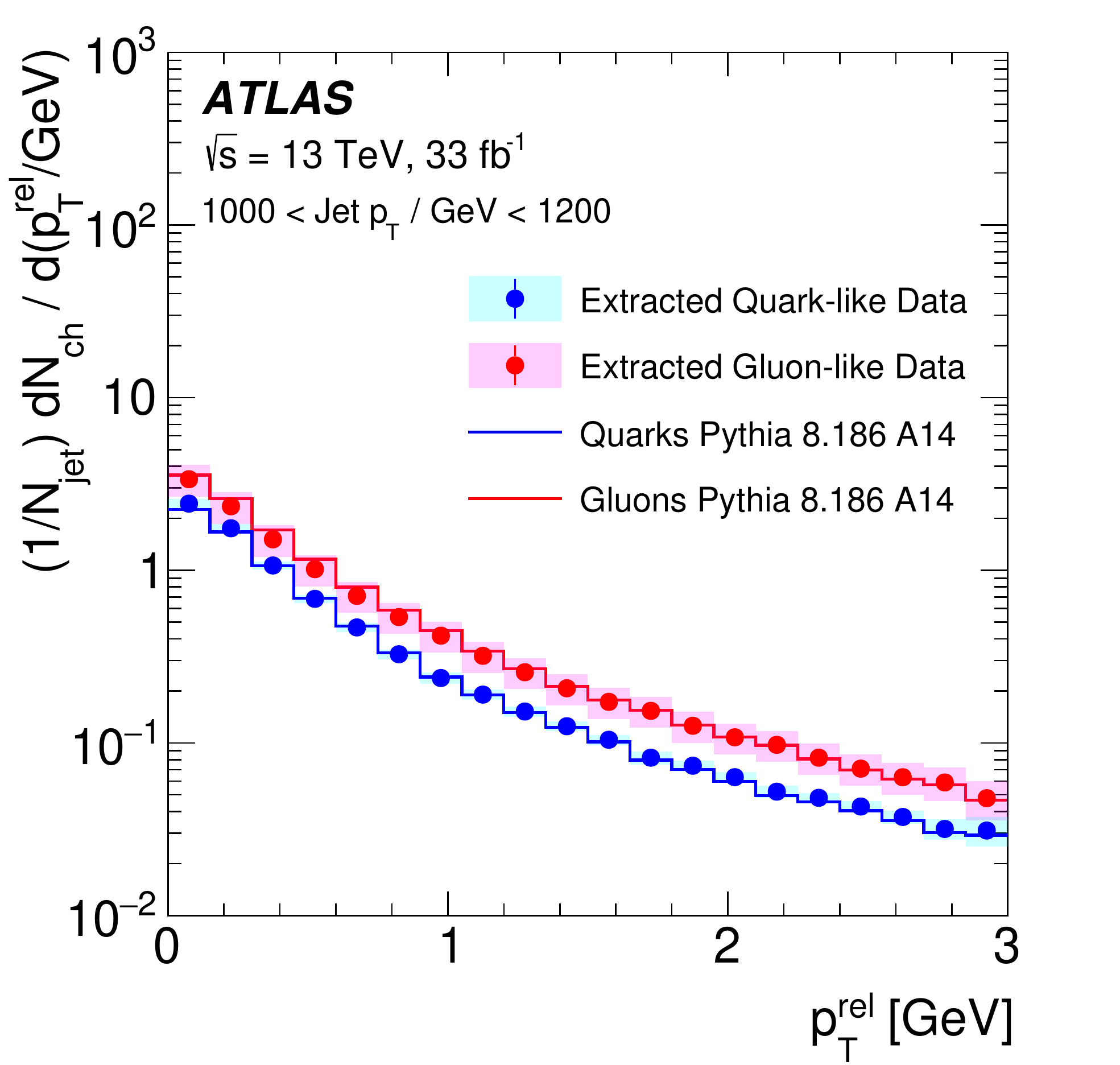}\includegraphics[width=0.5\textwidth]{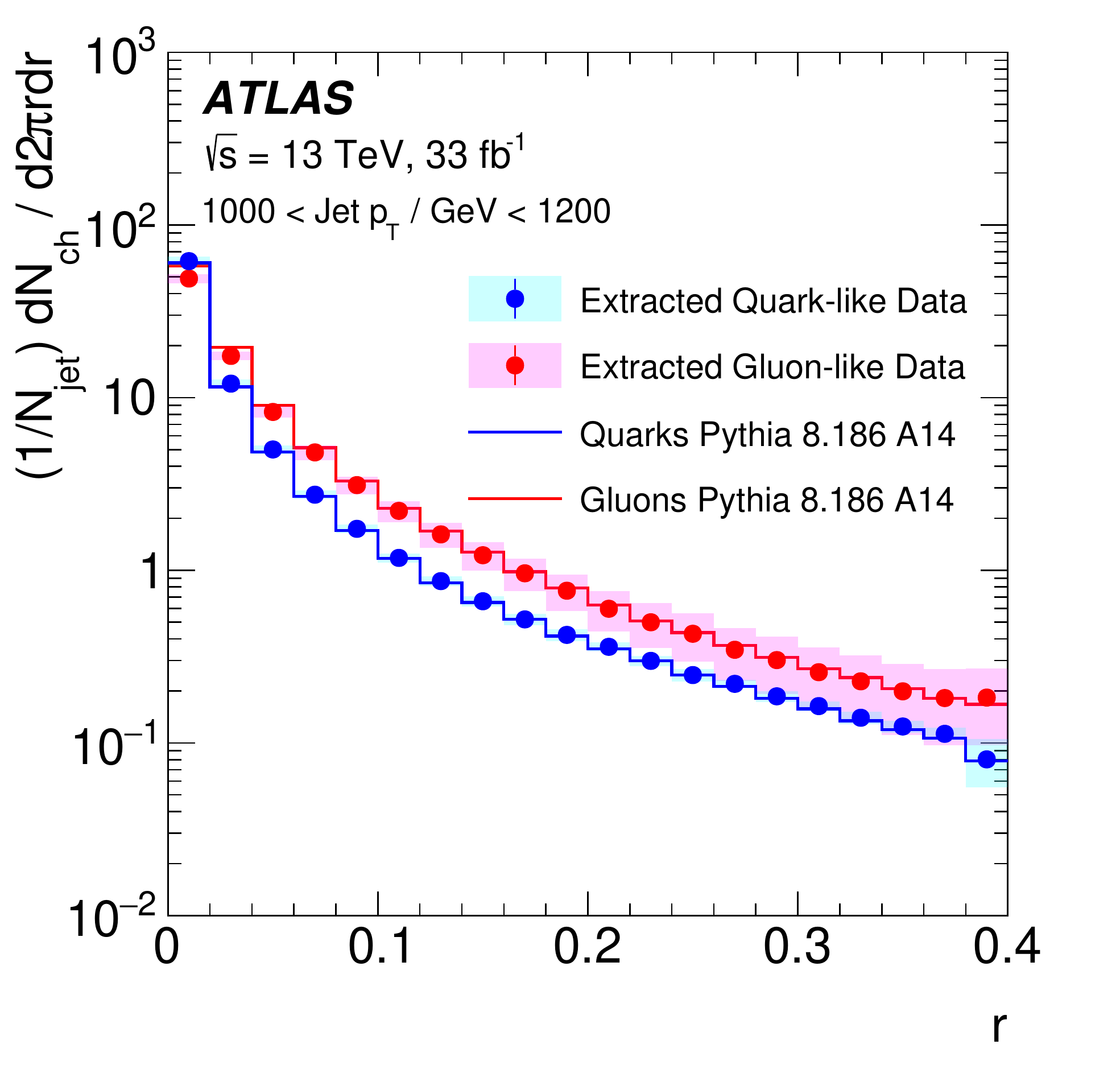}\\
\caption{The extracted quark- and gluon-like distributions of (top left) charged-particle multiplicity $n_\text{ch}$, (top right) transverse momentum fraction $\zeta$, (bottom right) transverse momentum $\pTrel$, and (bottom right) the radial profile in bins of the radial distance $r$ from the jet axis for jets with transverse momentum $1000$~\GeV\ $<\pTjet < 1200$~\GeV.  The quark- and gluon-jet distributions from \PYTHIA are also shown for comparison.  The uncertainty bands on the data include both statistical and systematic uncertainties while the error bars are due to statistical uncertainties only.}
\label{fig:extractedMCntrackdata}
\end{figure}
 
\begin{figure}[htbp]
\centering
\includegraphics[width=0.5\textwidth]{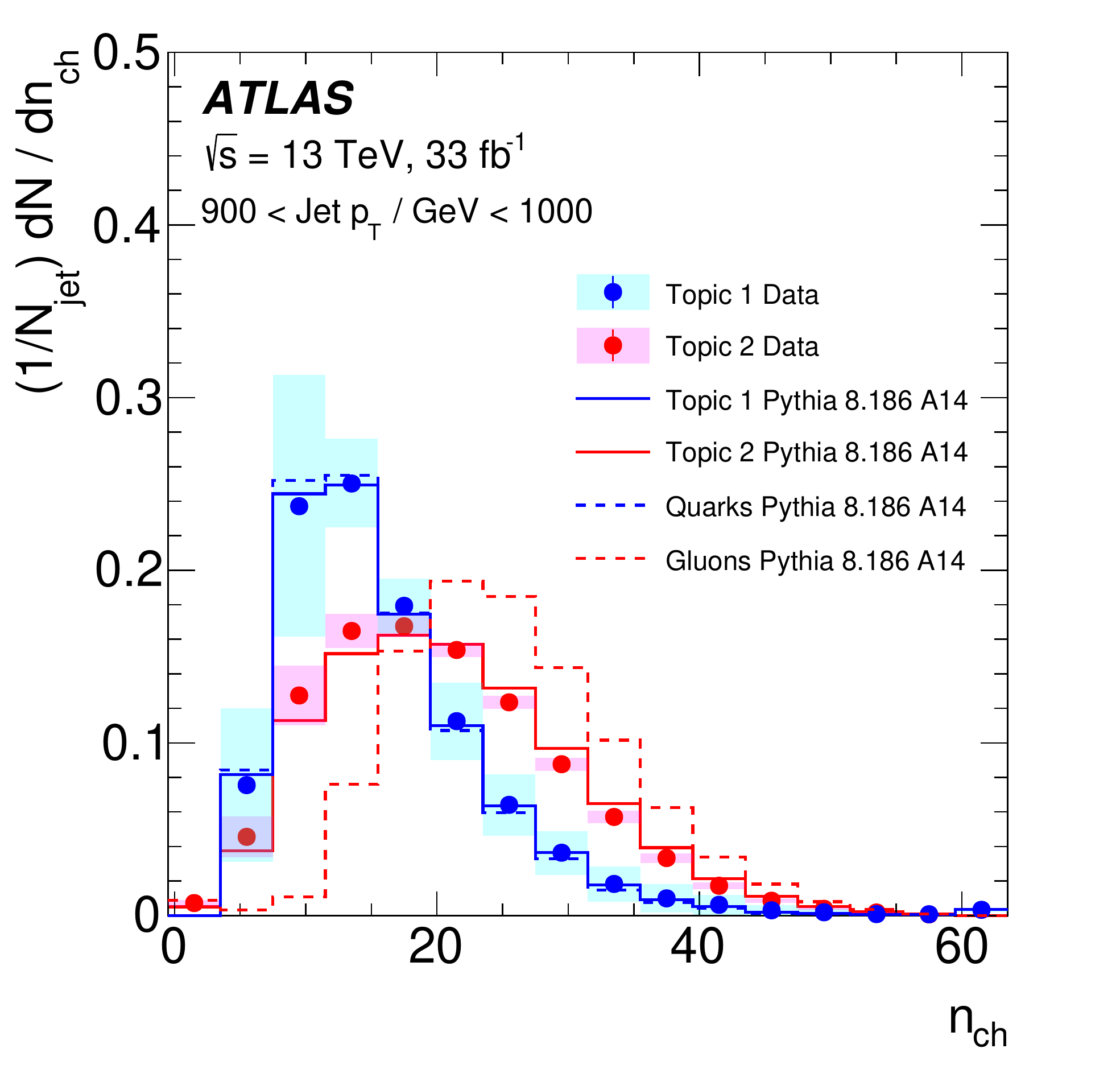}\includegraphics[width=0.5\textwidth]{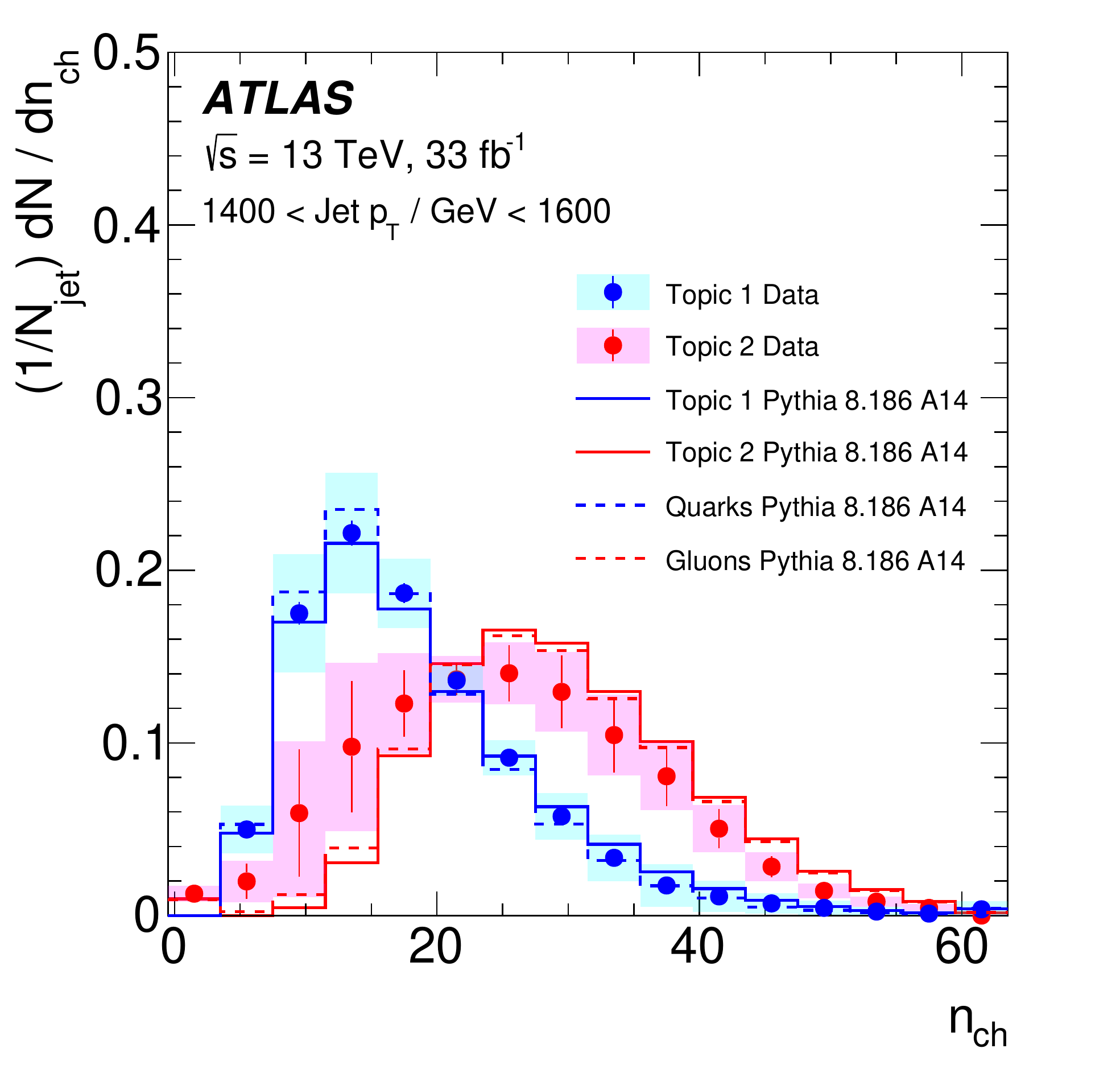}\\
\caption{The jet topics extracted using charged-particle multiplicity $n_\text{ch}$ for jets in two transverse momentum, $\pT$, bins together with the topics and quark and gluon distributions from \PYTHIA.  The uncertainty bands on the data include both statistical and systematic uncertainties while the error bars are due to statistical uncertainties only.}
\label{fig:extractedMCntrackdatatopics}
\end{figure}
 
While the full quark and gluon distributions presented in Figure~\ref{fig:extractedMCntrackdata} cannot be predicted from perturbative QCD, it is possible to model the $\pTjet$ dependence of the moments of the $\zeta$ distribution.  Positive moments of the fragmentation function have a perturbative evolution with a proper $\alphas$ power series via DGLAP-like equations.  In general, there are two terms that contribute to the right-hand side of Eq.~(\ref{eq:dglap}) that prevent an analytic solution: one term proportional to $D_g^h$ and one term proportional to $D_q^h$, where the coefficients for the $\kappa$ sums are the Mellin transforms $\tilde{P}_{p'\leftarrow p}(\kappa)=\int_0^1d\zeta\zeta^\kappa P_{p'\leftarrow p}(\zeta)$ for $p'=g$ and $p'=q$, respectively.  For gluon jets ($p=g$), the $g\rightarrow qq'$ splitting function is finite,\footnote{The splitting function $\tilde{P}_{q\leftarrow q}$ is also finite, but is not numerically small compared with $\tilde{P}_{g\leftarrow q}$ except when $\kappa$ is very small so this case is not considered further.} so $|\tilde{P}_{g\leftarrow g}|\gg|\tilde{P}_{q\leftarrow g}|$ for $\kappa\neq0.8$ where $\tilde{P}_{g\leftarrow g}$ switches sign.  Therefore, away from $\kappa\approx 0.8$ and in the modified leading-logarithm approximation (MLLA)\footnote{This means resummation that includes the leading-order splitting functions and the first-order running of the strong coupling.  A more refined calculation~\cite{Krohn:2012fg,Waalewijn2012sv} using SCET~\cite{Bauer:2001yt,Bauer:2001ct,Bauer:2000yr,Bauer:2000ew} and fragmenting jet functions~\cite{Procura:2011aq,Jain:2011xz,Procura:2009vm} is possible.  However, the deviations from this simple approach are higher-order corrections and do not qualitatively change the comparisons in this section.}~\cite{pinkbook,Khoze:2000iq,PerezRamos:2005nh},
 
\begin{align}
\label{eq:scalevaiolation}
\left\langle \sum_{i\in\text{jet}}\zeta_i^\kappa \right\rangle_\text{gluons}\hspace{-4mm}(\pT) \approxprop \log(\pT^2/\Lambda^2)^{2P_{g\leftarrow g}(\kappa)/\beta_0},
\end{align}
 
where $\beta_0$ is the first term in the QCD $\beta$-function and $\Lambda$ is a non-perturbative parameter (of order $\Lambda_\text{QCD}$).  The predictions are scaled to match the data in the sixth jet $\pT$ bin (referred to as the `anchor bin').  There is no a priori reason to select any particular bin as the anchor bin so one of the first bins after the lowest-threshold unprescaled jet trigger is selected.  Figure~\ref{fig:results:pgg} shows the distributions of the average $\sum_{i\in\text{jet}}\zeta_i^\kappa$ for $\kappa=0.5, 1.0$, and $2.0$ for gluon jets.  As mentioned above, $\tilde{P}_{g\leftarrow g}$ is predicted to change sign at $\kappa=0.8$, a trend which is supported by the data: for low $\kappa$, the average value increases with $\pT$ and when $\kappa$ is large, the average decreases with $\pT$.  For $\kappa=1$, momentum conservation and isospin symmetry predict that the average value of $\sum_{i\in\text{jet}}\zeta_i$ should be constant and approximately $2/3$, the ratio of charged pions to all pions\footnote{The measured value is not exactly $2/3$ because a jet's energy is only about 60\% due to pions.}.  The leading-logarithm (LL) calculation predicts $\tilde{P}_{g\leftarrow g}(1)\approx 0$ so the $\pT$ dependence is already negligible compared with the $\kappa=0.5$ and $\kappa=2$ cases.
 
\begin{figure}[htbp]
\centering
\includegraphics[width=0.48\textwidth]{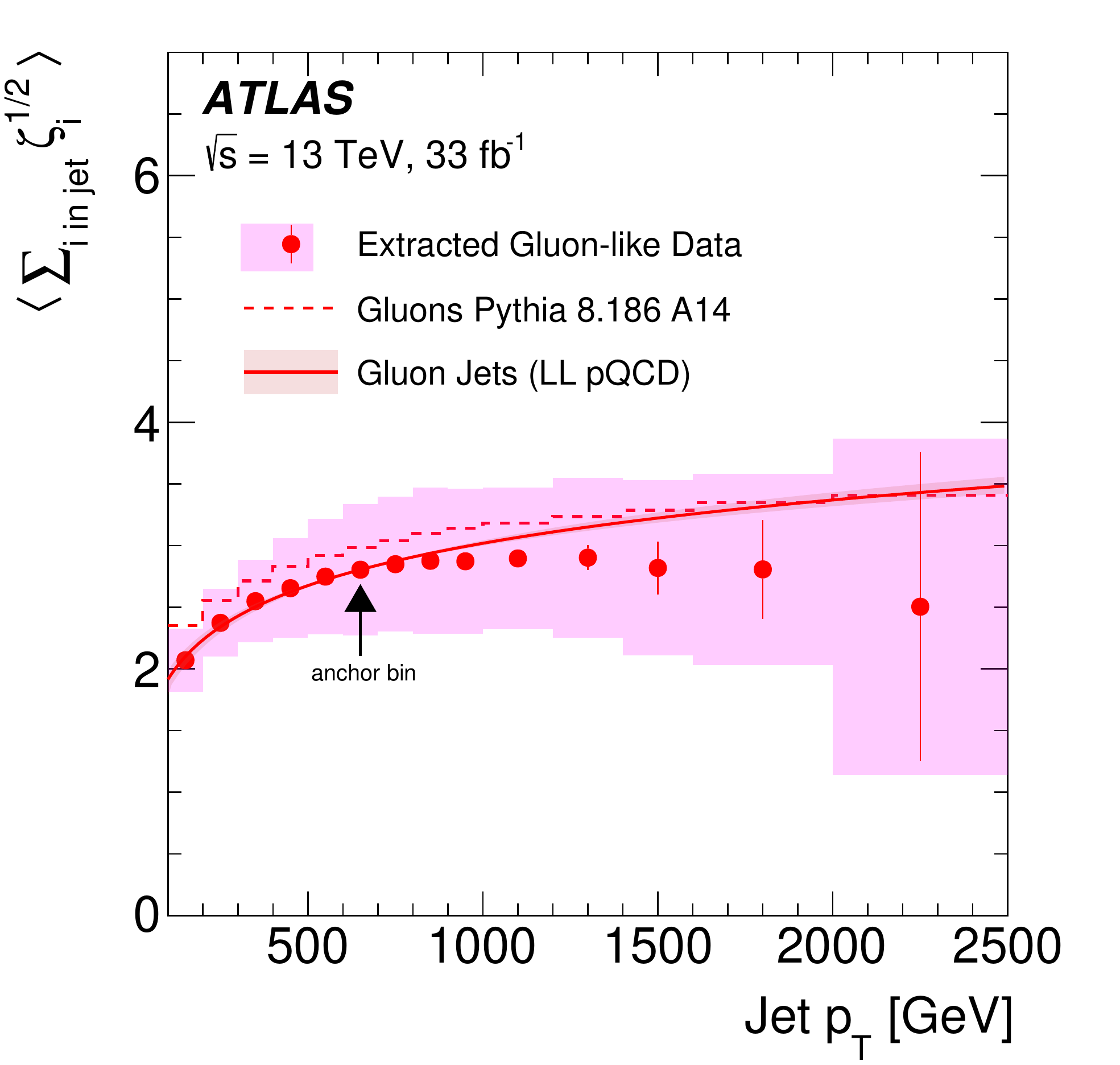}\includegraphics[width=0.48\textwidth]{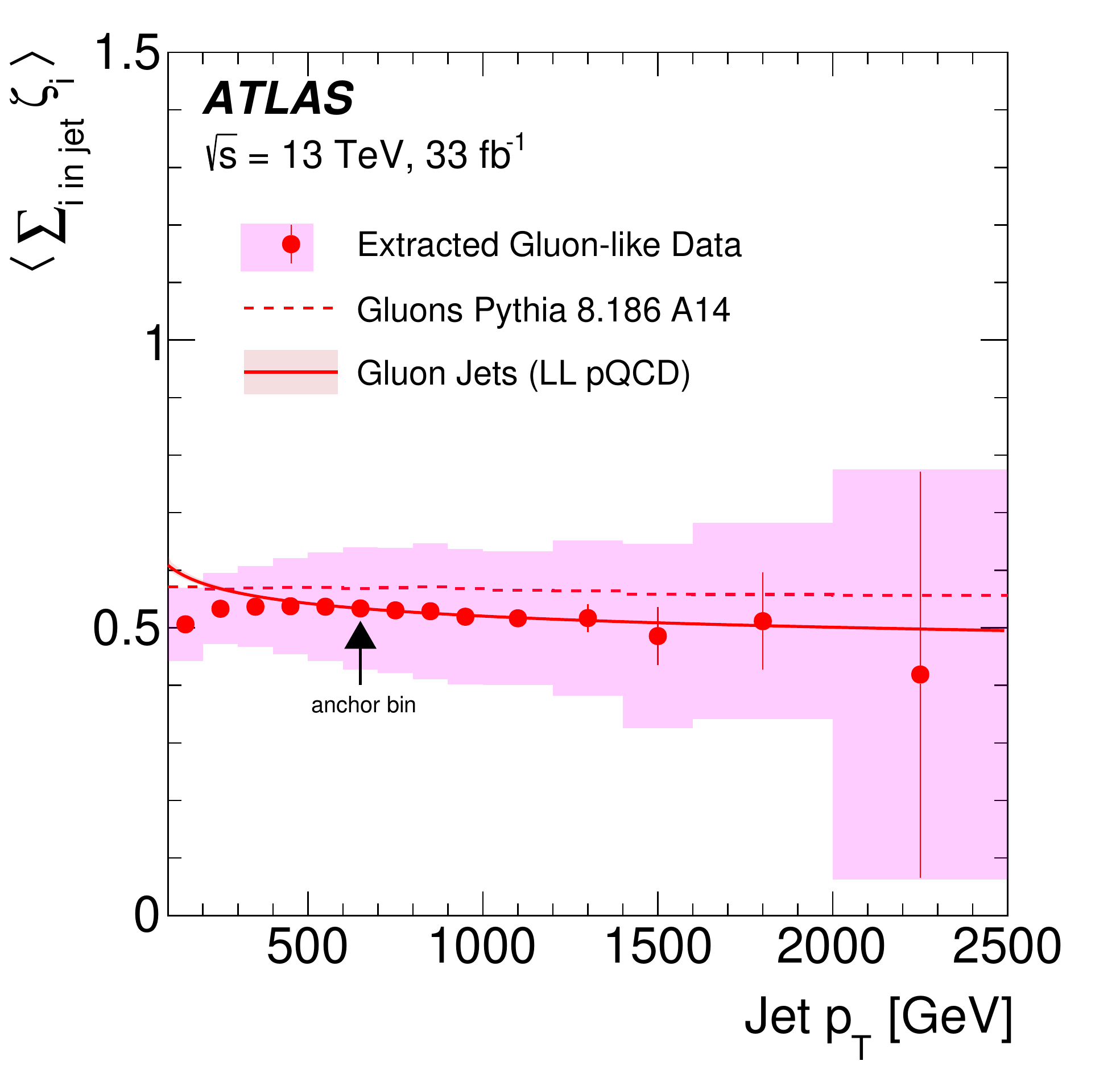}\\
\includegraphics[width=0.48\textwidth]{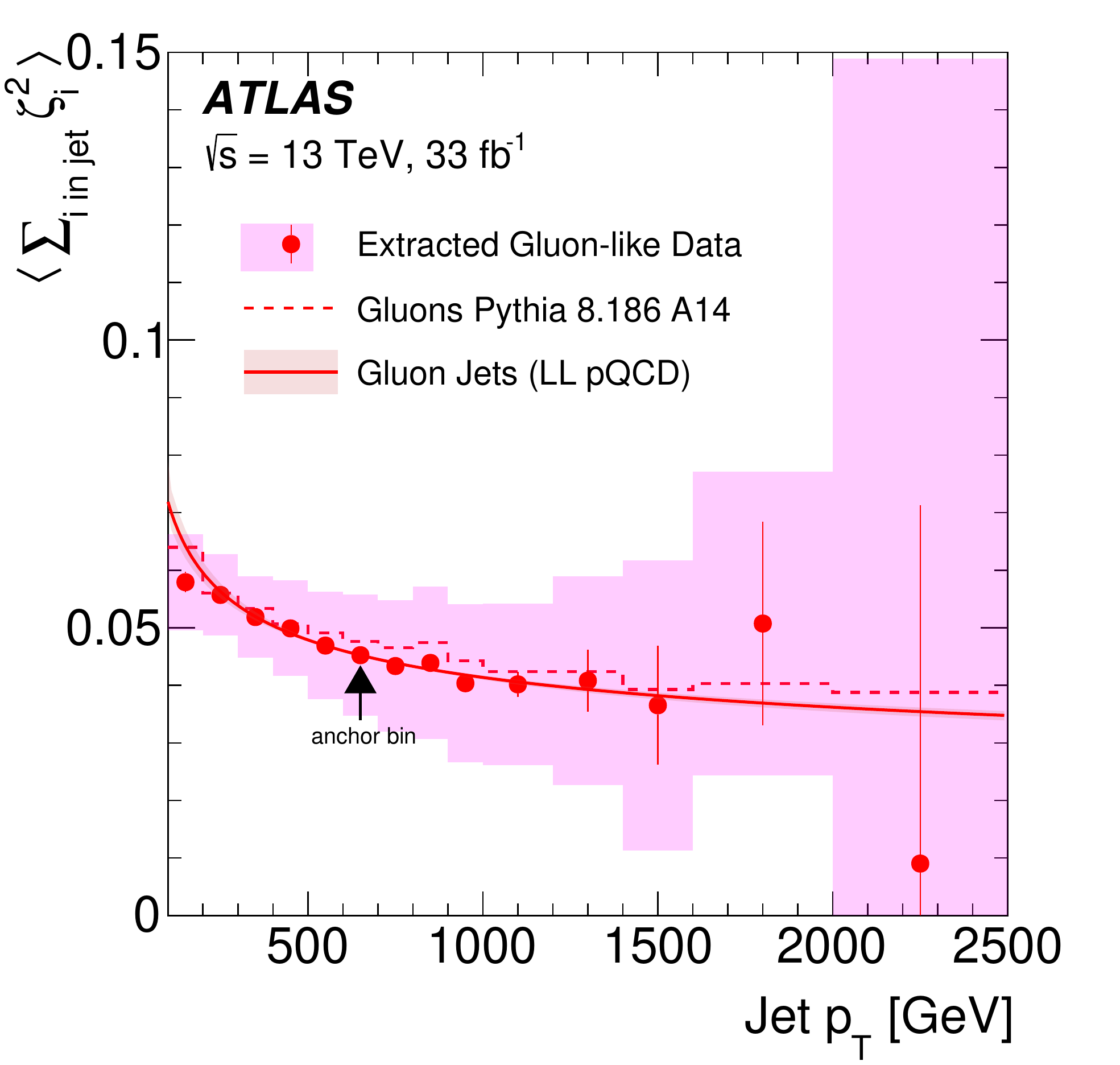}
\caption{The dependence on transverse momentum $\pT$ of the average extracted gluon-like transverse-momentum-fraction weighted sum $\sum_{i\in\text{jet}}\zeta_i^\kappa$ for (top left) $\kappa=0.5$, (top right) $\kappa=1.0$, and $\kappa=2.0$ (bottom).  For comparison, the results from \PYTHIA and a simple leading-logarithm (LL) calculation are also presented.  The prediction is normalized to the data in the sixth jet $\pT$ bin, called the anchor bin and indicated by an arrow.  The uncertainty band on the calculation is from varying $\Lambda$ in Eq.~(\ref{eq:scalevaiolation}) up and down by a factor of two from its nominal value of 400~\MeV\ (in most regions, this band is not much wider than the line width and thus not visible).  The uncertainty bands on the data include both statistical and systematic uncertainties while the error bars are due to statistical uncertainties only.}
\label{fig:results:pgg}
\end{figure}
 
When $\kappa\rightarrow 0$, both the quark and gluon fragmentation-function Mellin transforms diverge and so the analysis with Eq.~(\ref{eq:scalevaiolation}) is not accurate.  The $\kappa\rightarrow 0$ limit is $\langle n_\text{ch}\rangle$ and there is no known series in $\alphas$ to describe its $\pTjet$ dependence.  Despite this, the anomalous dimension for the $\pTjet$ dependence of $\langle n_\text{ch}\rangle$ has been calculated to `N$^3$LO' where the series is in $\sqrt{\alphas}$ instead of $\alphas$~\cite{Capella:1999ms,Dremin:1999ji}.  Figure~\ref{fig:extractedMCntrackmean} shows $\langle n_\text{ch}\rangle$ as a function of $\pTjet$ for both extracted quark-like and gluon-like jets as well as the topic extraction along with the prediction for the pure quark/gluon case.   Gluon jets from data deviate significantly from simulation and from the calculation at high jet $\pT$; this is also true to a lesser extent for quark jets, which seem to have a different slope that is most prominent at low jet $\pT$.  A similar trend was first observed in Ref.~\cite{Aad:2016oit}, albeit with lower precision in the highest $\pT$ bins.  There are several possibilities for this discrepancy, such as an unaccounted for potential source of bias in the quark/gluon jet fractions.  The data in the right panel of Figure~\ref{fig:extractedMCntrackmean} do not yet conclusively support or reject this hypothesis; with more data, it may be possible to determine if the data match topic 2 in \PYTHIA or deviate as is the case for gluons in the left panel of Figure~\ref{fig:extractedMCntrackmean}.
 
\begin{figure}[h!]
\centering
\includegraphics[width=0.45\textwidth]{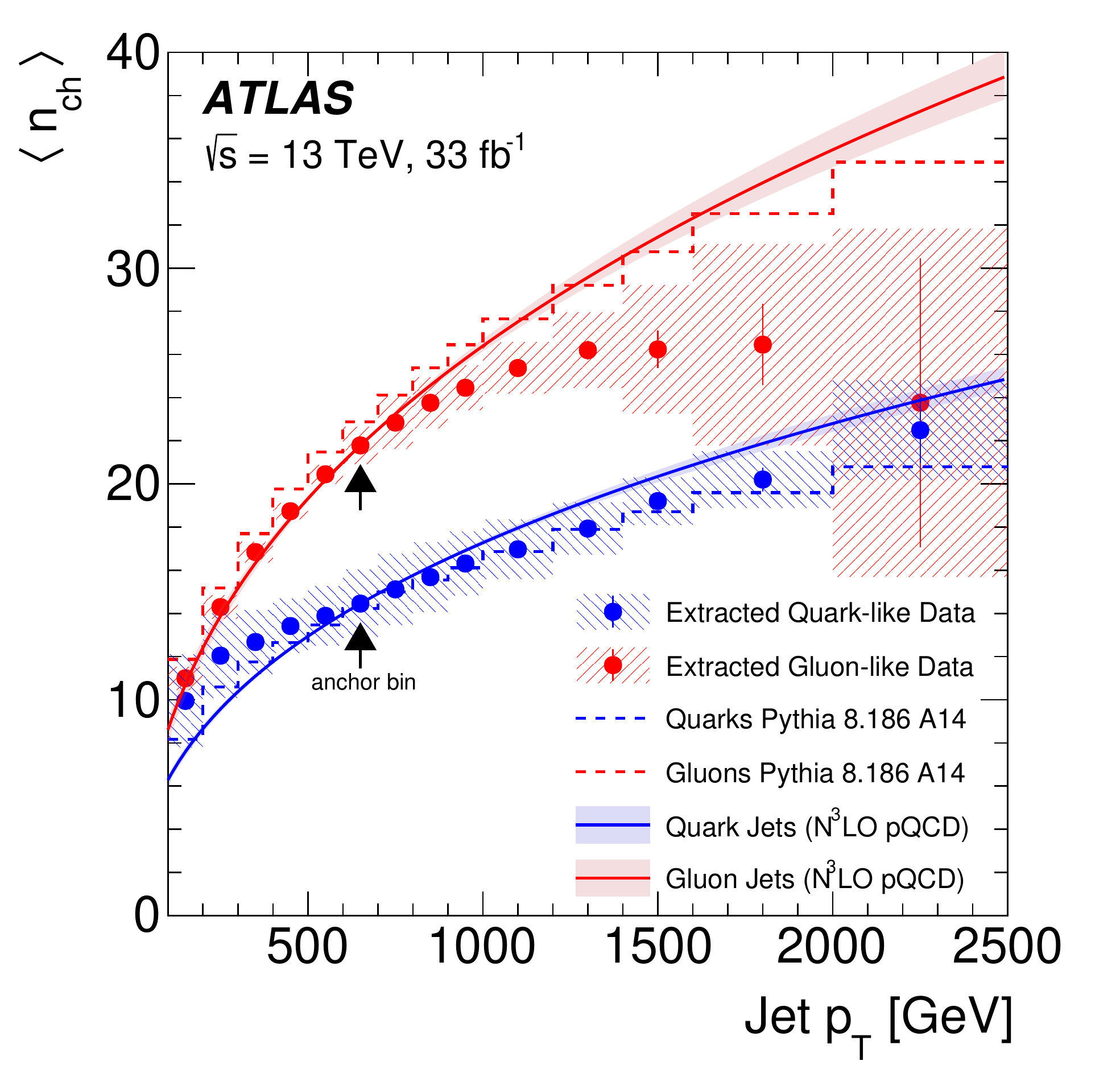}\includegraphics[width=0.45\textwidth]{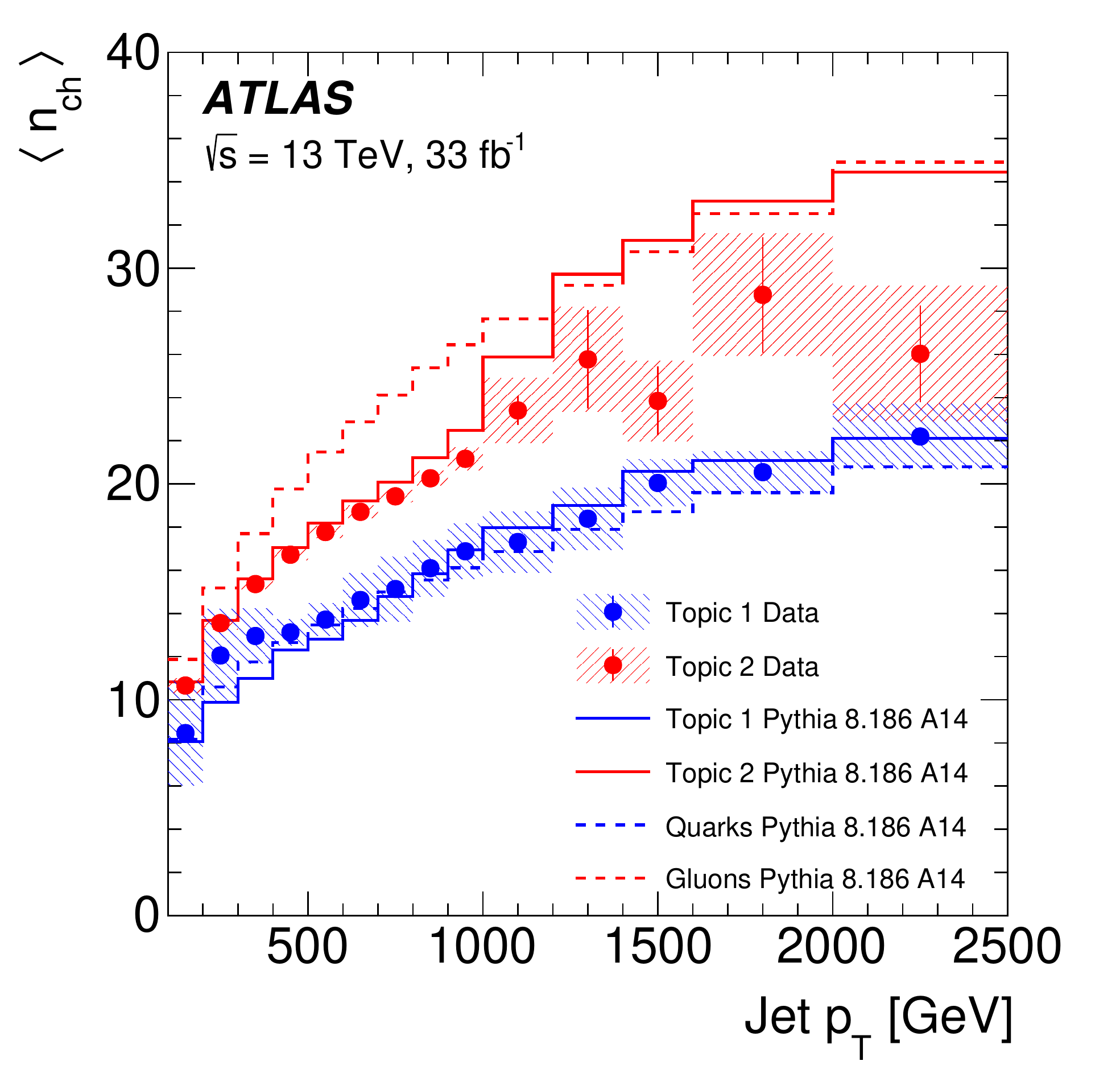}\\
\caption{Left (right): The dependence on jet transverse momentum $\pT$ of the mean charged-particle multiplicity $\langle n_\text{ch}\rangle$ for quark and gluon jets (topic 1 and topic 2) in data and in \PYTHIA as well as from a calculation using perturbative QCD.  The calculation cannot predict the overall normalization and therefore the prediction is normalized to the data in the sixth $\pT$ bin, called the anchor bin and indicated by an arrow.  The binning correction is not applied to the average topics, as this correction is very sensitive to fluctuations due to the limited number of simulated events.  The uncertainty bands on the data include both statistical and systematic uncertainties while the error bars are due to statistical uncertainties only.}
\label{fig:extractedMCntrackmean}
\end{figure}
 
The $\pT$ dependence of the average $\zeta$, $\pTrel$, and $r$ are shown in Figure~\ref{fig:extractedMCfragfuncmean}.  Gluon jets have more constituents than quark jets on average so their average $\zeta$ is lower.  For both quark and gluon jets, $\langle\zeta\rangle$ decreases with the jet $\pT$ in part because constituent multiplicity increases with $\pT$.  Gluon jets are wider than quark jets on average, but both quark and gluon jets become denser with increasing jet $\pT$.  The data show nearly the same trends as \PYTHIA in all cases.
 
\begin{figure}[h!]
\centering
\includegraphics[width=0.45\textwidth]{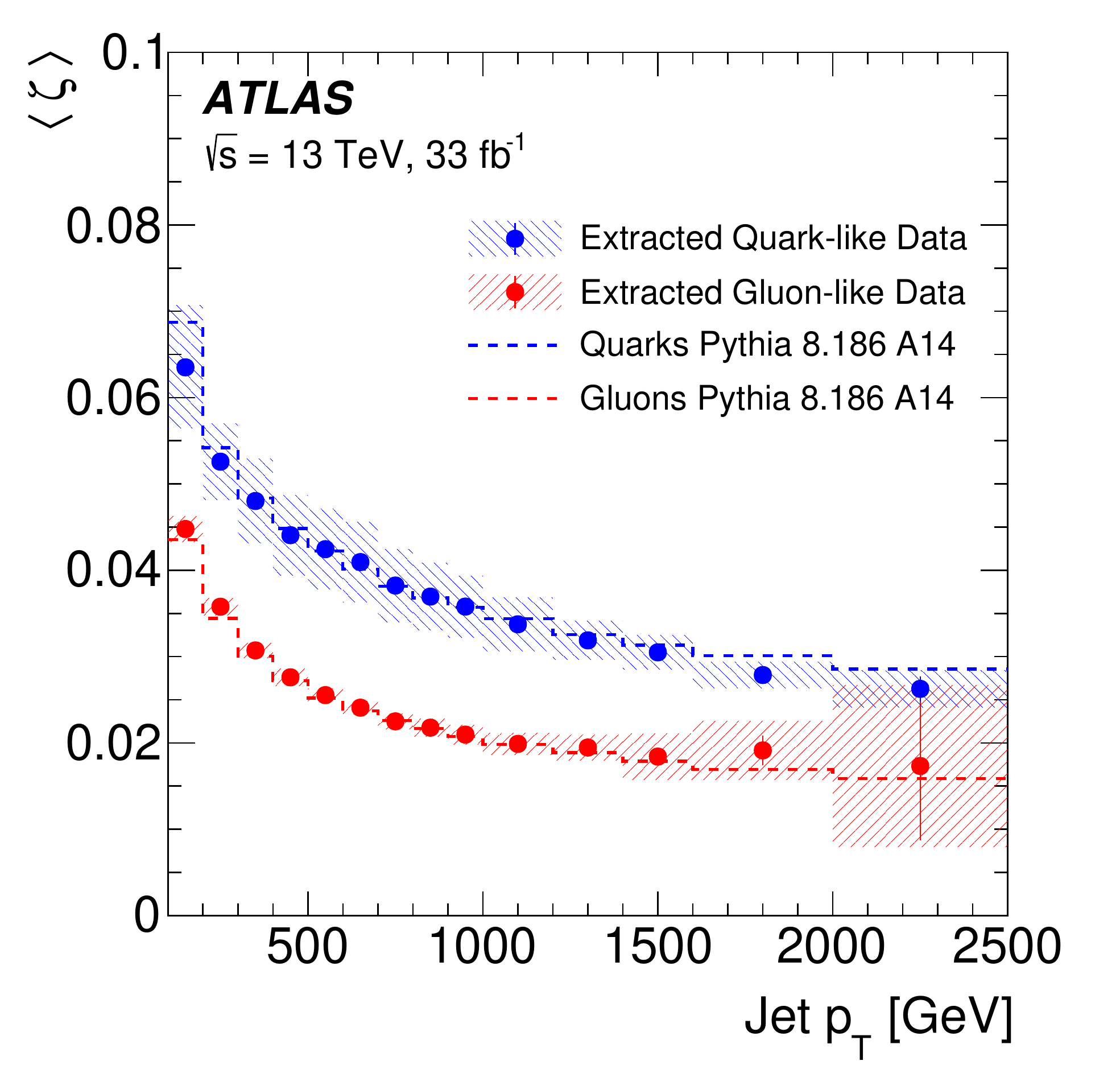}\includegraphics[width=0.45\textwidth]{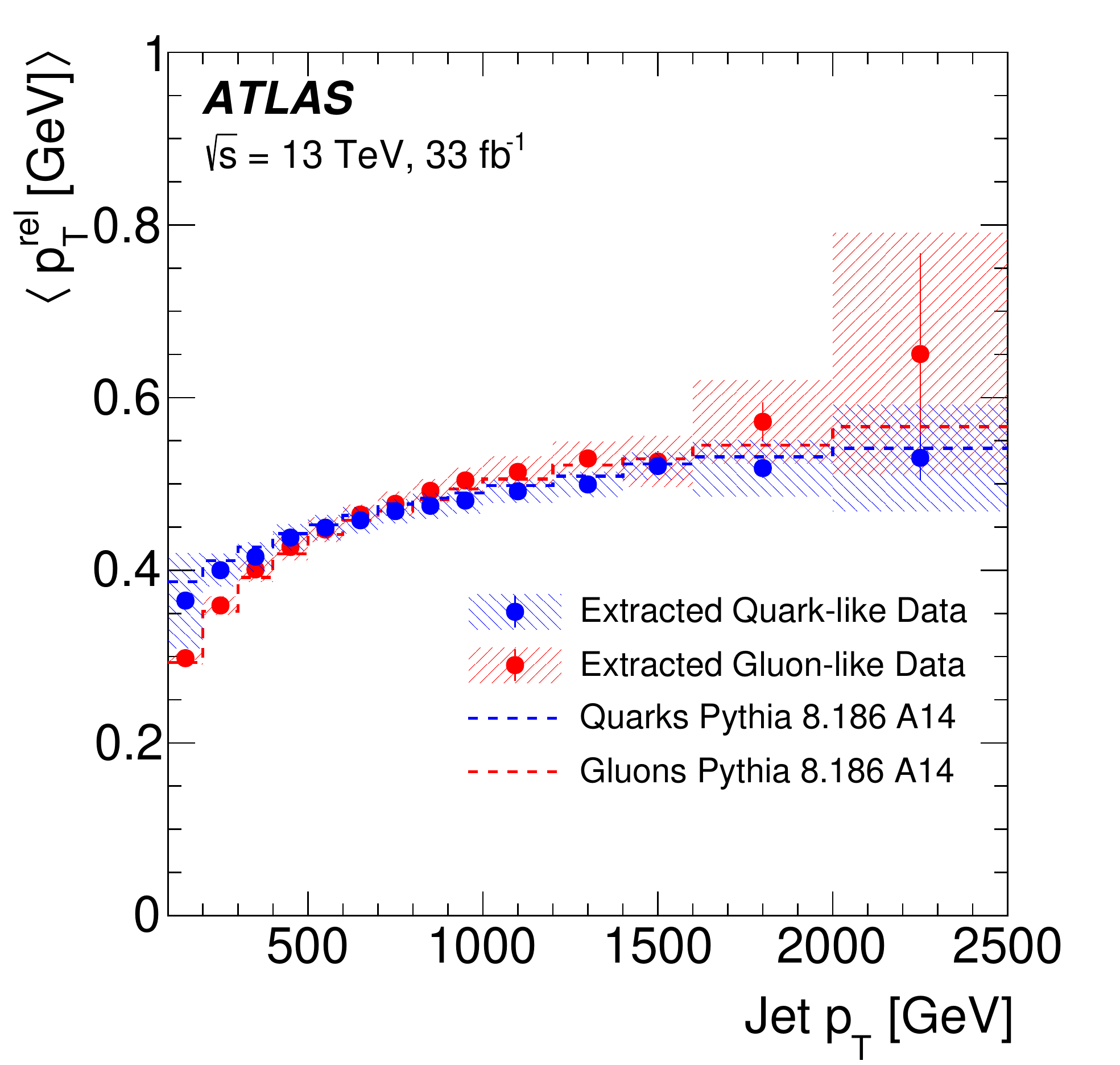}\\
\includegraphics[width=0.45\textwidth]{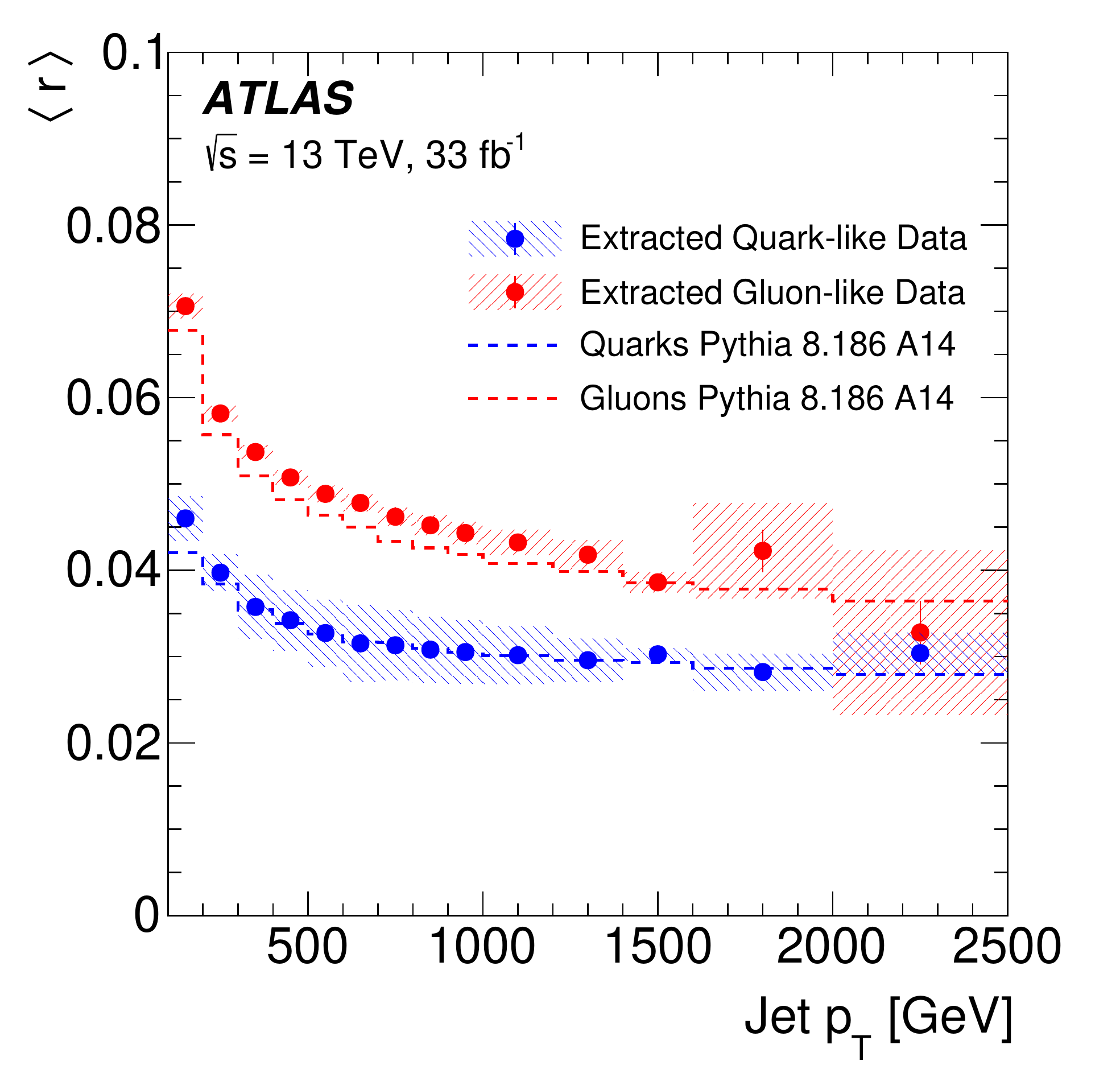}
\caption{The dependence on jet transverse momentum $\pT$ of the average extracted quark and gluon (top left) transverse momentum fraction $\zeta$, (top right) transverse momentum $\pTrel$, and (bottom) radial profile (per particle and not per jet). The uncertainty bands on the data include both statistical and systematic uncertainties while the error bars are due to statistical uncertainties only.}
\label{fig:extractedMCfragfuncmean}
\end{figure}
 
\FloatBarrier
 
\section{Conclusion}
\label{sec:conclusion}
 
This paper documents a measurement of track-based jet fragmentation functions in $pp$ collisions at $\sqrt{s} = 13$~\TeV.  The analysis uses a dataset corresponding to an integrated luminosity of 33~\ifb\ recorded by the ATLAS detector at the LHC.  Multiple jet properties, including the charged-particle multiplicity, the momentum fraction carried by charged particles, and angular properties of the radiation pattern inside jets are studied. There are key areas where there are significant disagreements between the ATLAS default MC simulation (\PYTHIA~8.2 with the A14 tune, \HERWIGpp, and \SHERPA) and the data, especially for the radial profiles and momentum distributions in \SHERPA.  The radial profile is systematically broader in data than in simulation, but the momentum transverse to the jet axis and the momentum fraction are well modeled within the precision of this measurement. Near 1~\TeV\ in jet $\pT$, these measurements have achieved percent-level uncertainties for a variety of observables. In addition to measuring the forward, central, and combined jet distributions, the forward and central jet spectra are considered separately to study quark- and gluon-like distributions.  A first measurement of topic modeling for the charged-particle multiplicity provides a promising alternative to traditional methods of extracting quark- and gluon-jet distributions that use input from simulation.  The simulations provide a reasonable description of the quark-like data across the jet $\pT$ range presented in this measurement, but the gluon-like data have systematically fewer charged particles than the simulations by about 10\%.
 
The unfolded data are made public through HepData to provide input to help improve both perturbative and non-perturbative aspects of fragmentation modeling in the future.
 
\section*{Acknowledgments}

We thank CERN for the very successful operation of the LHC, as well as the
support staff from our institutions without whom ATLAS could not be
operated efficiently.
 
We acknowledge the support of ANPCyT, Argentina; YerPhI, Armenia; ARC, Australia; BMWFW and FWF, Austria; ANAS, Azerbaijan; SSTC, Belarus; CNPq and FAPESP, Brazil; NSERC, NRC and CFI, Canada; CERN; CONICYT, Chile; CAS, MOST and NSFC, China; COLCIENCIAS, Colombia; MSMT CR, MPO CR and VSC CR, Czech Republic; DNRF and DNSRC, Denmark; IN2P3-CNRS, CEA-DRF/IRFU, France; SRNSFG, Georgia; BMBF, HGF, and MPG, Germany; GSRT, Greece; RGC, Hong Kong SAR, China; ISF and Benoziyo Center, Israel; INFN, Italy; MEXT and JSPS, Japan; CNRST, Morocco; NWO, Netherlands; RCN, Norway; MNiSW and NCN, Poland; FCT, Portugal; MNE/IFA, Romania; MES of Russia and NRC KI, Russian Federation; JINR; MESTD, Serbia; MSSR, Slovakia; ARRS and MIZ\v{S}, Slovenia; DST/NRF, South Africa; MINECO, Spain; SRC and Wallenberg Foundation, Sweden; SERI, SNSF and Cantons of Bern and Geneva, Switzerland; MOST, Taiwan; TAEK, Turkey; STFC, United Kingdom; DOE and NSF, United States of America. In addition, individual groups and members have received support from BCKDF, CANARIE, CRC and Compute Canada, Canada; COST, ERC, ERDF, Horizon 2020, and Marie Sk{\l}odowska-Curie Actions, European Union; Investissements d' Avenir Labex and Idex, ANR, France; DFG and AvH Foundation, Germany; Herakleitos, Thales and Aristeia programmes co-financed by EU-ESF and the Greek NSRF, Greece; BSF-NSF and GIF, Israel; CERCA Programme Generalitat de Catalunya, Spain; The Royal Society and Leverhulme Trust, United Kingdom.
 
The crucial computing support from all WLCG partners is acknowledged gratefully, in particular from CERN, the ATLAS Tier-1 facilities at TRIUMF (Canada), NDGF (Denmark, Norway, Sweden), CC-IN2P3 (France), KIT/GridKA (Germany), INFN-CNAF (Italy), NL-T1 (Netherlands), PIC (Spain), ASGC (Taiwan), RAL (UK) and BNL (USA), the Tier-2 facilities worldwide and large non-WLCG resource providers. Major contributors of computing resources are listed in Ref.~\cite{ATL-GEN-PUB-2016-002}.
 
\printbibliography
 
\clearpage % ATLAS Collaboration author list
% Reference date of STDM-2017-16 is 2019-03-05
% Author list last updated on date 19-JUN-19
% Data extracted on 19-Jun-2019 for paper reference STDM-2017-16
% at 9:47pm
 
\begin{flushleft}
{\Large The ATLAS Collaboration}

\bigskip

G.~Aad$^\textrm{\scriptsize 101}$,    
B.~Abbott$^\textrm{\scriptsize 128}$,    
D.C.~Abbott$^\textrm{\scriptsize 102}$,    
O.~Abdinov$^\textrm{\scriptsize 13,*}$,    
A.~Abed~Abud$^\textrm{\scriptsize 70a,70b}$,    
K.~Abeling$^\textrm{\scriptsize 53}$,    
D.K.~Abhayasinghe$^\textrm{\scriptsize 93}$,    
S.H.~Abidi$^\textrm{\scriptsize 167}$,    
O.S.~AbouZeid$^\textrm{\scriptsize 40}$,    
N.L.~Abraham$^\textrm{\scriptsize 156}$
% ...
\bigskip
\\

% ...
$^{13}$Institute of Physics, Azerbaijan Academy of Sciences, Baku; Azerbaijan.\\
% ...
$^{40}$Niels Bohr Institute, University of Copenhagen, Copenhagen; Denmark.\\
% ...
$^{53}$II. Physikalisches Institut, Georg-August-Universit\"{a}t G\"ottingen, G\"ottingen; Germany.\\
% ...
$^{70}$$^{(a)}$INFN Sezione di Pavia;$^{(b)}$Dipartimento di Fisica, Universit\`a di Pavia, Pavia; Italy.\\
% ...
$^{93}$Department of Physics, Royal Holloway University of London, Egham; United Kingdom.\\
% ...
$^{101}$CPPM, Aix-Marseille Universit\'e, CNRS/IN2P3, Marseille; France.\\
$^{102}$Department of Physics, University of Massachusetts, Amherst MA; United States of America.\\
% ...
$^{128}$Homer L. Dodge Department of Physics and Astronomy, University of Oklahoma, Norman OK; United States of America.\\
% ...
$^{156}$Department of Physics and Astronomy, University of Sussex, Brighton; United Kingdom.\\
% ...
$^{167}$Department of Physics, University of Toronto, Toronto ON; Canada.\\
% ...
$^{a}$ Also at CERN, Geneva; Switzerland.\\
$^{b}$ Also at CPPM, Aix-Marseille Universit\'e, CNRS/IN2P3, Marseille; France.\\
% ...
$^{*}$ Deceased

\end{flushleft}

% Created with Glance <Atlas.Glance@cern.ch>

\end{document}